\documentclass[%reprint,
nofootinbib,
twocolumn,
 amsmath,amssymb,
 aps,
]{revtex4-1}
\usepackage{amsmath}
\usepackage{amssymb}
\usepackage{bm}
\usepackage{color,graphicx}
\usepackage{slashed}
\usepackage{comment}
\usepackage{float}
\newcommand{\myprod}[2]{(#1\,#2)}

\newcounter{comment}

\begin{document}

\title{Theory of Deeply Virtual Compton Scattering off the Unpolarized Proton}

\author{Brandon Kriesten} 
\email{btk8bh@virginia.edu}
\affiliation{Department of Physics, University of Virginia, Charlottesville, VA 22904, USA.}

\author{Simonetta Liuti} 
\email{sl4y@virginia.edu}
\affiliation{Department of Physics, University of Virginia, Charlottesville, VA 22904, USA.}
%\affiliation{Laboratori Nazionali di Frascati, INFN, Frascati, Italy}

%\pacs{13.60.Hb, 13.40.Gp, 24.85.+p}

\begin{abstract}
Using the helicity amplitudes formalism, we study deeply virtual exclusive electron photoproduction  off an unpolarized nucleon target, $ep \rightarrow e' p' \gamma$, through a range of kinematics both in the fixed target setting with  initial electron energies of 6 GeV, 11 GeV and 24 GeV, and for an electron ion collider.  
We reformulate the cross section bringing to the forefront the defining features of the $ep \rightarrow e'p'\gamma$ process, where the observables are expressed as bilinear products of the independent helicity amplitudes which completely describe it in terms of the electric, magnetic and axial currents of the nucleon. 
%This structure is reflected in different ways in the Deeply virtual Compton scattering and Bethe-Heitler parts of the cross section.
%
These contributions are checked against the Fourier harmonics-based formalism which has provided so far the underlying mathematical framework to study Deeply virtual Compton scattering and related experiments. Using theoretical model calculations of the twist-two generalized parton distributions, $ H$, $E$, $\widetilde{H}$ and $\widetilde{E}$, 
we uncover large discrepancies between the harmonic series and our proposed framework. Most importantly, these numerical differences appear in the intermediate $Q^2$ range which represents a sweet spot for extracting generalized parton distributions from data. 
{We provide a framework that is ideal, on one side, to study  and compare the different conventions  that can be used to describe the leading order contribution to DVCS in QCD, while on the other, it facilitates a quantitative extraction of physically meaningful information from experiment  through traceable and controllable approximations in the intermediate $Q^2$ region.}
 
\end{abstract}

\maketitle

\section{Introduction}
\label{sec:intro}
Deeply Virtual Compton Scattering (DVCS) is measured through the exclusive process, $ep \rightarrow e'p' \gamma $, where, in the one photon exchange approximation, the virtual photon four-momentum squared, $Q^2$,  provides a hard scale for the process. Quantum Chromodynamics (QCD) factorization theorems allow us to single out the perturbative, short distance reaction from the non-perturbative, long distance
matrix elements described in terms of generalized parton distributions (GPDs) \cite{Ji:1996ek,Ji:1996nm}. 
%%%
%%%
\footnote{Detailed proofs of factorization for deeply virtual exclusive processes can be found in  \cite{Collins:1998be,Ji:1997nk,Ji:1998xh}).}
%%%
%%%
GPDs encode new information on the internal dynamics of the proton that will allow us, through a combined analysis of experimental data and lattice QCD results, to ultimately map out its 3D structure. These distributions enter the observables embedded in the Compton Form Factors (CFFs), which are convolutions over the longitudinal momentum variable $x$ with complex QCD Wilson coefficient functions (see reviews in \cite{Diehl:2003ny,Belitsky:2005qn,Kumericki:2016ehc}). 
At leading order, four quark chirality conserving GPDs giving eight CFFs,  describe all possible quark ($P_q$) proton ($P_p$) polarization configurations, $P_q P_p=UU, LL, UT, LT$, allowed by parity conservation, time reversal invariance and charge conjugation. 
%These $H (UU), E (UT), \widetilde{H} (LL), \widetilde{E} (LT)$. 
%The total set of kinematic variables that can be measured directly from experiment  $\xi$, $t$, and $Q^2$.
The eight CFFs appear simultaneously in all of the deeply virtual exclusive scattering experimental observables, independent of the specific beam-target polarization configuration.  
This poses a challenge for the extraction of CFFs from experiment which are affected by large theoretical uncertainties (see \cite{Kumericki:2016ehc} for a detailed list of experiments).

In Ref.\cite{Kriesten:2019jep} we introduced a 
new theoretical formulation of the cross section for the $ep \rightarrow e' p'\gamma$ process in all polarization configurations for the incoming electron and proton target.  
The main goal of Ref.\cite{Kriesten:2019jep}   {was to provide a formulation of the cross section in terms of CFFs}
%exact and complete calculation of all terms entering the cross section, 
that allowed one to evaluate {precisely} the impact of order $1/Q$ and higher power corrections of kinematic and dynamical origin.

Focusing on the unpolarized target case, in this paper we illustrate how the complete calculation of the cross section  leads to a more direct interpretation in terms of the electric, magnetic and axial current contributions to DVCS.
In our approach, the BH-DVCS interference term of the $ep \rightarrow e' p' \gamma$ cross section exhibits a structure analogous to the cross section for $ep$ elastic scattering  (see {\it e.g.} Refs.\cite{Perdrisat:2006hj,Gao:2003ag} and references therein),   preserving
the structure of the nucleon charge, magnetic, and axial current contributions to the cross section from the underlying helicity amplitudes configurations \cite{Arens:1996xw,Diehl:2005pc}.
 
{Previous studies, referring in particular to the widely adopted formalism of Refs. \cite{Belitsky:2000gz,Belitsky:2001ns,Belitsky:2010jw,Belitsky:2005qn,Belitsky:2012ch} (BKM) are organized, instead, in terms of harmonics of the azimuthal angle, $\phi$, and in kinematic powers of $1/Q$.}
On one side, the harmonics-based formalism  presents the appealing aspect of associating a dominant harmonic for each observable, therefore providing a simplified framework for experimental measurements. While this approach was perhaps needed in the pioneering analyses of the HERMES era \cite{Burkardt:2008jw}, 
at present, with Jefferson Lab @12 GeV and the future Electron Ion Collider (EIC) we are now entering a precision era where information on the physics content of the deeply virtual cross section, in particular its electroweak structure can be studied. 
{This structure, in turn, bears important consequences for building a phenomenological framework to quantitatively study the composition of angular momentum, mass and other mechanical properties of the proton in terms of quark and gluon contributions. 

Our new framework radically changes the extraction of CFFs from data \cite{Kriesten:2020apm}. In Refs.\cite{Belitsky:2001ns,Belitsky:2010jw}, according to the harmonics expansion,  the magnetic contribution containing the angular momentum related CFF combination $(H+E)$  
%and part of the axial contribution 
to the BH-DVCS interference term, was deemed as power-suppressed and, therefore, not included in the leading order formula. Following Ref.\cite{Kriesten:2020apm}, we introduce linear fits of the DVCS data that enable us to extract the magnetic contribution for the first time, with a relatively small size error.}

While our approach shares a common starting point with the initial studies of the cross section performed in Refs.\cite{Guichon:1998xv,Vanderhaeghen:1998uc,Vanderhaeghen:1999cas}, our results are in line with the recent study  in Refs.\cite{Braun:2011dg,Braun:2012hq,Braun:2014sta}, broadly labeled as ``finite $t$ and target mass corrections", in that we 
recognize the importance of kinematic contributions originating from the choice of reference frames where the QCD hadronic tensor and BH proton current are evaluated. 

  {A further advantage of the new formalism is that it provides compact expressions for the various kinematic coefficients expressed in terms of invariant four-vector products. 
From the practical point of view, we provide simplified expressions that can be readily used in the development of simulations and pseudo-data, for both fixed target and collider settings. A striking example is given by the form of the BH unpolarized cross section which reduces, in our case, to two lines of computation.} 
\footnote{Our results are consistent with a previous calculation for the unpolarized BH cross section given in terms  of Mandelstam invariants Ref.\cite{Guichon:1998xv,Vanderhaeghen:1998uc,Vanderhaeghen:1999cas}.}
The DVCS contribution is organized in terms of structure functions for the various polarization configurations for the lepton beam and nucleon target, in line with the general cross section formulations of Refs.\cite{Diehl:2005pc,Bacchetta:2006tn}.

The work presented in this paper is organized in the following points:

\noindent {\it i)} we discuss in detail the physics content of the expressions derived in Ref.\cite{Kriesten:2019jep}, including the origin of the phase dependence, and the polarization configurations for both the twist two and twist three contributions (Section \ref{sec:2});

\noindent {\it ii)} we perform a detailed numerical comparison with the formalism of BKM for the unpolarized cross section including the BH, DVCS and BH-DVCS interference terms.
The comparison is valid up to twist three, using the same model calculation of the GPDs for both the presents paper's and the BKM expressions. This ensures that the differences can be ascribed entirely to the formalism, (Section \ref{sec:3}); 

\noindent {\it iii)} we cover a range of kinematic regions, from Jefferson Lab @6 GeV and @12 GeV, to a hypothetical energy value of 24 GeV fixed target configuration \cite{Bogasz} to the EIcC \cite{Anderle:2021wcy} and EIC \cite{eicyellowreport}. All graphs are shown and discussed in Sec. \ref{sec:3}. 

\noindent Finally, we write our Conclusions and Outlook in Section \ref{sec:4}.
%

%%%%%%%%
%%%%%%%%
%%%%%%%% SECTION 2
\section{Unpolarized Scattering Cross Section}
\label{sec:2}
The cross section for the deeply virtual photon electroproduction process, $e(k) + p \rightarrow e^\prime(k^\prime) + p' + {\gamma'}(q')$, on an unpolarized proton, is derived from a
coherent superposition of the DVCS and Bethe-Heitler (BH) amplitudes, where
%%%%%%%%% BH
%%%%%%%%%%
%\subsection{Bethe-Heitler process, $\sigma_{UU}^{BH}$}
%\label{sec:BH}
the BH contribution arises when the final photon is emitted from either the initial or final electron. It is therefore a competing mechanism to the DVCS process evaluated at low momentum transfer, $-t<<Q^2$. One has, 
\begin{eqnarray}
\frac{d^5\sigma}{d x_{Bj} d Q^2 d|t| d\phi } &=&
\Gamma \,  
\big|T\big|^2 =  \Gamma |T_{\rm BH} + T_{\rm DVCS}|^2 \nonumber \\
&=& \Gamma \left( |T_{\rm BH}|^2 + |T_{\rm DVCS}|^2 + \mathcal{I} \right) ,
\label{eq:xs5foldgeneral}
\end{eqnarray}
the interference term, $\mathcal{I} $,  being defined as,
\begin{eqnarray}
\label{interf}
\mathcal{I} & = & T_{BH}^{*} T_{DVCS}
+ T_{DVCS}^{*} T_{BH} ,
\end{eqnarray}
where $\Gamma$ is the flux factor,
\begin{equation}
\label{eq:Gamma}
\Gamma = \frac{\alpha^3}{16\pi^2 (s-M^2)^2 \sqrt{1+\gamma^2}\, x_{Bj} } \, ,
\end{equation}
$\alpha$ is the electromagnetic fine structure constant; we define the relativistic invariants,
$Q^2 = -(k-k')^2$, $t=\Delta^2 =(p'-p)^2$, $x_{Bj} = {Q^2}/{2 (pq)}$, with $\nu = {(pq)}/{M}$, $s = (p+k)^2$; $\phi$ is the angle between the lepton and  hadron planes, $M$ being the proton mass. 
%%%%%
%%%%%

%%%%%%% Matrix elements
%%%%%%%
The amplitude for the DVCS process reads,
\begin{eqnarray}
\label{eq:TDVCScov}
T_{DVCS} & = &  e^3 j_{DVCS}^\mu \frac{\tilde{g}_{\mu \nu}}{q^2} J_{DVCS}^{\nu} 
\end{eqnarray}
where the lepton and hadron currents are respectively given by,
\begin{subequations}
\begin{eqnarray}
j_{DVCS}^\mu &= & \overline{u}(k',h)\gamma^\mu u(k,h) 
\\
J_{DVCS}^{\nu} & = & {\cal W}^{\mu \nu}(p,p') \left(\varepsilon^{\Lambda_{\gamma'}} _\mu(q') \right)^* \, .
\end{eqnarray}
\end{subequations}
 $W_{\mu \nu}$ is the DVCS hadronic tensor, $\varepsilon^{\Lambda_{\gamma'} \mu}(q')$ is the polarization vector of the outgoing photon, $\gamma'$.  
For BH one has,
\begin{eqnarray}
\label{eq:TDVCScov}
T_{BH} & = &  \frac{e^3}{\Delta^2}  j_{BH}^\mu  (J_{BH})_{\mu} \end{eqnarray}
with,
\begin{eqnarray}
j_{BH}^\mu & = &  \left(\varepsilon^{\Lambda_{\gamma'} \nu}(q')\right)^* L_{\mu \nu}^h(k,k',q')  
\\
(J_{BH})_\mu & = & \overline{U}(p',\Lambda')\Gamma_\mu U(p,\Lambda) 
\end{eqnarray}
where $L_{\mu \nu}^h$ is the tensor for electron scattering off the proton with the emission of a final photon; 
$\Gamma_\mu$ is the usual current operator given in terms of the Dirac and Pauli form factors, $F_1$ and $F_2$, as,
\begin{eqnarray}
\Gamma_\mu =\left(F_1+ F_2\right)\gamma_\mu 
  -\frac{(p+p')_\mu}{2M} F_2 
\end{eqnarray}
The DVCS amplitude involves the photon projection operator, $\tilde{g}_{\mu \nu}$, which is defined by the expansion \cite{Dmitrasinovic:1989bf,Boffi:1993gs},
\begin{eqnarray}
\tilde{g}_{\mu \nu} = \sum_{\Lambda_{\gamma^*}} (-1)^{\Lambda_{\gamma^*}} \left( \varepsilon_\mu^{\Lambda_{\gamma^*}} \right)^* \varepsilon_\nu^{\Lambda_{\gamma^*}}   .
\end{eqnarray}
Inserting the expansion in Eq.(\ref{eq:TDVCScov}) we obtain the following invariant expression,
\begin{eqnarray}
T_{DVCS} & = &  \frac{e^3}{q^2}  \left(j_{DVCS}^\mu \varepsilon_\mu^{\Lambda_{\gamma^*} \, *} \right) \left( J_{DVCS}^{\nu} \varepsilon_\nu ^{\Lambda_{\gamma^*}} \right) ,
\end{eqnarray}
where the photon polarization vector contracted with the hadron current is evaluated in the hadron scattering plane, and it is therefore rotated by a phase, 
\begin{equation}
    \varepsilon_\mu^{\Lambda_\gamma^*} (hadron) = e^{-i \Lambda_\gamma^* \phi} \, \varepsilon_\mu^{\Lambda_\gamma^*} (lepton) 
\end{equation}
This phase determines the $\phi$ dependence of the DVCS cross section. The BH cross section has only kinematic $\phi$ dependence through four-vector products of the type $(k \Delta)$, $(kq')$, ..., where $k$ lies in the lepton plane and $\Delta, q' ...$, lie in the hadron plane. The BH-DVCS interference term has both a phase dependence from the DVCS contribution, and a kinematic dependence on the angle $\phi$ through the BH contribution. All of these contributions were calculated explicitly in Ref.\cite{Kriesten:2019jep} carefully separating out the phase dependence from the kinematic one. The phase dependence is not made explicit in the harmonic expansion formalism of BKM, and we surmise that its different (or lack of) treatment is at the origin of the numerical discrepancies shown in this paper. 
We focus on the cross section for either an unpolarized or a polarized electron scattering off an unpolarized nucleon which are respectively given by,
\begin{eqnarray}
\label{eq:siguu}
\sigma_{UU} & = & \sigma_{UU}^{BH} + \sigma_{UU}^{DVCS}  + \sigma_{UU}^{\cal I}  \\
\label{eq:siglu}
\sigma_{LU} & = & \sigma_{LU}^{DVCS}  + \sigma_{LU}^{\cal I}
\end{eqnarray}
The detailed structure of the BH, DVCS, BH-DVCS interference contributions to the cross sections in Eqs.\eqref{eq:siguu},\eqref{eq:siglu} given in Ref.\cite{Kriesten:2019jep} in the Born approximation, read, 
%where $F_1$ and $F_2$ are the proton Dirac and Pauli form factors. 
\begin{widetext}
\begin{eqnarray}
\label{eq:siguuBH}
 \sigma_{UU}^{BH} =  \Gamma \big| T_{BH} \big|^2 
 = 
 \frac{\Gamma}{t^2} \, \left[ A_{BH} \left(F_1^2 + \tau F_2^2 \right)+ B_{BH} \, \tau G_M^2  \right]
 \end{eqnarray}
\begin{eqnarray}
\label{eq:siguuDVCS}
\sigma_{UU}^{DVCS} & = & \frac{\Gamma}{Q^2 (1-\epsilon)}\Big\{  F_{UU,T} + \epsilon F_{UU,L}+ \epsilon \cos 2\phi F_{UU}^{\cos 2 \phi}  
 +  \sqrt{\epsilon(\epsilon+1)} \cos \phi F_{UU}^{\cos \phi} \Big\} \\
 \label{eq:sigluDVCS}
\sigma_{LU}^{DVCS} & = & \frac{\Gamma}{Q^2 (1-\epsilon)} (2h)   \sqrt{2\epsilon(1-\epsilon)} \,\,  \sin \phi \, F_{LU}^{\sin \phi}
\end{eqnarray}
%%%
\begin{eqnarray}
\label{eq:siguuI}
\sigma_{UU}^{\cal I}  &= & e_l
 \, \frac{\Gamma}{ Q^2 \mid t \mid} \left\{  A_{UU}^{\cal I}  \Re e \big(F_1 \mathcal{H} + \tau F_2  \mathcal{E} \big)   + B_{UU}^{\cal I}    G_M \Re e \big( \mathcal{H}+ \mathcal{E} \big) 
 + C_{UU}^{\cal I}   
G_M \Re e\mathcal{ \widetilde{H}} 
 + \frac{\sqrt{t_0-t}}{Q} \, F_{UU}^{{\cal I}, tw 3} \right\} \\
 %%%%%
 \label{eq:sigLUI}
\sigma_{LU}^{\cal I} &= & e_l
 \, \frac{\Gamma}{ Q^2 \mid t \mid} \left\{A_{LU}^{\cal I} \Im m  \big(F_1 \mathcal{H} + \tau F_2  \mathcal{E} \big)  + B_{LU}^{\cal I}  G_M \, \Im m \big( \mathcal{H}+ \mathcal{E} \big) 
 +  C_{LU}^{\cal I} G_M \Im m \mathcal{ \widetilde{H}} \, ,
  +  \frac{\sqrt{t_0-t}}{Q} \, F^{{\cal I}, tw 3}_{LU} \right\}.
\end{eqnarray}
\end{widetext}
%%%%%%%
%%%%%%%
where $e_l$ is the lepton charge, $y = {(qp)}/{(kp)}$, $\tau = -t/4M^2$, $t_0$ is the minimum $t$ value allowed by taking the transverse four-momentum transfer $\Delta_T \geq 0$; 
%\begin{eqnarray}
%t_0&
$\epsilon$, the ratio of longitudinal to transverse virtual photon flux in DVCS is given by,
\[
\epsilon\equiv \frac{1-y-\frac{1}{4}y^2\gamma^2}{1-y+\frac{1}{2}y^2 +\frac{1}{4}y^2\gamma^2} \]
$F_1$ and $F_2$ are the Dirac and Pauli form factors; $A_{BH}$ and $B_{BH}$ are  kinematic coefficients expressed in terms of four-vector products involving all the relevant four-momenta: the initial and final electron momenta, $k$ and $k'$, the final photon momentum, $q'$, $\Delta$, and the average proton momentum $P=(p+p')/2$. In Eq.(\ref{eq:siguuBH}) we wrote their kinematic dependence on the relevant kinematic variables fore the process, $y$, $x_{Bj}$, $t$, $Q^2$ and $\phi$. Their detailed expressions are given in Ref.\cite{Kriesten:2019jep}.
%%%% CFFs
The CFFs are defined, in the QCD factorization framework, as convolutions of the GPDs for each quark flavor, $q$, with the Wilson coefficients functions. At leading order we have for,
${\cal F}_q$ = (${\cal H}_q$, ${\cal E}_q$), and $\widetilde{\cal F }_q$= ($\widetilde{\cal H}_q$ $\widetilde{\cal E}_q$), respectively,
\begin{eqnarray}
\mathcal{F}_q(\xi,t)  & = & {\cal C} \left( C^+ \, F_q \right) \equiv \int_{-1}^{1} dx   C^+(x,\xi) F_q(x,\xi,t), 
\nonumber \\ \\
\mathcal{\widetilde{F}}_q(\xi,t) & = & {\cal C} \left( C^- \, \widetilde{F}_q \right) \equiv \int_{-1}^{1} dx   C^-(x,\xi) \widetilde{F}_q(x,\xi,t) . \nonumber \\
%\widetilde{\mathcal{F}} = \int_{-1}^{1} dx  \widetilde{F}(x,\xi,t)  C^-(x,\xi) ,
\label{CFF}
\end{eqnarray}
\noindent with the leading order coefficients functions given by,
 \begin{equation}
 { C^\pm(x,\xi) = \frac{1}{x-\xi - i \epsilon} \mp \frac{1}{x+\xi - i \epsilon} . } %C^\pm(x,\xi) = \frac{1}{x-\xi - i \epsilon} \pm \frac{1}{x+\xi - i \epsilon},
 \end{equation}
 The GPDs observe crossing symmetry relations with respect to $x \rightarrow -x$, which allow us to introduce valence (symmetric) and quark singlet (anti-symmetric) distributions (for a detailed discussion see Refs.\cite{GolecBiernat:1998ja,Goldstein:2012az}). In DVCS the proton GPD is written in terms of the quark GPDs as,
\begin{eqnarray}
H & = & \sum_q e_q^2 \, H_q 
\end{eqnarray}
$e_q$ being the quark charge. The neutron GPD can be obtained using isospin symmetry.  

\vspace{0.3cm}
\noindent Although the full structure of the cross section was already given in Ref.\cite{Kriesten:2019jep}, to facilitate data analyses and interpretations, we make the following observations:
\vspace{0.3cm}

\noindent $\bullet$ The BH cross section is cast in a form similar to $ep$ elastic scattering. However, due to the additional photon radiated from the electron in either the initial or final state,  the virtual photon exchanged with the target is aligned along $\Delta$ at an angle $\phi$, and the kinematic coefficients, $A_{BH}$ and $B_{BH}$, multiplying the form factors acquire a complicated dependence in $\phi$ Ref.\cite{Kriesten:2019jep}.
A more physical interpretation of these terms can also be obtained by writing the coefficients combination, 
\begin{equation}
\epsilon_{BH} = \left( 1+ \frac{B_{BH}}{A_{BH}}(1+\tau) \right)^{-1}
\end{equation}
 measuring the exchanged virtual photon's longitudinal polarization relative to the transverse. Notice that $\epsilon_{BH} \neq \epsilon$.

\noindent $\bullet$ The DVCS structure functions are bilinear functions of the CFFs multiplied by kinematic coefficients that are directly related to the helicity structure for the process. 
We introduced a similar notation as in Refs.\cite{Diehl:2005pc,Bacchetta:2006tn} defining $F_{UU,T}$, $F_{UU,L}$, $F_{UU}^{\cos \phi}$, $F_{UU}^{\cos 2 \phi}$, and $F_{LU}^{\sin \phi}$, where the first and second subscript define the polarization of the beam and target, respectively, the third subscript defines the polarization of the virtual photon; the superscripts refer to the azimuthal angular dependence associated with each structure function. Notice the structure of the multiplicative factors in each structure function: $F_{UU,T}$ has no $\sqrt{t_0-t}/Q$ factor and is therefore the dominant term at high $Q^2$; $F_{UU}^{\cos \phi}$, $F_{UU}^{\sin \phi}$ contain one helicity flip factor $\propto \sqrt{t_0-t}/Q$; $F_{UU}^{\cos 2 \phi}$ is a leading twist contribution associated with a double helicity flip and it is both proportional to $(t_0-t)/M^2$, and suppressed by a factor $\alpha_S$ from the gluon coupling; finally, $F_{UU,L}$, containing only twist-three GPDs, is given by the product of two single-flip terms yielding a multiplicative factor of $(t_0-t)/Q^2$.

%What renders DVCS analyses complicated with respect to inclusive and semi-inclusive deep inelastic scattering (SIDIS) experiments, is that owing to the fact that the CFFs appear in quadratic form, different polarization configurations do not cancel out when forming the various beam-target polarization structures. For instance, $F_{UU,T}$, the unpolarized structure function, contains CFFs from all GPDs, $H,E$,  $\widetilde{H},\widetilde{E}$ (similar results are obtained for the other observables).
%On the contrary, in SIDIS one has a one to one correspondence between the beam-target polarization configurations and TMDs. For instance, the structure function $F_{UU,T}$ in SIDIS measures the convolution of $f_1$ and the corresponding unpolarized fragmentation function, $D_1$, similarly, $F_{LL}$ measures the convolution of $g_{1L}$ and $D_1$, and so on \cite{Bacchetta:2006tn,Mulders:1995dh}. 
%

\noindent $\bullet$ 
The BH-DVCS interference contribution is expressed in terms of linear combinations of products of CFFs  elastic form factors, $F_1$ and $F_2$, with the coefficients, $A_{UU}^{\cal I}$, $B_{UU}^{\cal I}$, $C_{UU}^{\cal I}$, which are functions of $(Q^2, x_{Bj}, t, y, \phi)$ \cite{Kriesten:2019jep} .  
Similar to the DVCS contribution, there is no obvious connection between the various beam/target polarization configurations and  the GPDs contributing to the structure functions. 
On the other hand, similar to the BH term, 
%The $\phi$ dependence in this case is both of kinematic origin and from the phase in the virtual photon polarization vector.  
one can single out the contributions, 
\[ F_1 \mathcal{H} + \tau F_2  \mathcal{E}, \quad   
G_M (\mathcal{H}+ \mathcal{E} ),  \quad G_M \mathcal{ \widetilde{H}}, \]
where the electric and magnetic properties of the cross section are clearly separated out, and, in addition, one has the equivalent of an axial charge term. The latter  appears {similarly to the parity violating term in elastic scattering. It is, however, parity conserving in DVCS-BH interference because of the presence of the extra photon emitted at the proton vertex.}

For completeness we list the expressions for both the BH and BH-DVCS interference terms in Appendix \ref{app0}.

\subsection{Comparison with previous formulations}
%%%%%
  {In exclusive unpolarized scattering processes from the proton it is expected that the magnetic contribution will appear suppressed with respect to the electric one due to the phase dependence of the spin flip amplitude describing this term.
Indeed, in Ref.\cite{Kriesten:2019jep} we found that this holds specifically for DVCS, where the magnetic contribution to $\sigma_{UU}^{\cal I}$ appears multiplied by a smaller coefficient, $B_{UU}^{\cal I}$, than the electric form factor. 
This contribution represents, however, the most interesting term of the unpolarized cross section: since it contains the combination of CFFs, ${\cal H} + {\cal E}$, it brings us closer to {getting a quantitative hold of angular momentum} \cite{Ji:1996ek}, and a goal of this paper is to help galvanize the  efforts to extract information on this important quantity. Notice, in fact, that, as shown in Section \ref{sec:3}, in our formalism the coefficient of the magnetic term, $B_{UU}^{\cal I}$, is larger than the coefficient of the axial term, $C_{UU}^{\cal I}$, thus bringing this term within current experimental grasp. 
In the BKM formalism, on the contrary the coefficient $C_{UU}^{\cal I}$ is larger than $B_{UU}^{\cal I}$. The analytic forms of the coefficients $A_{UU}^{\cal I}$, $B_{UU}^{\cal I}$ and $C_{UU}^{\cal I}$, evaluated using the BKM formalism are also given in Appendix \ref{appa}. 
%
%More details on the extraction of ${\cal H} + {\cal E}$ with our method can be found in the detailed analysis of Ref.\cite{Kriesten:2020apm}. 
}
  {
Another difference is in the twist three GPD contributions, written in detail in the next Section, which we define along the lines of the GPD decomposition of the correlation function of Ref.\cite{Meissner:2009ww}. This allows us for the first time to give a physical interpretation of the various twist three contributions in terms of orbital angular momentum and spin orbit contributions.}

  {
%\left(F_1 \mathcal{H} + \tau F_2  \mathcal{E} \right)   + B_{UU}^{\cal I}    G_M \Re e \left( \mathcal{H}+ \mathcal{E} \right)
% + C_{UU}^{\cal I}   
%G_M \Re e\mathcal{ \widetilde{H}}
The numerical evaluation of the differences with the BKM formalism presented in   Refs.\cite{Belitsky:2000gz,Belitsky:2001ns,Belitsky:2010jw} is presented in Section \ref{sec:3}  in various kinematic regimes.}
%%%%%%%%%%%%%%%%%%%%%%%%%%%
%%%%%%%%%%%%%%%%%%%%%%%%%%%

\subsection{Twist three}
We discuss the structure of the BH-DVCS interference term at twist three, in view of the fact that it contains GPDs describing {the longitudinal component of orbital angular momentum, $\widetilde{E}_{2T}$, the spin orbit term $(2 \widetilde{\mathcal{H}}'_{2T} +  \mathcal{E}_{2T}')$ \cite{Raja:2017xlo,Rajan:2016tlg}, and terms related to transverse angular momentum, ${H}_{2T}$,}
%%% tw 3
\begin{widetext}
\begin{eqnarray}
\label{eq:Int_FUU3}
F_{UU}^{{\cal I}, tw 3} &= &  A^{ (3) \cal I}_{UU}  \Big[ F_{1} \Big( \Re e (2   \mathcal{\widetilde{H}}_{2T} +   \mathcal{E}_{2T}) - \Re e (2  \widetilde{\mathcal{H}}'_{2T} +  \mathcal{E}_{2T}') \Big)
 +  F_{2} \Big( \Re e  (  \mathcal{H}_{2T} + \tau  \mathcal{\widetilde{H}}_{2T}) - \Re e( \mathcal{H}_{2T}' + \tau  \widetilde{\mathcal{H}}_{2T}' ) \Big) \Big] 
 \nonumber \\
 &+ &  B^{(3) \cal I}_{UU} G_M \,  ( \Re e \widetilde{\mathcal{E}}_{2T} - \Re e \widetilde{\mathcal{E}}_{2T}' ) \nonumber 
 \\
 &+ &  C^{(3) \cal I}_{UU}  G_M \,   \Big[2\xi   (\Re e  \mathcal{H}_{2T} - \Re e \mathcal{H}_{2T}')- \tau \Big( \Re e  (\widetilde{\mathcal{E}}_{2T}  - \xi \mathcal{E}_{2T})  - \Re e \, (\widetilde{\mathcal{E}}_{2T}'    
       -\xi \mathcal{E}_{2T}' )  \Big)\Big]
\end{eqnarray}
\end{widetext}
For a polarized electron beam we obtain a structure analogous to the unpolarized case, where  the $\Re e$ parts of the CFFs are replaced with with the $\Im m$ parts, namely,
\begin{widetext}
\begin{eqnarray}
F^{\cal I}_{LU} & = & A^{ (3) \cal I}_{LU}  \Big[ F_{1} \Big( \Im m (2   \mathcal{\widetilde{H}}_{2T} +   \mathcal{E}_{2T}) - \Im m (2  \widetilde{\mathcal{H}}'_{2T} +  \mathcal{E}_{2T}') \Big)
 +  F_{2} \Big( \Im m  (  \mathcal{H}_{2T} + \tau  \mathcal{\widetilde{H}}_{2T}) - \Im m( \mathcal{H}_{2T}' + \tau  \widetilde{\mathcal{H}}_{2T}' ) \Big) \Big] 
 \nonumber \\
 &+ &  B^{(3) \cal I}_{LU} G_M \,  ( \Im m \widetilde{\mathcal{E}}_{2T} - \Im m \widetilde{\mathcal{E}}_{2T}' ) \nonumber 
 \\
 &+ &  C^{(3) \cal I}_{LU}  G_M \,   \Big[2\xi   (\Im m  \mathcal{H}_{2T} - \Im m \mathcal{H}_{2T}')- \tau \Big( \Im m  (\widetilde{\mathcal{E}}_{2T}  - \xi \mathcal{E}_{2T})  - \Im m \, (\widetilde{\mathcal{E}}_{2T}'    
       -\xi \mathcal{E}_{2T}' )  \Big)\Big]
\end{eqnarray}
\end{widetext}
The coefficients, $A^{ (3) \cal I}_{UU}$, $B^{ (3) \cal I}_{UU}$, $C^{ (3) \cal I}_{UU}$ and $A^{ (3) \cal I}_{LU}$, $B^{ (3) \cal I}_{LU}$, $C^{ (3) \cal I}_{LU}$, are written in terms of four-vector products involving all relevant variables, for the electron, $k$, $k'$, final photon, $q'$, momentum transfer, $\Delta$, and average proton momentum, $P$. Similarly to the twist-two case, they can be expressed in terms of the set of variables ($Q^2$, $x_{Bj}$, $t$, $y$, $\phi$). Their specific expressions are given for the first time in Ref.\cite{Kriesten:2019jep}.   {A comparison with BKM cannot be performed due to the inherently different structure of their dynamic twist three expressions. Note for the twist three CFFs, ${\cal H}_{2T}, {\cal E}_{2T}...$, we use the same notation as for the twist two case, by defining them through the convolution with the leading order  Wilson coefficient functions given in Eqs.(\ref{CFF}). The possible role of an explicit qgq term is beyond the scope of this paper.} 

The notation for the twist three GPDs is illustrated in Table \ref{tab:GPD} where we show along with our our notation, their quark-proton polarization configuration, the corresponding notation in the TMD sector, and the notation from Ref.\cite{Meissner:2009ww}
(see also Table I in Ref.\cite{Kriesten:2019jep}). The notation follows the one adopted for TMDs \cite{Jaffe:1992ra,Mulders:1995dh}, namely,
\begin{itemize}
\item $H$  and $f$  correspond to the vector coupling in the parametrization of the quark-proton correlation function \item$\widetilde{H}$  and $g$   correspond to axial-vector coupling 
\item the $\perp$ superscript indicates an unsaturated transverse momentum index in the correlation function's coefficient \cite{Raja:2017xlo}
\item the subscript $L(T)$ involves the amplitude for a longitudinally (transversely) polarized target. 
\end{itemize} 

\begin{table}[htp]
\centering
\begin{tabular}{|c|c|c|c|}
\hline
GPD &  $P_q P_p$ & TMD & Ref.\cite{Meissner:2009ww} \\
\hline 
%%% tw 3
$H^\perp$ &  UU & $f^\perp$ &  $ 2\widetilde{H}_{2T} + E_{2T} $ \\
\hline 
$\widetilde{H}_L^\perp$ &  LL & $g_L^\perp$ & $2\widetilde{H}_{2T}' + E_{2T}' $ \\
\hline
$ {H_L^\perp} $&  UL & $f_L^{\perp \, {\bf (*)}}$ & $ \widetilde{E}_{2T} - \xi E_{2T}$ \\
\hline
$\widetilde{H}^\perp $ &  LU & $g^{\perp \, {\bf (*)}}$ &  $ \widetilde{E}_{2T}' - \xi E_{2T}'$\\
\hline
  {$E^\perp$} &  UT & $f_T^{\bf (*)}$ & $ H_{2T} + \, \tau \widetilde{H}_{2T} $
\\
\hline
  {$\widetilde{E}^\perp$} &  LT & $g_T'$ & $ H_{2T}' + \, \tau \widetilde{H}_{2T}'$  \\
 \hline
\end{tabular}
\caption{Twist-three GPDs and their helicity content. In the first column we show the GPDs notation for this paper; the second column shows the quark and proton polarizations; the third column shows the analogous configurations in the TMD sector; finally, the fourth column shows the corresponding notation from Ref.\cite{Meissner:2009ww}. The asterisk denotes naive T-odd twist-three TMDs (we define $\tau=\frac{t_o-t}{4M^2}$).} 
\label{tab:GPD}
\end{table}

%%%%%%%%
%%%%%%%% Phase dependence
\subsection{Azimuthal angular dependence}
\label{sec:phase} 
The azimuthal angular, $\phi$, dependence is a key feature of the cross section, appearing with different capacities in the description of the BH, DVCS, and BH-DVCS interference contributions. In electron scattering exclusive reactions the cross section assumes a characteristic dependence on the {\it phase}, $\phi$, which originates from rotating the virtual photon polarization vector, $\varepsilon^{\Lambda_\gamma^*}_\mu$, from the leptonic to the hadronic plane
(see {\it e.g.} Refs.\cite{Boffi:1993gs,Donnelly:1985ry} and the detailed reiteration for deeply virtual scattering in Ref.\cite{Arens:1996xw,Diehl:2005pc}). 
This dependence allows us in general to single out the contributions from the overlap of different transverse and longitudinal amplitudes ({\it e.g.} $\sigma_T, \sigma_{TT}, \sigma_{LT}...$) describing different physics content. 

In the specific case of exclusive photoproduction, we distinguish between the BH, and DVCS processes. In BH there are only two structure functions, {\it i.e.} the proton elastic form factors, therefore, similar to elastic scattering we do not organize the cross section by writing out the various virtual photon polarization components. The dependence on $\phi$ is purely kinematic. 

For the DVCS and BH-DVCS contributions we set the virtual photon for DVCS, $q$,  along the $z$ axis in the laboratory frame, while the virtual photon, $\Delta$, for the BH process is along $\Delta$, at an angle $\phi$ with respect to $q$. 

As shown  below, this mismatch results in a much more complicated dependence of the polarization vector products contributing to the cross section, generating substantial $t$ and target mass corrections \cite{Kriesten:2019jep}.

\subsubsection{Phase dependence of pure DVCS contribution}
%%%%%%%%
%%%%%%%% Phase dependence
The polarization vectors for the virtual photon of momentum $q$ along the {negative} $z$-axis in the laboratory frame are defined as,
\begin{eqnarray}
\label{eq:eps}
 \varepsilon^{\Lambda_{{\gamma^*}}=\pm 1}   & \equiv & \frac{1}{\sqrt{2}} (0; \mp 1, {i}, 0), \, \, \\
\varepsilon^{\Lambda_{{\gamma^*}}=0} &\equiv &\frac{1}{Q} (\mid {\vec q} \mid; 0, 0 ,q_0) =\frac{1}{\gamma} (\sqrt{1+\gamma^2}; 0, 0, 1), \nonumber \\ 
\label{eps0} 
 \end{eqnarray}
 Notice that the DVCS helicity amplitudes are evaluated in the CoM frame of the final photon-hadron system, which defines the hadron plane at an angle $\phi$ with respect to the lepton plane \cite{Kriesten:2019jep}. 
The cross section is evaluated by transforming to the lepton plane rotating the polarization vectors defining the helicty amplitude, $f_{\Lambda\Lambda'}^{\Lambda_\gamma^* \Lambda_\gamma'}$, by  $-\phi$ about the $z$ axis. Another way to express this is that the lepton produces a definite helicity virtual photon which we take along the  $z$ axis in the lepton plane; however, the virtual photon's interaction with the target occurs in the hadron plane which is rotated through an azimuthal angle $\phi$. 
The phase dependence of the DVCS contribution to the cross section is a consequence of such a rotation about the axis where the virtual photon lies \cite{Dmitrasinovic:1989bf,Boffi:1993gs}. 
 The  $\phi$ rotation about the $z$-axis changes the phase of the transverse components, and leaves the longitudinal polarization vector unchanged as,
\begin{eqnarray}
\label{eq:epsrot1}
\varepsilon^{\Lambda_{{\gamma^*}}   =  \pm 1} & \rightarrow & \frac{e^{-i\Lambda_{{\gamma^*}}\phi}}{\sqrt{2}} (0, \mp 1, {i}, 0) 
\label{eps_hadron}
%\varepsilon^{\Lambda_{\gamma'} =  \pm 1} 
%& \rightarrow &
% \frac{e^{-i\Lambda_{\gamma}'\phi}}{\sqrt{2}} ( 0, \mp 1 , i,0)   + (0, 0_T, \pm \sin\theta)
\label{eq:epsprime1} 
\end{eqnarray} 

The ejected (real) photon polarization vectors read,
%\begin{subequations} 
\begin{eqnarray}
&&\varepsilon^{\Lambda_\gamma^\prime = \pm 1}  \equiv \frac{1}{\sqrt{2}} \big( 0; \mp \cos\theta \cos\phi +i \sin\phi, 
\nonumber \\ 
&& \hspace{2cm} \mp \cos\theta \sin\phi {+ i} \cos\phi , \pm \sin\theta\big), 
\label{eq:epsprime} 
\end{eqnarray}
$\varepsilon^{\Lambda_\gamma^\prime}$ in principle also undergoes a phase rotation, however, this phase rotation does not contribute to the cross section due to the completeness relation obtained summing over the physical (on-shell) states \cite{Gastmans},
\begin{eqnarray}
\label{eq:polLambda'}
\sum_{\Lambda_\gamma'} \left( \varepsilon^{\Lambda_\gamma^\prime}_\mu (q') \right)^* \varepsilon^{\Lambda_\gamma^\prime}_\nu(q') & = &  {-g_{\mu\nu}  }
\end{eqnarray}
%
%%%%%%% BH-DVCS
\subsubsection{Phase dependence of BH-DVCS interference term}
%twist-two and twist-three contributions to
%
The same treatment described above is applied to the BH-DVCS interference term. Here the different polarizations allow us to distinguish the twist-two and twist-three terms as,
%As a result, we obtain two different configurations, for a transversely and longitudinally polarized virtual photon, respectively, defining the twist two and twist three structures,
\begin{eqnarray}
\label{eq:gT}
twist \quad 2 \rightarrow \sum_{\Lambda_{\gamma}^{*}  =   \pm 1} \Big(\varepsilon_{\mu}^{\Lambda_{\gamma}^{*}}\Big)^{*}\varepsilon_{\nu}^{\Lambda_{\gamma}^{*}} & = & \cos{\phi} \, g^{T}_{\mu \nu} - \sin{\phi} \, \varepsilon^{T}_{\mu \nu}
\nonumber \\ \\
\label{eq:gL}
twist \quad 3 \rightarrow  \Big(\varepsilon_{\mu}^{\Lambda_{\gamma}^{*}=0}\Big)^{*} \varepsilon_{\nu}^{\Lambda_{\gamma}^{*}=0} & = & g_{\mu \nu}^L  
\end{eqnarray}
where the components of $g_{\mu \nu}^L$ are: $g_{0 0}^L  = 1+ \nu^2/Q^2$, $g_{0 3}^L = g_{3 0}^L = \nu \sqrt{\nu^2 + Q^2}/Q^2$, and  $g_{3 3}^L = \nu^2/Q^2$.
%%%%%%%%%%%%%%%
%%% FIGURE 1
\begin{figure}
\includegraphics[width=7cm]{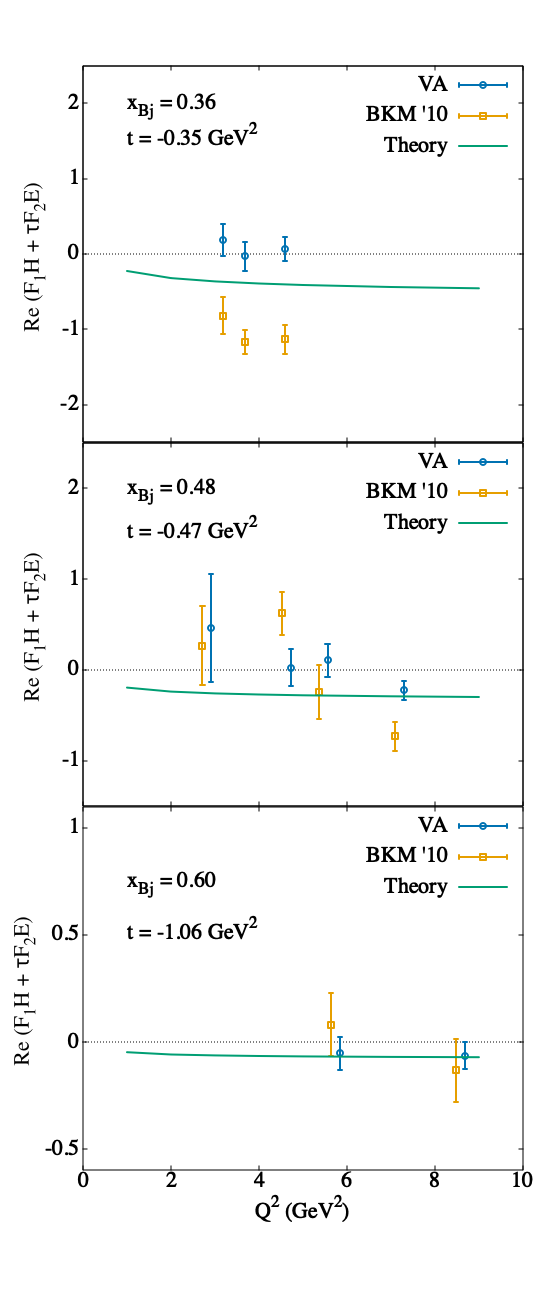}
\caption{  {The CFFs combination, $F_1(t) \Re e {\cal H} + \tau F_2(t) \Re e {\cal E}$, corresponding to the first term in Eq.(\ref{eq:siguuI}), plotted vs $Q^2$. This quantity was extracted from data \cite{Georges:2018kyi} using the formalism presented in this paper and in Ref.\cite{Kriesten:2019jep} (blue circles), and using the formalism of Ref.\cite{Belitsky:2010jw} (orange squares).} The curves in the figure illustrate the effect of perturbative QCD evolution, calculated using the GPD parametrization in Refs.\cite{GonzalezHernandez:2012jv,Kriesten:2021sqc}.}
\label{fig:Q2dep}
\end{figure}

%%%%%%%%%%%%%
%%% FIGURE 2
%%%%%%%%%%%%%%
\begin{figure}
\begin{center}
\includegraphics[width=6cm,height=6.25cm]{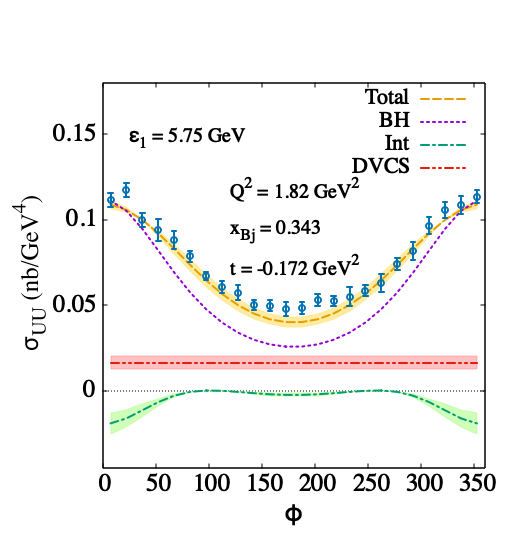}
\includegraphics[width=6cm,height=6.25cm]{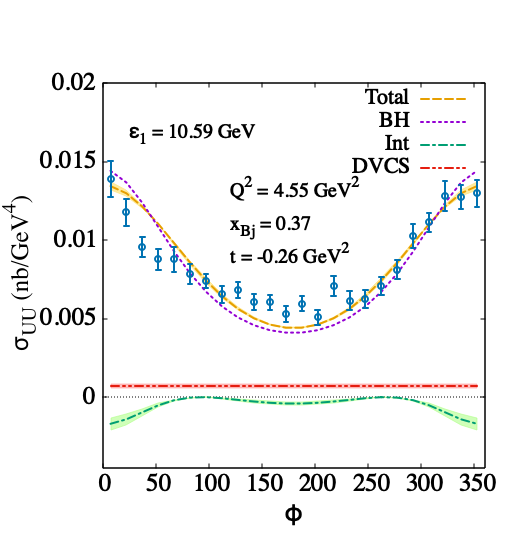}
\includegraphics[width=6cm,height=6.25cm]{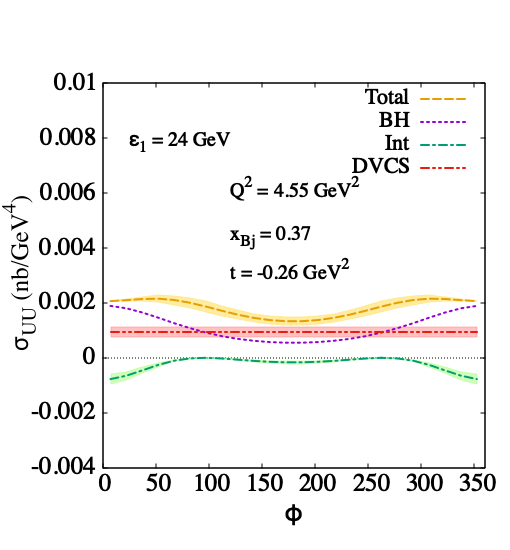}
\end{center}
\caption{The cross section $\sigma_{UU}$, Eq.\eqref{eq:siguu} for the kinematic bins from Ref.\cite{Defurne:2015kxq} with initial electron energy $\epsilon_1=$ 5.75 GeV, $Q^2= 1.8$ GeV$^2$, $t= -0.172$ GeV$^2$, $x_{Bj} = 0.34$ (top panel), Ref. \cite{Georges:2018kyi} $\epsilon_1=$11.5 GeV, $Q^2= 4.5$ GeV$^2$, $t= -0.29$ GeV$^2$, $x_{Bj} = 0.37$ (second panel), and for projected values of a fixed target experiment at $\epsilon_1=$24 GeV (third panel). The curves correspond to the  contributions from: $\sigma_{UU}$ Eq.\eqref{eq:siguu}, $\sigma_{UU}^{BH}$, \eqref{eq:siguuBH}, $\sigma_{UU}^{DVCS}$, \eqref{eq:siguuDVCS}, and $\sigma_{UU}^{\cal I}$, \eqref{eq:siguuI}, calculated in the laboratory system.}
\label{fig:xsx_va}
\end{figure}

%%%%%%%%%%%%%%%%%
%%%%%%% figure 3
%%%%%%%%%%%%%%%%%%
\begin{figure}
\includegraphics[width=6cm,height=6.5cm]{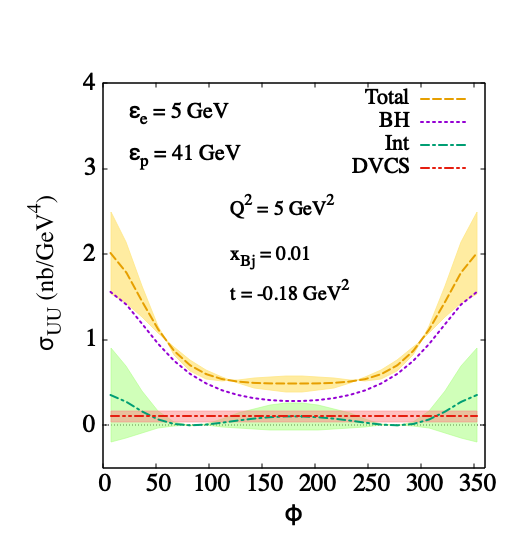}
\includegraphics[width=6cm,height=6.5cm]{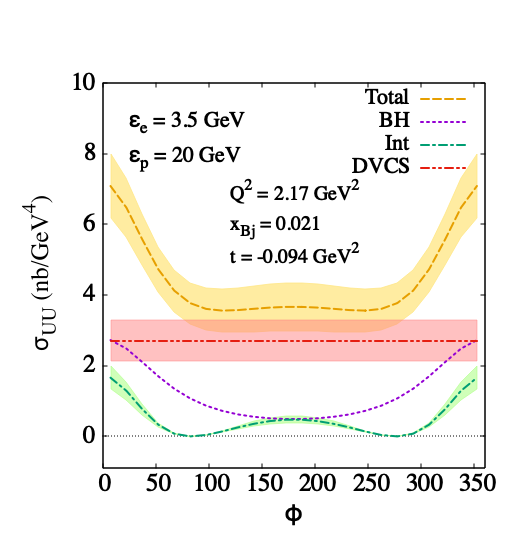}
\caption{The cross section $\sigma_{UU}$ in a collider setting kinematics for an EIC with initial electron energy $\epsilon_e= 5$ GeV, initial proton energy $\epsilon_p= 41$ GeV, and kinematic bin $x_{Bj} = 0.01$, $t = -0.18$ GeV$^{2}$, and $Q^{2} = 5$ GeV$^{2}$ of the US-based EIC \cite{eicyellowreport} (top panel), and initial electron energy $\epsilon_e = 3.5$ GeV, initial proton energy $\epsilon_p = 20$ GeV, and kinematic bin $x_{Bj} = 0.021 $, $t = -0.094$ GeV$^{2}$, and $Q^{2} = 2.17$ GeV$^{2}$ of the China-based EIC \cite{Anderle:2021wcy} (bottom panel).}
\label{fig:xsx_eic}
\end{figure}

%%%%%%%%%%%%%%%%
%%% FIGURE 4
%%%%%%%%%%%%%%%%%
\begin{figure}
\includegraphics[width=7cm]{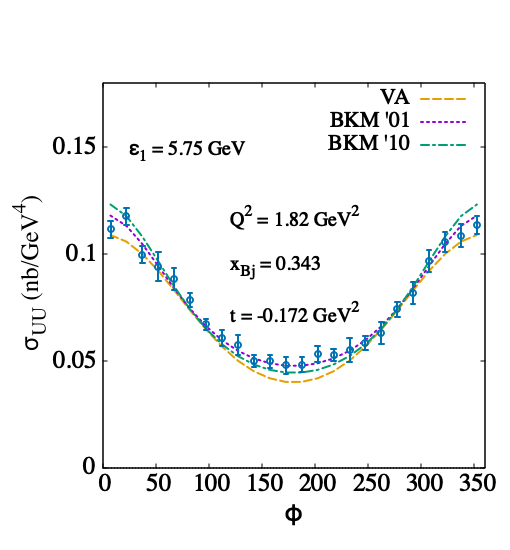}
\includegraphics[width=7cm]{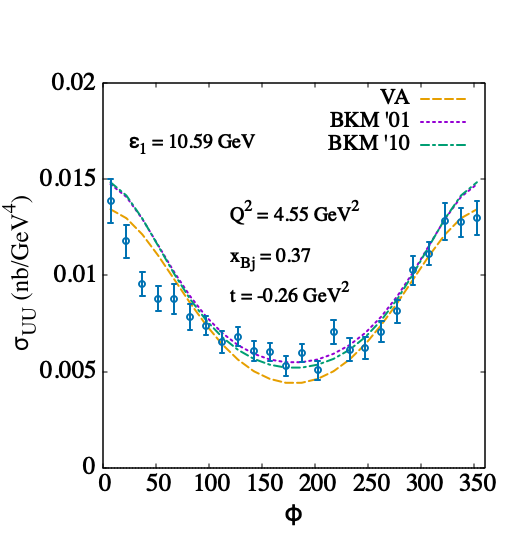}
\caption{{Total unpolarized cross section with VA theory CFF's: (top) $\epsilon_1 = 5.7 \,\, \text{GeV}$, $Q^{2} = 1.8 \,\, \text{GeV}^{2}, \, x_{Bj} = 0.34, \, t = -0.17 \,\, \text{GeV}^{2}$; (bottom)  $\epsilon_1 = 10.6 \,\, \text{GeV}$ $Q^{2} = 4.5 \,\, \text{GeV}^{2}, \, x_{Bj} = 0.37, \, t = -0.26 \,\, \text{GeV}^{2}$ .}}
\label{fig:int_comp_bkm5}
\end{figure}

In addition to the phase dependence, differently from the pure DVCS term where the azimuthal angular dependence resides entirely in the phase factors, the BH-DVCS contribution contains a $\phi$ dependence of kinematic origin. The kinematic $\phi$ dependence arises from the orientation of the $\Delta$ vector which lies at an angle $\phi$ in the hadronic plane and generates a $\phi$ dependence {through factors of 4-vector products $(k\Delta)$} in the coefficients of the structure functions similar to the BH case. 
As a result, we single out an overall multiplicative $\cos \phi$ term in the twist-two contribution to $\sigma_{UU}^{\cal I}$, and a  $\sin \phi$ term in the twist-two contribution to $\sigma_{LU}^{\cal I}$, which originate from the phase dependence arising from the helicity amplitudes of the hadronic current. The coefficients contain also the kinematic $\phi$ dependence as explained above.  A similar situation is found at twist-three where the phase dependence cancels out  since for the virtual photon, $e^{-i\Lambda_{\gamma^{*}}\phi}$ is 1 as a consequence of its longitudinal polarization, thus the $\phi$ dependence is entirely of kinematic origin.   

This result seems at variance with the representation given in the original harmonics expansion of approach Refs.\cite{Belitsky:2001ns,Belitsky:2010jw}. However, one should notice that in these papers the distinction between the $\phi$ dependence from the phase of the polarization vectors and the $\phi$ dependence from the kinematics is not evident. This point of departure of the two formalisms is important as also addressed in Refs.\cite{Braun:2011dg,Braun:2012hq,Braun:2014sta}, where the question of $t$ and target mass corrections resulting from the different choices of the orientation of $\Delta$ has been also studied.
%{The harmonics expansion does not allow for a decomposition of the cross section into various subleading twist components through $\phi$-dependence from the ``phase" of the helicity amplitudes resulting from the hadronic current. 
In particular, it is important to establish a pathway in the decomposition of subleading twist terms which in our case clearly result from the $\phi$-dependence from the ``phase" of the hadronic current. This ensures that what enters into our twist-three cross section terms are directly due to twist-three GPDs and not to kinematically suppressed terms. These issues seem to have  prevented so far a clean extraction of information from data \cite{Defurne:2017paw}. 

%%%%%%%%%%%%%%
%%% FIGURE 5
%%%%%%%%%%%%%%
\begin{figure*}
\begin{center}
\includegraphics[width=6.5cm]{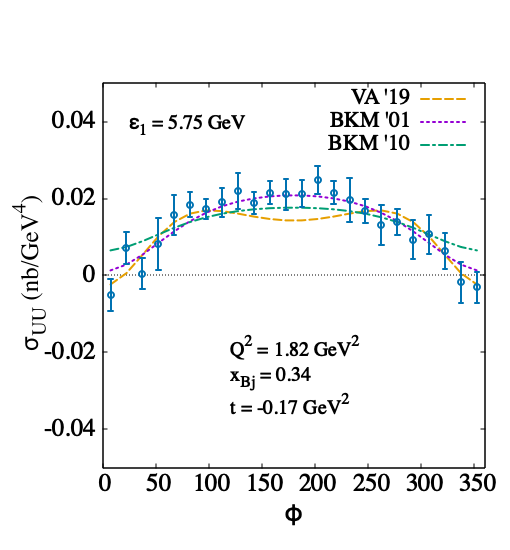}
\includegraphics[width=6.5cm]{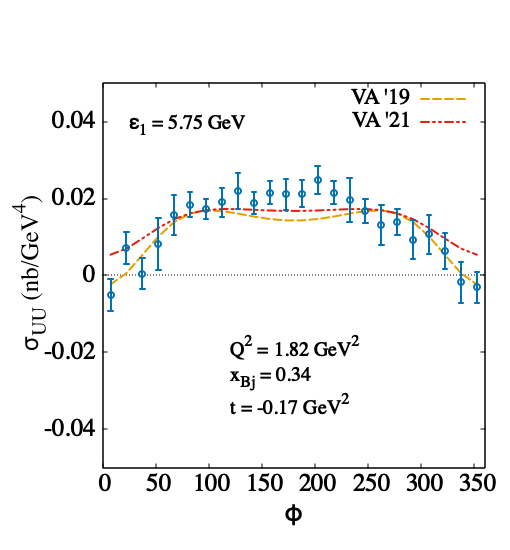}\\
\includegraphics[width=6.5cm]{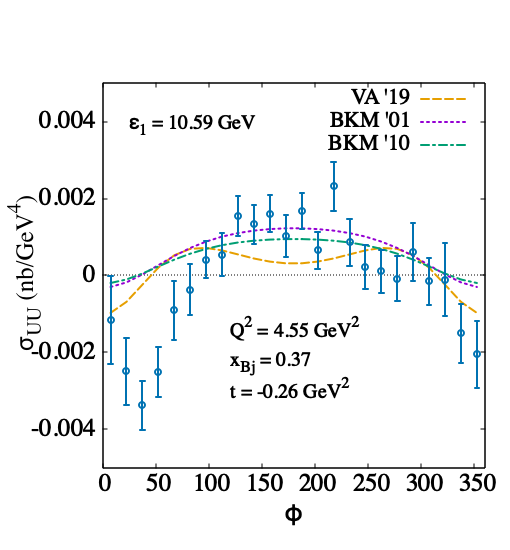}
\includegraphics[width=6.5cm]{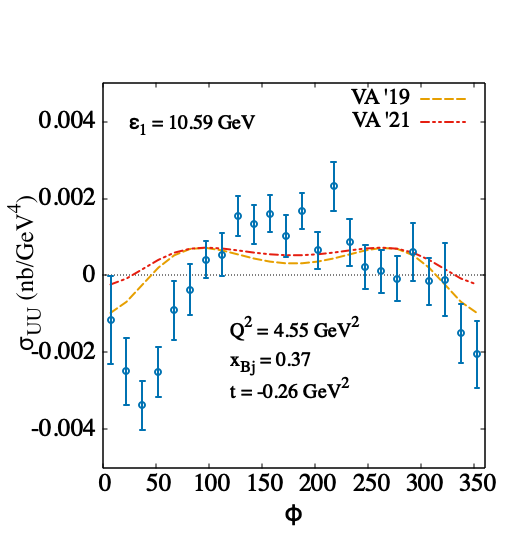}\\
\includegraphics[width=6.5cm]{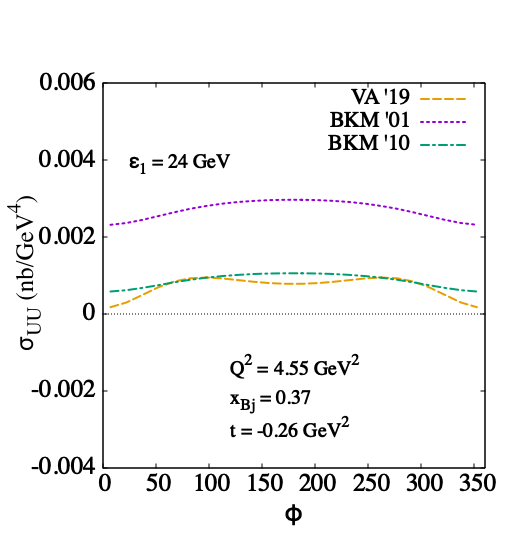}
\includegraphics[width=6.5cm]{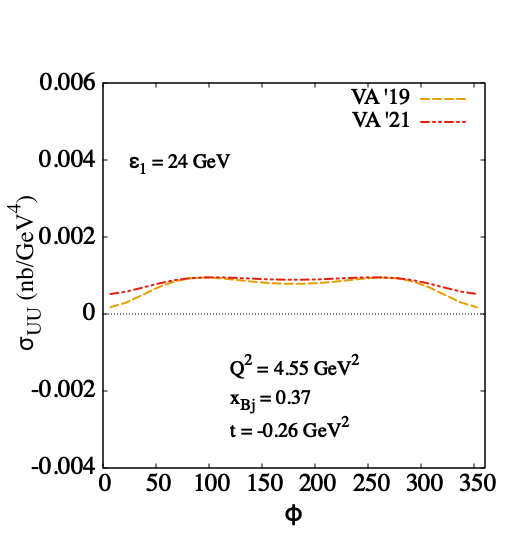}
\end{center}
\caption{Comparison of the interference plus DVCS contributions to the cross section, $\sigma_{UU}^{\cal I} + \sigma_{UU}^{DVCS}$, calculated using the frameworks from the VA and  BKM groups, respectively.   {{\it Left panels}: same notation and kinematic bins as in Fig.\ref{fig:xsx_va}; {\it Right panels}: Comparison between the the formalism of Ref.\cite{Kriesten:2019jep}, VA'19, the same formulation including the higher order gauge fixing coefficients given in Appendix \ref{appa}, VA'21, and BKM'10.} }
\label{fig:int_comp_bkm_va}
\end{figure*}

%%%%%%%%%%%%%%%%
%%% FIGURE 6
%%%%%%%%%%%%%%%%
\begin{figure}
\includegraphics[width=8cm]{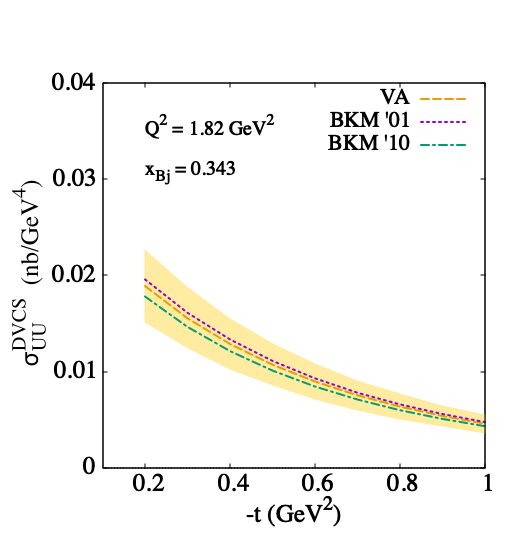}
\caption{Comparison of the unpolarized cross section, $\sigma_{UU}^{DVCS}$, Eq.(\ref{eq:siguuDVCS}), obtained in the VA and BKM frameworks, respectively. $\sigma_{UU}^{DVCS}$ is plotted vs. $-t$ using the CFFs from the reggeized diquark model. The error band represents a theoretical error from the reggeized diquark model fit.}
\label{fig:DVCS}
\end{figure}
%%%%%%%%%
%%%%%%%%%
%%%%%%%%% SECTION III
%%%%%%%%%
%%%%%%%%%
\section{Numerical Results}
\label{sec:3}
In this Section we present numerical evaluations of the BH, DVCS and BH-DVCS interference terms evaluated in Ref.\cite{Kriesten:2019jep}, emphasizing the feature of our new formalism as compared to previous approaches discussed in Section \ref{sec:2}. 

We compare results obtained both in the helicity amplitudes and in the BKM formulations  \cite{Belitsky:2001ns,Belitsky:2010jw}, for various observables entering the unpolarized cross section at kinematic settings ranging from recent measurements at Jefferson Lab \cite{Defurne:2015kxq,Georges:2018kyi}, to a 24 GeV fixed target scenario \cite{Bogasz}, and the EIC \cite{Anderle:2021wcy,eicyellowreport}. 
Since the goal of this paper is to highlight the new features of the framework for exclusive electoproduction as compared to BKM, we restrain from discussing  issues involving fits to the cross section, the extraction of the CFFs from experimental data, and the modeling of GPDs based on DVCS data. These topics will be discussed in upcoming publications.  
%

%%%% FIGURE 1  description%%%%
We start from showing in Figure \ref{fig:Q2dep} what is perhaps the biggest consequence of our new framework for deeply virtual exclusive scattering on an unpolarized proton: the dependence on the scale, $Q^2$, of the   {dominant contribution to the BH-DVCS interference matrix element, $\Re e (F_1{\cal H} + \tau F_2 {\cal E})$. The extraction of this quantity in our formalism is consistent with a slow $Q^2$ dependence as predicted by the perturbative QCD evolution equations for GPDs \cite{Ji:1998xh,GolecBiernat:1998ja,Musatov:1999xp}. In the BKM formalism, on the contrary, oscillations appear which could indicate spurious $Q^2$ dependence resulting from the approximations taken in the cross section coefficients.}

%%%
Notice that, although BKM use the harmonics based formulation where, in particular, the $\sigma_{UU}^{\cal I}$ cross section is not organized in terms of $A_{UU}^{\cal I}$, $B_{UU}^{\cal I}$, $C_{UU}^{\cal I}$, one can retrieve equivalent expressions by rearranging the various harmonics contributions. These expressions are displayed in Appendix \ref{appa}. We note that even by doing so, the analytic comparison between the two formulations represents a formidable task due to the   {inherent complications arising from the different choices of variables for the lengthy coefficients.}
%%% 

%% FIGURE 7
\begin{figure*}
\begin{center}
\includegraphics[width=6.5cm]{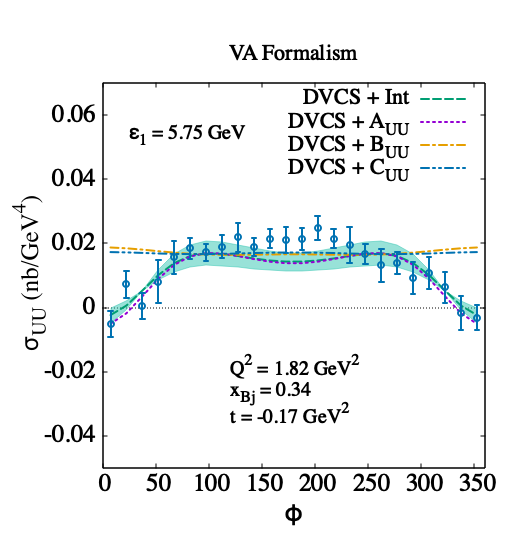}
\includegraphics[width=6.5cm]{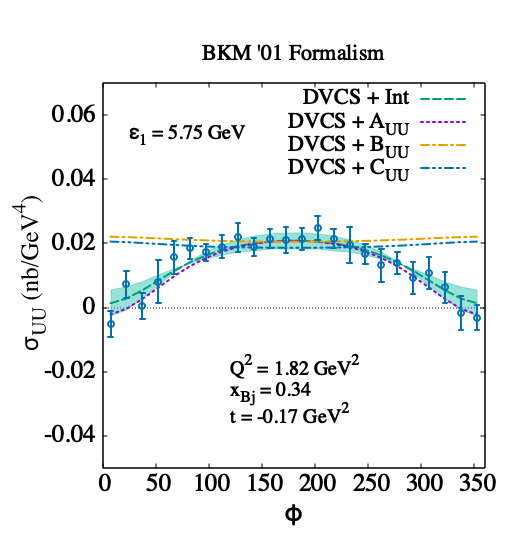}
\includegraphics[width=6.5cm]{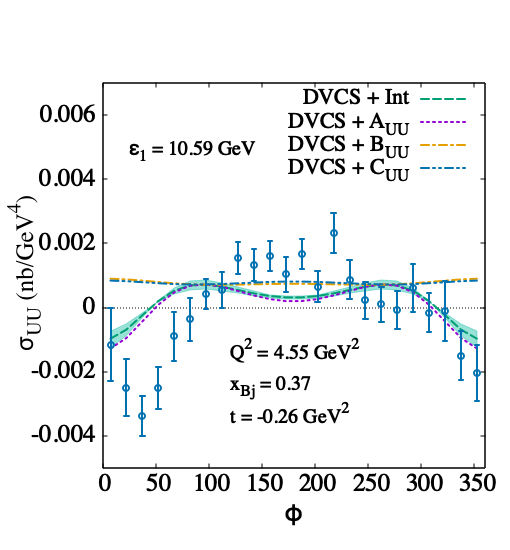}
\includegraphics[width=6.5cm]{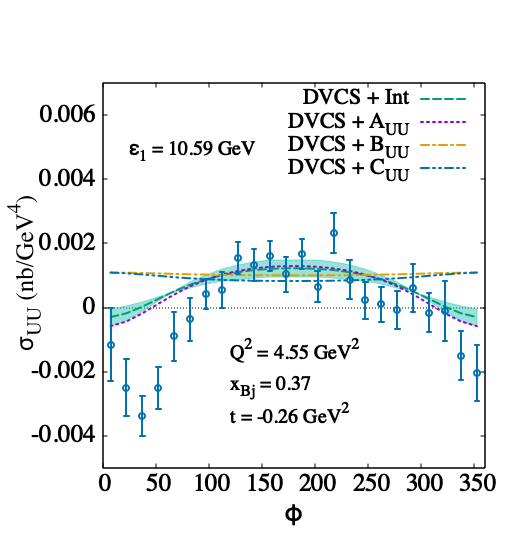}
\includegraphics[width=6.5cm]{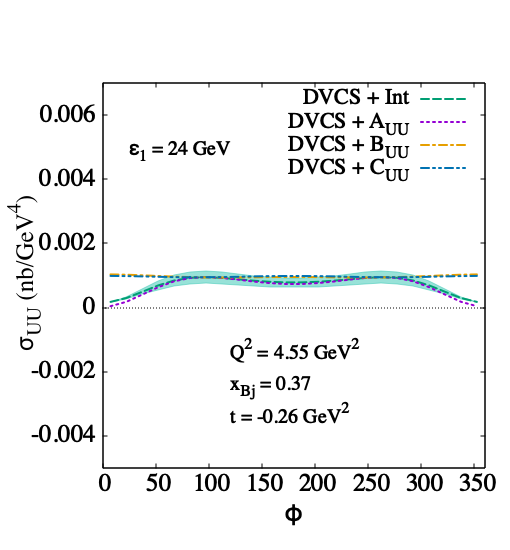}
\includegraphics[width=6.5cm]{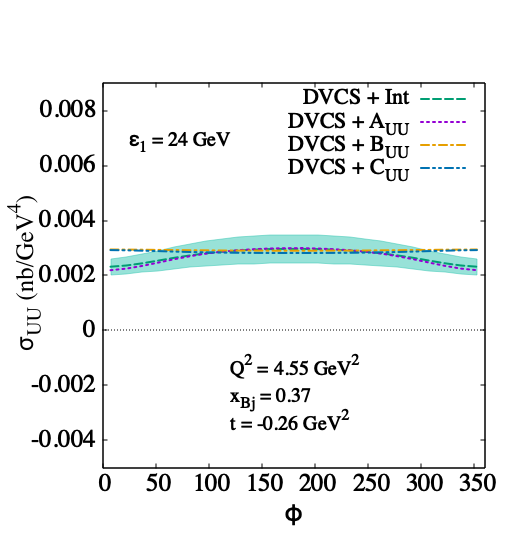}
\end{center}
%%%
\caption{{BH-DVCS interference contribution to the cross section  $\sigma_{UU}^{\cal I}$, Eq.\eqref{eq:siguuI} in the VA (left) and BKM (right) formalism. The initial electron energy is $\epsilon_1=$ 5.75 GeV from Ref.\cite{Defurne:2015kxq}  (top panels), $\epsilon_1=$ 10.5 GeV Ref.\cite{Georges:2018kyi}  (middle panels), and for a projected value of a fixed target experiment at $\epsilon_1=$24 GeV (bottom panels)}. The curves correspond to the calculation at twist-two using the reggeized diquark model \cite{GonzalezHernandez:2012jv} for the electric current  which appears in the cross section multiplied by $A_{UU}$, the magnetic term with coefficient $B_{UU}$, and the axial term, $C_{UU}$.}
\label{fig:int_va}
\end{figure*}

%%%%%%%%%%%%%
%%% FIGURE 8
%%%%%%%%%%%%%
\begin{figure}
\begin{center}
\includegraphics[width=5.5cm]{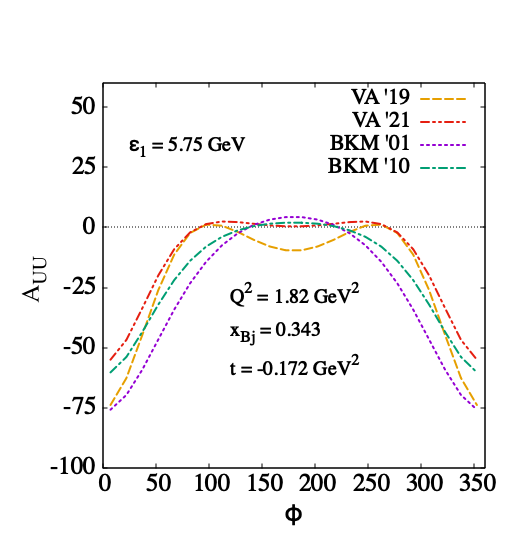}
\includegraphics[width=5.5cm]{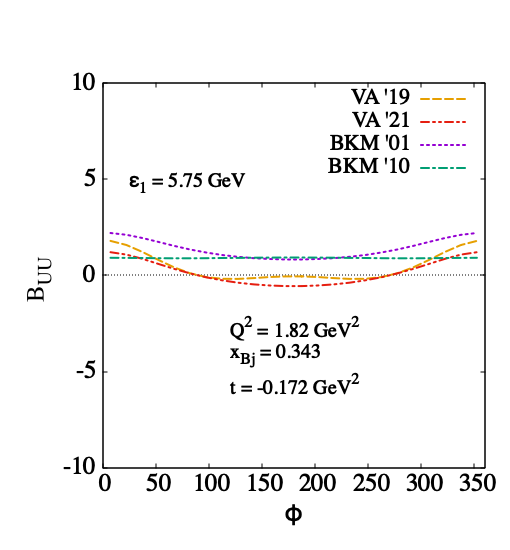}
\includegraphics[width=5.5cm]{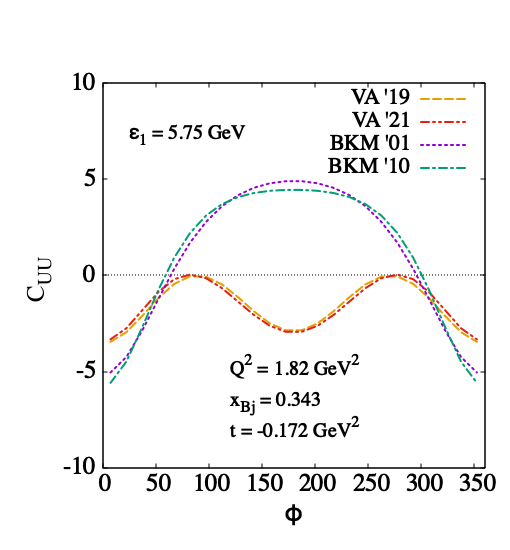}
\end{center}
\caption{Comparison of the kinematic coefficients, $A_{UU}$ (top panel), $B_{UU}$ (middle panel), and $C_{UU}$ (bottom panel), at the kinematics point $Q^{2} = 1.82 \,\, \text{GeV}^{2}, \, x_{Bj} = 0.34, t = -0.17 \,\, \text{GeV}^{2}, \,\epsilon_1 = 5.75 \,\, \text{GeV}$. }
\label{fig:int_comp_bkm2}
\end{figure}
%%%%%%%%%%%%

%%%%%%%%%%%%%%%%
%%% FIGURE 9
%%%%%%%%%%%%%%%
\begin{figure*}
\begin{center}
\includegraphics[width=7cm]{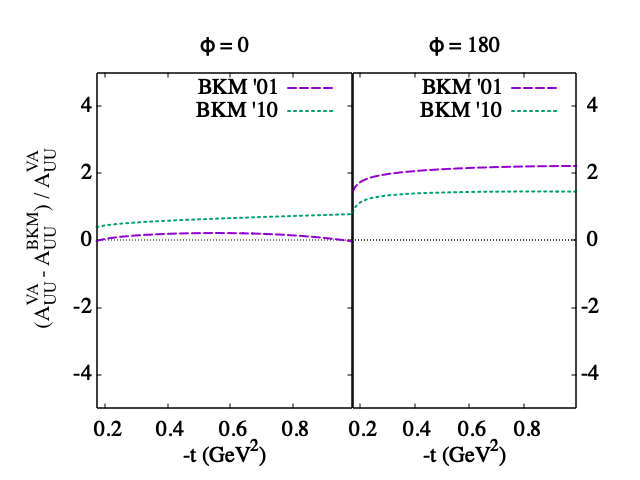}
\includegraphics[width=7cm]{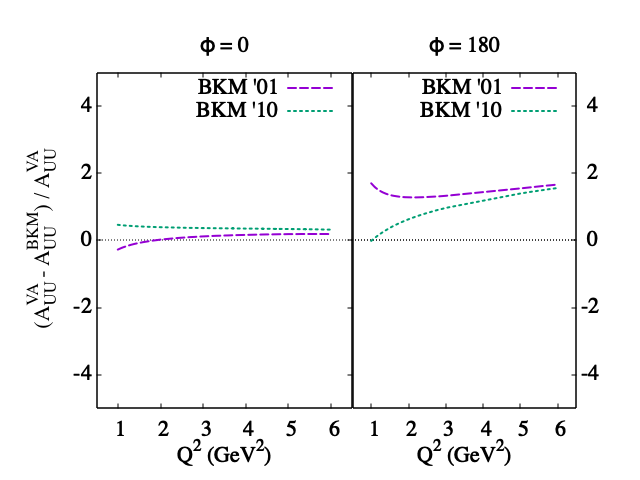}
\includegraphics[width=7cm]{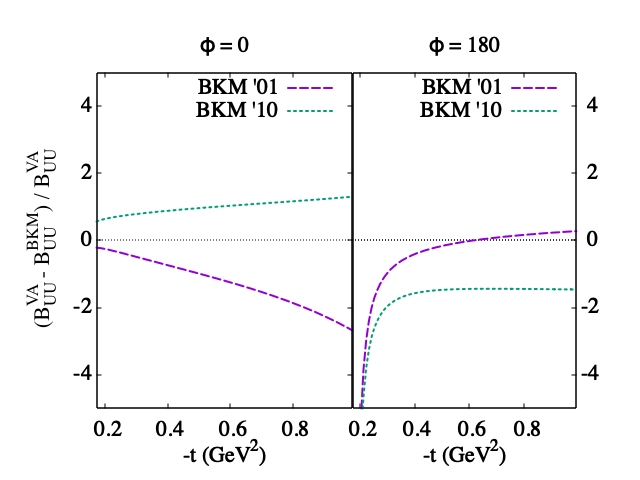}
\includegraphics[width=7cm]{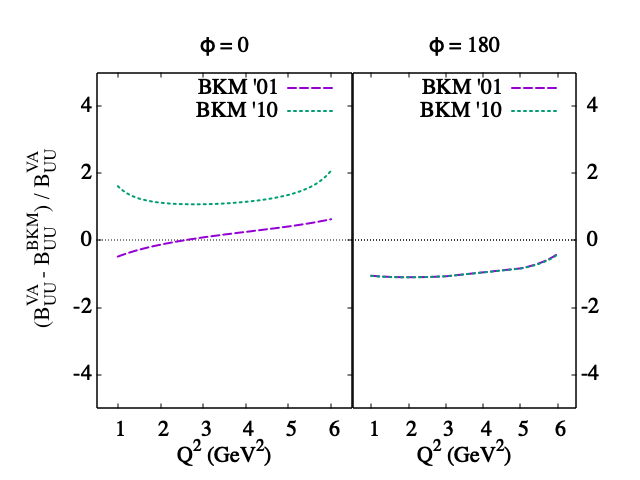}
\includegraphics[width=7cm]{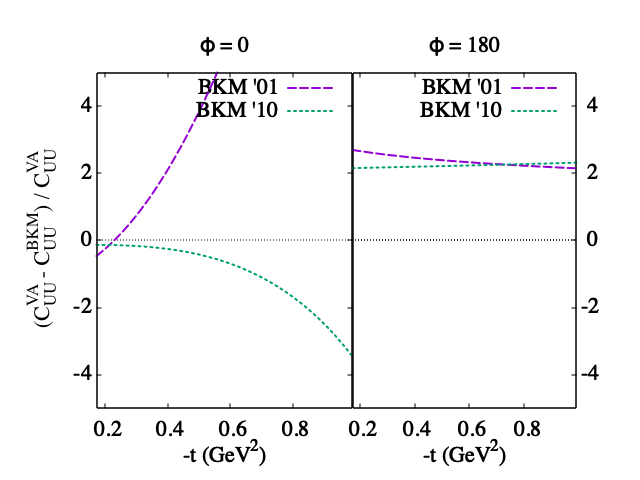}
\includegraphics[width=7cm]{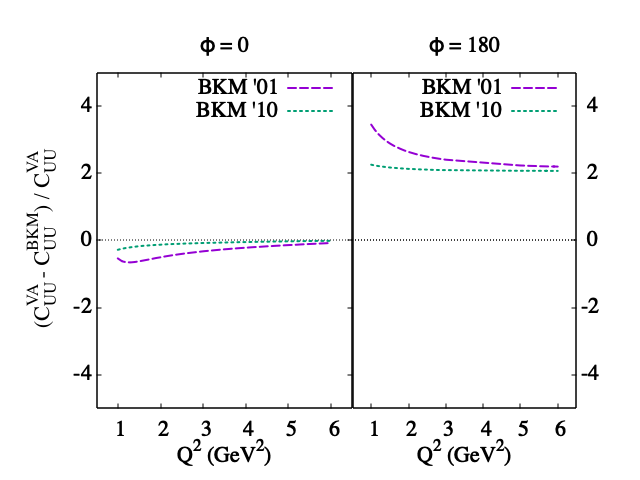}
\end{center}
\caption{{Coefficients $A_{UU}^{\cal I}$ (top panels), $B_{UU}^{\cal I}$ (middle panels), $C_{UU}^{\cal I}$ (bottom panels). On the {\it lhs} the coefficients are compared to the formulation from Ref.\cite{Belitsky:2001ns} (BKM'01) and \cite{Belitsky:2010jw} (BKM'10), plotted  vs. $-t$ at two different values of $\phi$ for the kinematic setting: $\epsilon_1 = 5.7$ GeV, $x_{Bj} = 0.34$, $Q^2=1.8$ GeV$^2$;  
%{\it Right}: 
on the {\it rhs}  we show the percentage deviations between the BKM'01, BKM'10 calculations and the formalism presented in this paper plotted vs. $Q^2$,  for $\epsilon_1 = 10.591 \,\, \text{GeV}$, $ x_{Bj} = 0.34, -t = 0.17 \,\, \text{GeV}^{2}$.}}
\label{fig:int_comp_bkm3}
\end{figure*}

%%%%%%%%%%%%%%%%%%%%
%%% Figure 10
%%%%% twist three
\begin{figure}
\includegraphics[width=7.5cm]{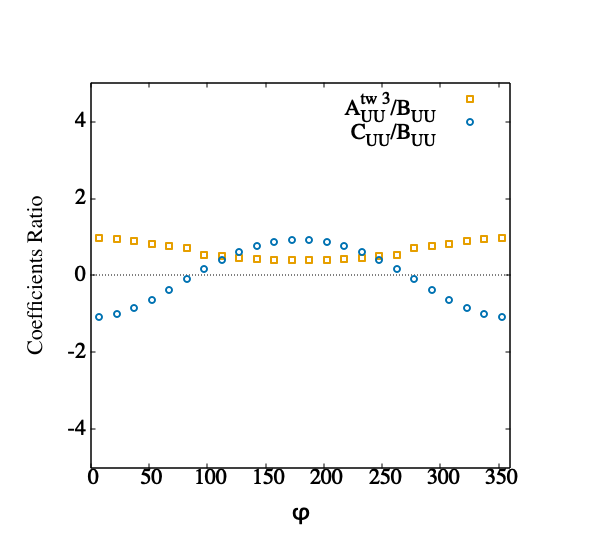}
\caption{  {Ratio of the kinematic coefficient  of the twist three term, $A_{UU}^{(3)\cal{I}}$, in Eq.\eqref{eq:Int_FUU3}, over the twist two coefficient $B_{UU}$ in Eq.\eqref{eq:siguuI}, plotted vs. $\phi$ for $\epsilon_1 = 5.75$ GeV, $Q^2 = 1.820$ GeV$^2$, $x_{Bj}= 0.34$ and $t= -0.17$ GeV$^2$. For comparison we also plot the ratio of the kinematically subdominant twist two coefficient,  $C_{UU}$ over $B_{UU}$.}}
\label{fig:twist3}
\end{figure}

%We present results for both the cross section $\sigma_{UU}$, and the beam polarized cross section, $\sigma_{LU}$,  using the  GPD
%model,  
%\begin{itemize}
%    \item  the dependence on the variables, $t$, $Q^2$ and $x_{Bj}$, separating out the kinematic dependence on $Q^2$ from the twist expansion one
%    \item  the contribution of the electric, magnetic and axial currents to the interference term 
%\item the comparison to BKM 
%    \item the twist-three contribution in WW approximation
%\end{itemize}

\noindent $\bullet$ 
%%% Figure 2 description
%%%%%%%%%
In Figure \ref{fig:xsx_va} we present the cross section $\sigma_{UU}$, Eq.\eqref{eq:siguu}, with the separate contributions, $\sigma_{UU}^{BH}$, \eqref{eq:siguuBH}, $\sigma_{UU}^{DVCS}$, \eqref{eq:siguuDVCS}, and $\sigma_{UU}^{\cal I}$, \eqref{eq:siguuI}, calculated in the VA  framework. We consider three different settings in the laboratory system with electron beam energies: $(k_{e})_o = \epsilon_1 = $ 5.75, 11.5, 24 GeV, and correspondingly increasing $Q^2$ values. The experimental data are from Ref.\cite{Defurne:2015kxq} (top panel) and Ref.\cite{Georges:2018kyi} (middle panel). The 24 GeV setting \cite{Bogasz} is becoming an exciting possibility that will allow further explorations of GPDs and the 3D structure of the nucleon in a wide kinematic range. In particular, the access to larger $Q^2$ values in the valence region will allow us to settle many issues related to power corrections and the onset of QCD factorization.   
The theoretical predictions {for DVCS and the DVCS-BH interference} were calculated at leading twist using GPDs from the spectator model in \cite{Goldstein:2010gu,GonzalezHernandez:2012jv,Kriesten:2021sqc}, summarized in Section \ref{sec:GPDmodel}. We underline that these are predictions, not fits, where the model parameters were fixed using constraints from experiments other than DVCS. The latter include recent nucleon form factor and PDF measurements. These curves show a realistic picture of the   {relative sizes} of the various contributions.
The BH term is known to high precision, since its calculation is based on QED, the only unknowns being the nucleon form factors at low four-momentum transfer, $t$, where their uncertainty is small. The uncertainty band in the figure refers to the error from the fit  in  \cite{Goldstein:2010gu,GonzalezHernandez:2012jv}.

%%% Figure 3 description %%%%
\noindent $\bullet$   {The various contributions to the unpolarized cross section are shown in Figure \ref{fig:xsx_eic} for collider configurations at the EIC typical kinematic setting (upper panel) \cite{eicyellowreport}, and at EIcC kinematics (lower panel)\cite{Anderle:2021wcy}.}
%Using our formalism, since $\sigma_{UU}^{\cal I}$ is proportional to $\cos \phi$ and $\sigma_{UU}^{\cal DVCS}$ is constant in $\phi$, we can separate out the interference and DVCS terms by first subtracting  the BH calculation, and then evaluating the DVCS term from the value of the cross section at $90$ and $270$ degrees (where the interference term should be equal to zero), respectively.  \textcolor{blue}{we do this for the Rosenbluth separation but what is shown here is the calculation of the DVCS using VA theoretical CFFs and their respective errors.} 

%%% Figure 4 description
\noindent $\bullet$ In Figure \ref{fig:int_comp_bkm5} we compare {three cross section formulations: the present framework (VA), the formulations of Refs.\cite{Belitsky:2001ns} (BKM'01), and Ref.\cite{Belitsky:2010jw} (BKM'10), respectively. It should be noticed that the quantity plotted, $\sigma_{UU}$, is the sum of BH, DVCS and BH-DVCS interefernce contributions. All three calculations use the same BH cross section, and the DVCS and DVCS-BH interference terms use the same CFFs values displayed in Table \ref{tab:CFF}, but they differ in the analytic form of the coefficients from the three different formulations. One can see sensible discrepancies between the VA and BKM calculations. It should be noticed that the differences are suppressed, {\it i.e.} they appear smaller in value owing to the fact that the cross section is dominated by the BH contribution which is the same in both the VA and BKM formulations. The differences will, however, affect the extraction of the CFFs. }
%Notwithstanding, discrepancies among the frameworks will strongly affect quantitative fits to the data.

%%% Figure 5 desciption
\noindent $\bullet$  
%where we plot our extraction from the data of , along with theoretical predictions in the reggeized diquark model for the same kinematic choices as Fig.\ref{fig:xsx_va}, 
%for initial electron energy $\epsilon_1=$ 5.75 GeV (top panel)  Ref.\cite{Defurne:2015kxq},  $\epsilon_1=$11.5 GeV (middle panel) Ref.\cite{Georges:2018kyi}, and for a projected value of a fixed target experiment with $\epsilon_1=$ 24 GeV. 
To better describe the size of the difference between the two formalisms we compare the $\sigma_{UU}^{\cal I} + \sigma{_UU}^{DVCS}$ terms in Fig. \ref{fig:int_comp_bkm_va}.  One can see clear discrepancies between the two frameworks, that do not seem to decrease with increasing electron energy. Notice that the VA formalism has different features of the $\phi$ modulations characterizing the cross section at central values of $\phi$. 
%These numerical differences persist in the $1.5 < Q^2 < 10$ GeV $^2$ range, and up to $\epsilon_1$= 24 GeV. 

  {This feature shows up clearly in the {\it lhs} panels.
%Without that overall cos phi the curves almost agree, and this is shown in the same graphs (adding a curve). 
On the {\it rhs} the effect of the extra terms originating from gauge invariance preserving coefficients is displayed. The impact of these terms tends to disappear with larger $Q^2$. 
%On the rhs we show both the theoretical prediction and a calculation done fitting the CFFs to the data. 
It should be stressed that none of the curves shown in the figure corresponds to a fit of the DVCS data in that the values of the CFFs are theoretical predictions.}

%%% Figure 6 description
%%%%%%%%%%%
\noindent $\bullet$ The $t$ dependence of the DVCS contribution, $\sigma_{UU}^{DVCS}$, is presented in Figure \ref{fig:DVCS}, for the kinematic bin, $\epsilon_1=$ 5.75 GeV, $Q^2 = 1.8$ GeV$^2$, $x_{Bj}=0.34$  (other kinematics display a similar trend). One can see that for this term, the improved calculation of Ref.\cite{Belitsky:2010jw} brings the VA and BKM evaluations closer.

%%%% Figure 7 and 8 description
  {\noindent $\bullet$ To interpret the origin of the BKM-VA discrepancies, in Figures \ref{fig:int_va} and \ref{fig:int_comp_bkm2}
we juxtapose calculations using the VA formalism (left panels) to the BKM formalism (right panels). Fig.\ref{fig:int_va} shows the same quantity,  $\sigma_{UU}^{DVCS}+ \sigma_{UU}^{\cal I}$, as in Fig.\ref{fig:int_comp_bkm_va} displaying 
the contributions to the latter from the three terms, $A_{UU}^{\cal I} (F_1 {\cal H} + \tau F_2 {\cal E})$,  $B_{UU}^{\cal I}\propto G_M ({\cal H} + {\cal E}$), and $C_{UU}^{\cal I} \propto G_M \widetilde{H}$ (Eq.(\ref{eq:siguuI}). From the figure one can see that the term proportional to $A_{UU}^{\cal I}$ dominates the cross section. What is striking is the different weight that the  $A_{UU}^{\cal I}$, $B_{UU}^{\cal I}$, and $C_{UU}^{\cal I}$ terms carry, respectively, in the VA and BKM frameworks.
The differences with the VA formalism are particularly striking for the axial term, $C_{UU}^{\cal I}$, which is both smaller in size and has a complex $\phi$ modulations for the VA case. These differences persist in the kinematic range of Jlab @ 12 GeV Ref.\cite{Georges:2018kyi} (not shown in the figure). 
While the $A_{UU}^{\cal I}$ term dominates the VA cross section, the contribution from $C_{UU}^{\cal I}$ is important in the BKM case, especially with increasing energy. 
The uncertainty bands in the figure   {represent the theoretical error evaluated using the model in Ref. \cite{GonzalezHernandez:2012jv}}. }

%%%% Figure 9
\noindent $\bullet$  In Figure  \ref{fig:int_comp_bkm3} we compare in detail the coefficients, $A_{UU}^{\cal I}$, $B_{UU}^{\cal I}$, $C_{UU}^{\cal I}$  for the BKM and VA formulations.
%%%
%%%
%
In order to understand whether the discrepancies are due to terms proportional to $M^2/Q^2$, $t/Q^2$, we studied the behavior of the coefficients vs. $t$ and $Q^2$. 
The results shown in Figure \ref{fig:int_comp_bkm3} represent  the percentage deviations of the BKM'01 and BKM'10  coefficients from the VA ones, $A_{UU}$ (top), $B_{UU}$ (middle), and $C_{UU}$ (bottom), evaluated at $\phi=0$ (left panel) and $\phi=180$ degrees (right panel). On the {\it lhs} we plot the percentage deviations as a function of $-t$,  at $Q^2=1.8$ GeV$^2$; on the {\it rhs} they are plotted vs. $Q^2$ at $-t= 0.17$ GeV$^2$. 
Notice that the differences among the approaches tend to subside at small $t$, and in the large $Q^2$ limit. Our findings substantiate the hypothesis that the treatment of $t$-dependent  and target mass corrections is important, although no systematic effect can be singled out. 
%Fig.\ref{fig:int_comp_bkm3} also illustrates clearly how the discrepancies between BKM'01 and BKM'10 diminish in the high $Q^2$ limit for $\phi = 0$, while at $\phi=180$ degrees the differences persist. 

%Summarizing, in the range $1.5 < Q^2 < 10$ GeV$^2$, which represents a sweet spot for extracting CFFs in the Jlab @ 12 GeV program and elsewhere, a precise determination of the $t$ and target mass dependence of the coefficients is mandatory for a meaningful extraction of the CFFs.
%%% Figure 6 Kumericki

%%%% Figure 10 description
\noindent {$\bullet$ We also evaluated the impact of the coefficients of the twist three contributions  These are presented in Figure \ref{fig:twist3}. Our estimate shows that twist three terms are small, of the same size of the   {$B_{UU}^{\cal I}$, $C_{UU}^{\cal I}$ } terms (see Figs.\ref{fig:int_va}, \ref{fig:int_comp_bkm2} ).}

%%%%%%%%%%%%%%%%
%%% FIGURE 11
\begin{figure}
\includegraphics[width=7cm]{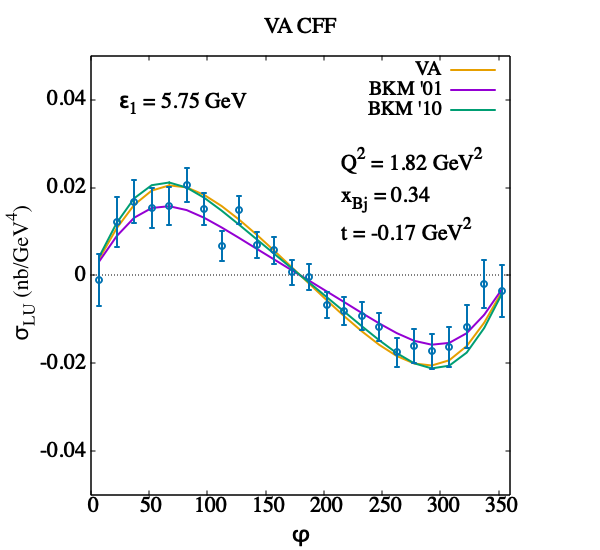}
\includegraphics[width=7cm]{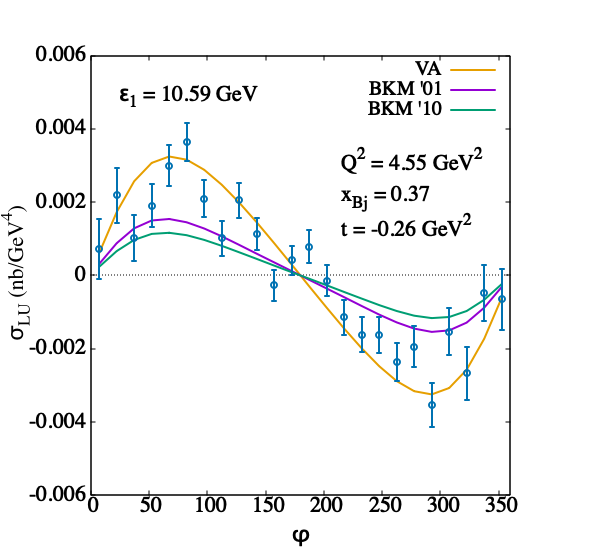}
\includegraphics[width=7cm]{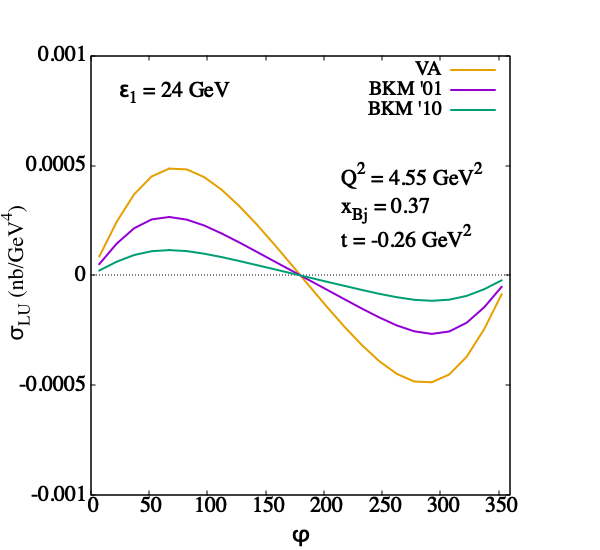}
\caption{Total LU  Cross Section with VA theory CFF's at kinematic bins: $Q^{2} = 1.82 \,\, \text{GeV}^{2}, \, x_{Bj} = 0.34, \, t = -0.17 \,\, \text{GeV}^{2}, \, \epsilon_1 = 5.75 \,\, \text{GeV}$ (top);  $Q^{2} = 4.55 \,\, \text{GeV}^{2}, \, x_{Bj} = 0.37, \, t = -0.26 \,\, \text{GeV}^{2}, \, \epsilon_1 = 10.591 \,\, \text{GeV}$ (middle);  $Q^{2} = 4.55 \,\, \text{GeV}^{2}, \, x_{Bj} = 0.37, \, t = -0.26 \,\, \text{GeV}^{2}, \, \epsilon_1 = 24 \,\, \text{GeV}$ (bottom).}
\label{fig:LU_BKM_VA}
\end{figure}

%%%%%%%%%%%%%%%%
%%% FIGURE 12
\begin{figure}
\includegraphics[width=7cm]{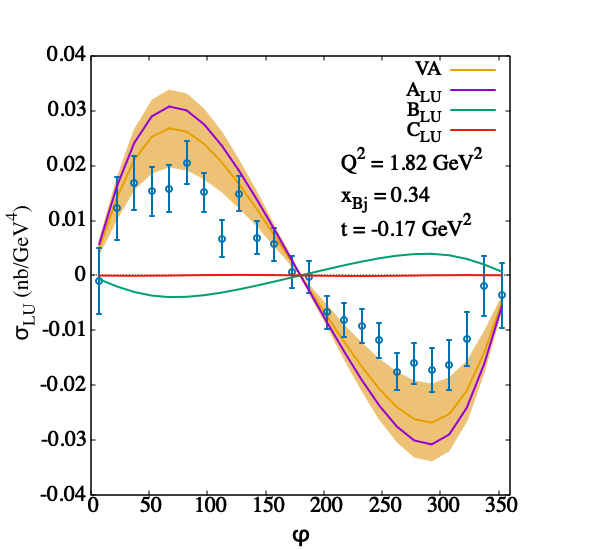}
\includegraphics[width=7cm]{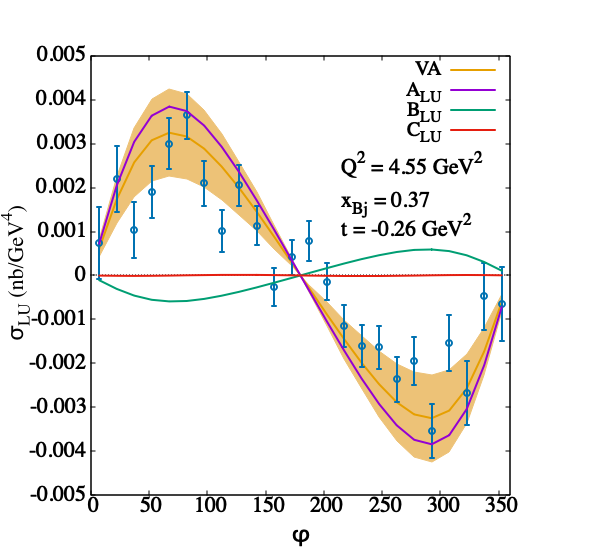}
\includegraphics[width=7cm]{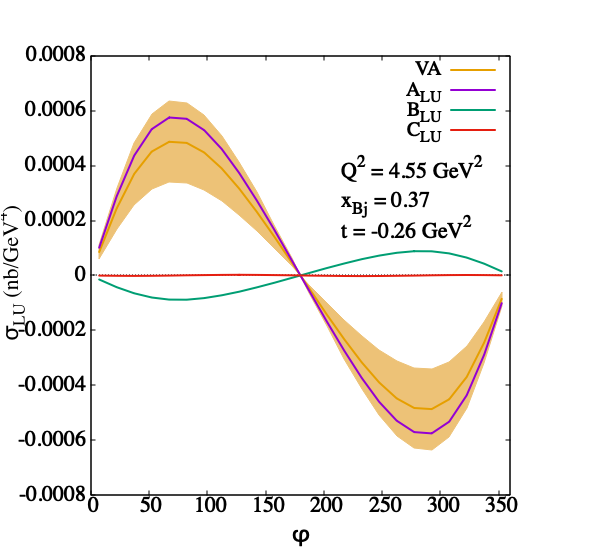}
\caption{Total LU  Cross Section with VA theory CFF's at the same kinematic bins as Fig.\ref{fig:LU_BKM_VA}. The different contributions from the $A_{LU}, B_{LU}$ and $C_{LU}$ terms are shown cross section are shown.}
\label{fig:LU_VA}
\end{figure}

%%%%Figure 13
%%%%%
\begin{center}
\begin{figure}
\includegraphics[width=7.5cm]{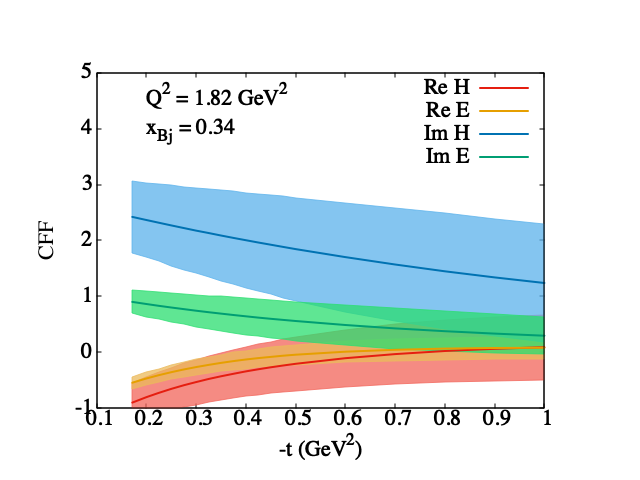}
\caption{VA Compton form factors $H$ and $E$ calculated in the reggeized diquark model, Eq.\eqref{eq:model}.}
\label{fig:CFF}
\end{figure}
\end{center}

%%%%%%%%%%%%%
%%% FIGURE 14
%%%%%%%%%%%%%
\begin{figure*}
\begin{center}
%\includegraphics[width=7.cm]{Brandon New Plots/UU plots/int_xsx_kumericki_errors_q21p82xbjp34tp17.png}
%\includegraphics[width=7.cm]{Brandon New Plots/UU plots/int_xsx_kumericki_bkm_q21p82xbjp34tp17.png}
%%%
%\includegraphics[width=7.cm]{Brandon New Plots/UU plots/int_xsx_kumericki_errors_q24p55xbjp37tp26.png}
%\includegraphics[width=7.cm]{Brandon New Plots/UU plots/int_xsx_kumericki_bkm_q24p55xbjp37tp26.png}
%%%
%\includegraphics[width=7.cm]{Brandon New Plots/UU plots/int_xsx_kumericki_q24p55xbjp37tp26eb24gev.png}
%\includegraphics[width=7.cm]{Brandon New Plots/UU plots/int_xsx_kumericki_bkm_q24p55xbjp37tp26eb24gev.png}
\includegraphics[width=7.cm]{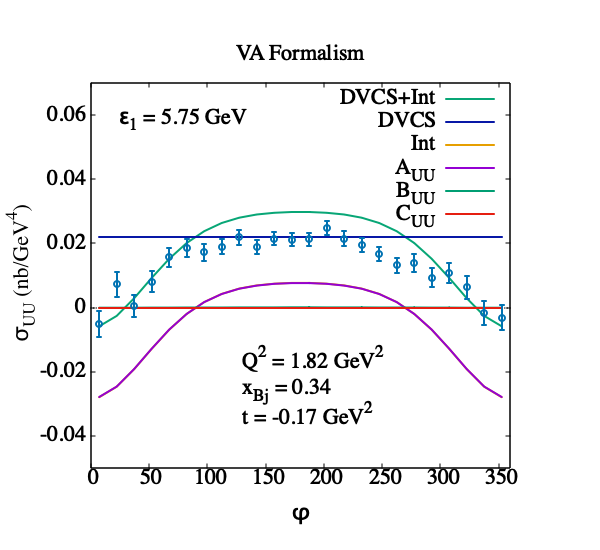}
\includegraphics[width=7.cm]{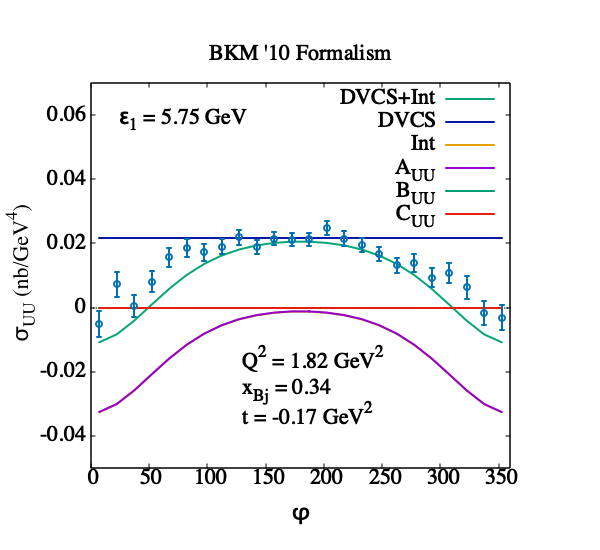}
%%%
\includegraphics[width=7.cm]{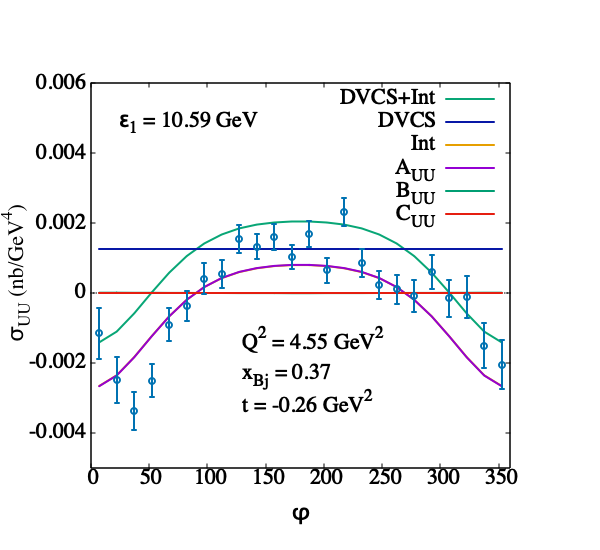}
\includegraphics[width=7.cm]{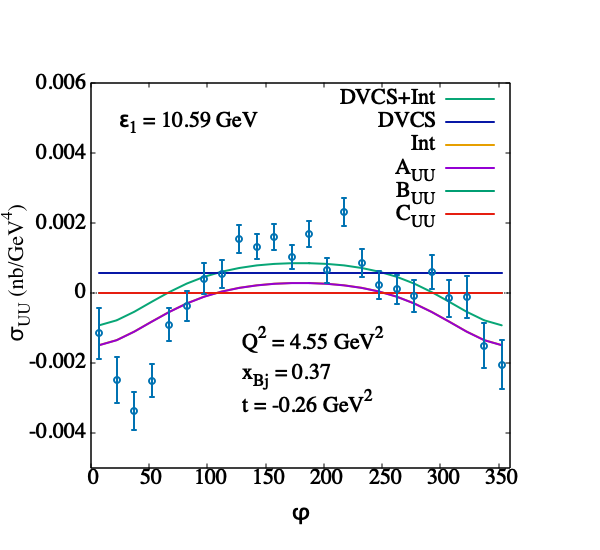}
%%%
\includegraphics[width=7.cm]{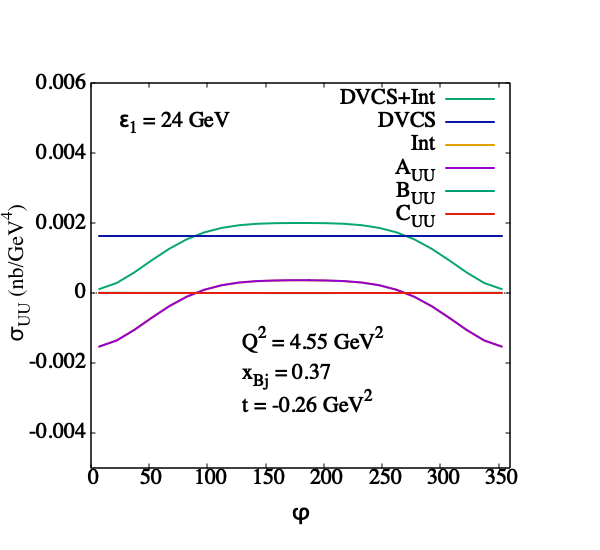}
\includegraphics[width=7.cm]{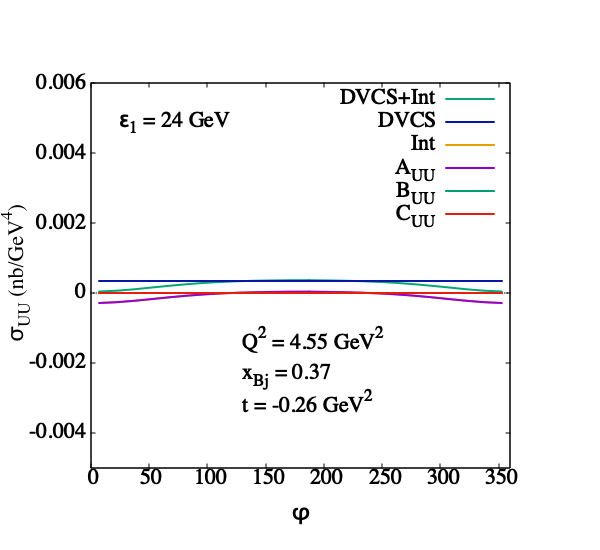}
\end{center}
\caption{Sum of contributions $\sigma_{UU}^{\cal I}+\sigma_{UU}^{DVCS}$ using CFF values from Kumericki et al. \cite{Kumericki:2011zc} in two different formalisms for the cross section: VA formalism (left), BKM10 formalism (right). }
\label{fig:int_comp_kumericki}
\end{figure*}
%

%%%% Figures 11 and 12 description
\noindent $\bullet$ Finally, in Figures \ref{fig:LU_BKM_VA}, \ref{fig:LU_VA}, we present our results for the polarized beam cross section, $\sigma_{LU}$, Eq.\eqref{eq:siglu}. 
In Fig. \ref{fig:LU_BKM_VA} we show the comparison with BKM'01 and BKM'10. Notice that with the VA formalism one can see that the twist-two CFFs do not describe quantitatively the cross section at $Q^2=1.8$ GeV$^2$, while the agreement improves increasing $Q^2$ to $4.5$ GeV$^2$. 
Fig.\ref{fig:LU_VA} displays the different contributions from the  $A_{LU}^{\cal I}$, $B_{LU}^{\cal I}$, $C_{LU}^{\cal I}$ terms, Eq.\eqref{eq:sigLUI}. Notice, in this case, the smallness of the axial contribution, $\propto G_M \Im m \widetilde{H}$. 

%%%%%%
\subsection{GPD Model}
\label{sec:GPDmodel}
To compare the VA and BKM cross section frameworks we used the  parametrization from Refs. \cite{Goldstein:2012az,GonzalezHernandez:2012jv,Kriesten:2021sqc} which is based on the reggeized diquark model.
In the DGLAP region, $x>\xi$, the parametric form for $F_q = H_q, E_q, \widetilde{H}_q, \widetilde{E}_q$,  at the initial scale $Q_o^2 \approx 0.1$ GeV$^2$, reads,
\begin{equation}
\label{eq:model}
    F_q(x,\xi,t) = {\cal N}_q \, x^{[\alpha_q +\alpha_q'(1-x)^{p_q} \, t]} \, F_{diq}(x,\xi,t) \, .  \quad\quad q=u,d
\end{equation}
where $F_{diq}$, is obtained from a diquark calculation with mass parameters, $m_q$ (quark mass), $M_\Lambda$ (dipole cut-off mass), $M_X^q$ (spectator diquark mass);  $\propto x^{[\alpha_q +\alpha_q'(1-x)^{p_q} \, t]}$ accounts for the Regge behavior at low $x$. In the ERBL region, $-\xi < x< \xi$, we use a simple parametric form constrained by parity conservation and charge conjugation.  
The parameters for the twist-two GPDs are constrained from experimental data on the nucleon elastic form factors and PDFs, using: 

\vspace{0.3cm}\noindent {\it i)} the GPD normalization conditions,
\begin{eqnarray}
\label{eq:constrain1}
\int_{-1}^{1} H_q(x,\xi,t; Q^2)\, dx  & = & F_1^q(t), \nonumber  \\ \int_{-1}^{1} E_q(x,\xi,t; Q^2) \, dx  &  = & F_2^q(t) \nonumber \\
\int_{-1}^{1} \widetilde{H}_q(x,\xi,t; Q^2) \, dx & = & G_A^q(t) \nonumber \\  \quad  \int_{-1}^{1} \widetilde{E}_q(x,\xi,t; Q^2)\, dx  &  = &  G_P^q(t) \, ,
\end{eqnarray}
where we used the flavor separated data on the elastic nucleon form factors, $F_1^q$ and $F_2^q$ \cite{Cates:2011pz}, and the nucleon axial \cite{Schindler:2006jq} and pseudoscalar \cite{Gorringe:2002xx} form factor parametrizations; 

\vspace{0.3cm}\noindent {\it ii)} the forward limit conditions,
\begin{eqnarray}
\label{eq:constrain2}
H_q(x,0,0; Q^2) = q(x,Q^2), \quad  \widetilde{H}_q (x,0,0; Q^2) = \Delta q(x,Q^2) \nonumber \\
\end{eqnarray}
with the unpolarized PDF, $q(x)$, and the helicity distribution, $\Delta q(x)$, being evaluated using current nucleon PDFs parametrizations (details are in Refs.\cite{Ahmad:2006gn,Ahmad:2007vw}).
To compare with data, the GPDs are perturbatively evolved at leading order to the scale, of the data, $Q^2$ \cite{GolecBiernat:1998ja,Ji:1997nk,Ji:1998xh,Musatov:1999xp}.
%

%%%% Figure 13 description
  {In Table \ref{tab:CFF} we present the values of the CFFs, ${\cal H},  {\cal E},  \widetilde{\cal H}, \widetilde{\cal E}$, calculated at  $x_{Bj} = 0.34, 0.37$, $-t= 0.17, 0.26$ GeV$^2$ and $Q^2 = 1.8, 4.5$ GeV$^2$, compared with the values from the analyses in Refs.\cite{Kumericki:2011zc,Kumericki:2015lhb}.} The imaginary and real components of ${\cal H}$ and ${\cal E}$ from the VA model are also shown in Figure \ref{fig:CFF} in a similar kinematic range. The uncertainty bands in the figure represent the theoretical error of the parametrization \cite{GonzalezHernandez:2012jv}.

\begin{figure}
\includegraphics[width=7cm]{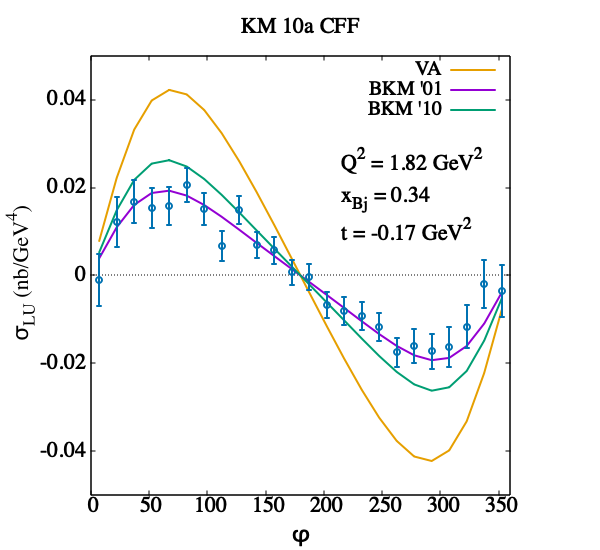}
\includegraphics[width=7cm]{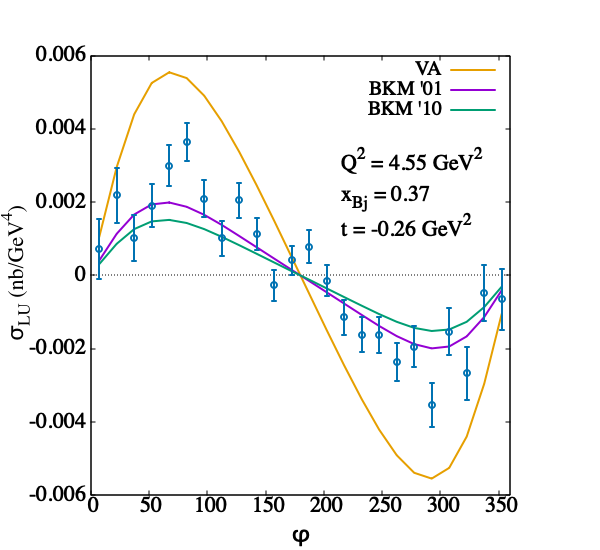}
\includegraphics[width=7cm]{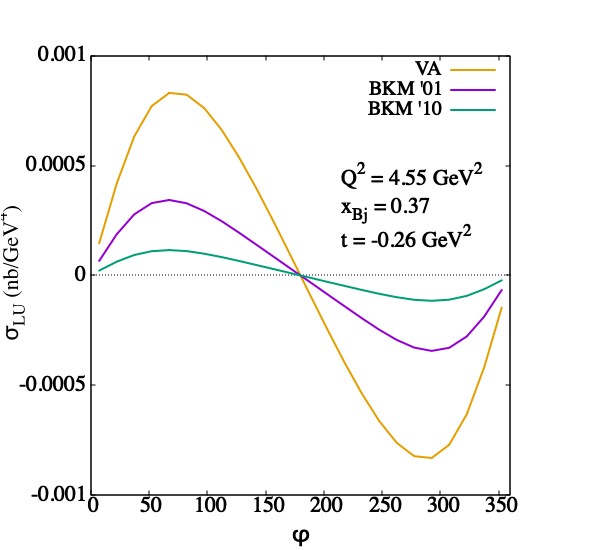}
\caption{Total LU  Cross Section with Kumericki CFF's at kinematic bin {\it Left:} $Q^{2} = 1.82 \,\, \text{GeV}^{2}, \, x_{Bj} = 0.34, \, t = -0.17 \,\, \text{GeV}^{2}, \, \epsilon = 5.75 \,\, \text{GeV}$ and {\it Right:}  $Q^{2} = 4.55 \,\, \text{GeV}^{2}, \, x_{Bj} = 0.37, \, t = -0.26 \,\, \text{GeV}^{2}, \, \epsilon = 10.591 \,\, \text{GeV}$. Our prediction for a 24 GeV beam energy.}
\label{fig:LU_Kumericki}
\end{figure}

%%%%%%%%%%%%%%%%%%%%%%%%%%%%%%%%%%%%%%%%%%%%%%%%%%%%%%%%%%%
%%% Figure 14 description
{In Figure \ref{fig:int_comp_kumericki} we show the potential impact of the new formalism on extracting the values of the CFFs from experiment. On the {\it rhs} we show $\sigma_{UU}^{\cal I}+ \sigma_{UU}^{DVCS}$ using the CFFs extracted from the fit in Ref.\cite{Kumericki:2011zc} using the BKM formalism.   {On the {\it lhs} we show the same quantity evaluated using the same CFFs from Ref.\cite{Kumericki:2011zc}, but with the VA formalism. The fact that the data on the {\it lhs} can no longer be fitted, quantifies once more the discrepancies between the two frameworks.}} Notice, in particular, that the value of the KM15 $\Re e \widetilde{E}$ contributing to $\sigma_{UU}^{DVCS}$, makes this term three times larger than our value.
(All values of the form factors used in the plots are displayed in Table \ref{tab:CFF}. )
%and thus the data sit very low as shown in the first row of
%Fig.\ref{fig:int_comp_kumericki}.  
%
%%%%% Figure 15
Figure \ref{fig:LU_Kumericki} shows a similar trend for $\sigma_{LU}$. 
\begin{widetext}
\begin{table*}
\centering
\begin{tabular}{|c|c|c|c|c|c|c|c|c|c|c|c|}
\hline
\,\, CFF \,\, & \,\,$x_{Bj}$ \,\,&  \,\, $-t \,\, (\text{GeV}^{2})$ \,\, & \,\, $Q^{2} \,\, (\text{GeV}^{2}) $ \,\, &  \,\,  Re $H$  \,\, &  \,\, Re $E$  \,\, &  \,\, Re $\widetilde{H}$  \,\, &  \,\, Re $\widetilde{E}$  \,\, &  \,\, Im $H$  \,\, &  \,\, Im $E$  \,\, & \,\,  Im $\widetilde{H}$  \,\, & \,\,   Im $\widetilde{E}$  \,\,  \\
 \hline
    VA & 0.34 & 0.17 & 1.82 & -0.897 & -0.541  & 0.244 & 2.207  & 4.842  & 1.806  & 1.131  & 5.383  \\
 \hline 
 VA & 0.37 & 0.26 & 4.55 &
   -0.884  & -0.424  & 0.312  & 2.900  & 3.702  & 1.298  & 0.911  & 3.915  \\
 \hline \hline
  KM15 & 0.34 & 0.17 & 1.82 &
    -2.254 &  2.212 &  1.399 & 141.362  &  3.506 & -  & 1.565  & -  \\
 \hline
   KM15 & 0.37 & 0.26 & 4.55 &
    -2.143 &  1.990 &  1.098 & 87.385  &  2.793 & -  & 1.371  & -  \\
 \hline \hline 
    KM10a & 0.34 & 0.17 & 1.82 &
    -1.513 &  1.583 &  - & 40.863  &  3.783 & -  & -  & -  \\
 \hline
     KM10a & 0.37 & 0.26 & 4.55 &
    -1.574 &  1.518 &  - & 22.146  &  3.147 & -  & -  & -  \\
 \hline
\end{tabular}
\caption{Value of Compton Form Factors using Virginia reggeized spectator model, global extraction from KM 15 \cite{Kumericki:2015lhb}  and KM 10a \cite{Kumericki:2011zc}.} 
\label{tab:CFF}
\end{table*}
\end{widetext}

%%%%%%%%%%%%%%%%
%%% FIGURE 12

%%%%%%%%%%%%%%%%%%%%%%%%%%%%%
%%%%%%%%%%%%%%%%%%%%%%%%%%%%%
%%%%%%%%%%%%%%%%%%%%%%%%%%%%%
\section{Conclusions and Outlook}
\label{sec:4}
In order to extract information on the QCD matrix elements of deeply virtual exclusive electron scattering processes, one needs to first understand the detailed structure of the cross section. 
In deeply virtual exclusive photoproduction, in particular, 
%the structure functions containing information on GPDs enter 
%the cross section in expressions with lengthy coefficients 
tracking analytically the dependence in the high  $Q^2$ limit on the invariants $x_{Bj}$ and  $t$,  as well as on the angle $\phi$ between the lepton and hadron planes, has constituted a challenge which has been hampering, so far,   a clean extraction of the various contributions to the cross section. 

  {In our study of unpolarized scattering, differently from previous approaches, we argue that by organizing the cross section for the $ep \rightarrow e'p'\gamma$  scattering process according to its electromagnetic structure allows us to clearly separate the contributions of the various twist two CFFs as well as the twist three components.}

{Furthermore the cross section is described in terms of manageable and streamlined structures for the BH and DVCS contributions  up to twist three. In particular, the
%For the BH process we obtain an expression similar to elastic scattering experiments where the coefficients of the electric and magnetic proton elastic form factors squared acquire a $\phi$ dependence due to the emission of a real photon, and the tilt of the exchanged virtual photon off the lepton plane (along $\Delta$).
%
%On the other hand, the DVCS contribution is described with the virtual photon, $q$, aligned along the $z$-axis, cast in a form similar to the semi-inclusive $ep$ scattering process, where the various structure functions depend on bilinear combinations of CFFs multiplied by simple expressions in the variable $\epsilon$ (the longitudinal to transverse virtual photon polarization ratio). Twist-three contributions can be unambiguously separated out by the phase, $\phi$, dependence . 
%
DVCS-BH interference term is described  by:}
\begin{itemize}
\item 
{an electric contribution, $(F_1-\tau F_2)({\cal H} - \tau {\cal E})$}
\item 
{a magnetic contribution, $(F_1+F_2)({\cal H}+{\cal E})$ containing the combination of GPDs necessary to extract angular momentum,
}
\item {an axial contribution, $(F_1+F_2) \, \widetilde{\cal H}$ }
\end{itemize}
{The latter is reminiscent of the $G_M G_A$ contribution in elastic scattering,} but it is now allowed without violating parity conservation because of the extra degree of freedom provided by the outgoing photon with momentum $q'\neq q$. 

   The kinematic coefficients are lengthy but straightforward to calculate functions of $\phi$ evaluated to all orders in $1/Q$. In addition to a kinematic $\phi$ dependence, we explain how the virtual photon phase $\phi$ dependence originates in a clearly distinguishable way. {This distinction is important to for disentangling terms of different twist both in DVCS and in related processes including {\it e.g.} timelike Compton scattering.}
 
%The new framework allows us to represent the interference  cross section in a linear form in the CFF combinations, ${\cal H}-\tau {\cal E}$ and ${\cal H}+{\cal E}$, and to perform a Rosenbluth separation.
%A quantitative extraction of the CFFs from available data through such a linear fit of the reduced cross section will be presented in upcoming work. 

%We write the cross section in a form $\phi$ dependence, and the occurrence of inverse power terms in the hard scale, $Q^2$. Our main finding is that one cannot simply disregard power suppressed terms, not even in the high energy limit, but all of these contribute to the coincidence scattering process which resembles elastic electron-nucleon scattering in this aspect. 
%%%
%%%%
The reorganization of the cross section also uncovers substantial discrepancies with the harmonics decomposition of BKM. We find several discrepancies in both the pure DVCS, and DVCS-BH interference terms, while the BH contribution turns out to be numerically equivalent. These discrepancies are important and can affect considerably the extraction of CFFs from data, and consequently any conclusion on the behavior of angular momentum, pressure or shear forces inside the proton. We tracked the differences between the BKM and VA formalism numerically, as a function of the various kinematic variables involved. We conclude that while the differences tend to be reduced at high $Q^2$ for some of the observables, this is not a general rule. 
%The inherent reason behind the discrepancies is that in coincidence reactions, as it is well known from {\it e.g.} $ep$ elastic scattering, one obtains misleading results by organizing  the cross section coefficients according to their  $t/Q^2 \rightarrow 0$ dependence, since this is not motivated in this case by the QCD twist expansion. As expected, making use of decades old experience in coincidence scattering experiments,  the full $Q^2$ and $t$ dependence needs to be considered.   
%This is what our framework provides.    
%Why these differences? Because it in coincidence reactions it is wrong to organize with a precise $Q^2 \rightarrow \infty$ in mind.  
%\textcolor{blue}{It seems that while the total cross section may be minimally affected by our changes, when one looks at the interference cross section we see that emphasis has been placed on different Compton form factors than in our formalism (rough draft). }

{What is the inherent reason behind such discrepancies? Our results might be compared in principle, with the studies in Refs.\cite{Braun:2011dg,Braun:2012hq,Braun:2014sta} which result in both  $t$-dependent and target mass corrections, albeit using the harmonics decomposition.  
A common point of view is that the choice of the leading twist decomposition of the DVCS hadronic tensor is not unique. The consequences of this ambiguity results in a different structure of power corrections.} 

Future work in this direction, including numerical evaluations of twist-three three CFFs, as well as a straightforward extension of our framework to timelike Compton scattering, will help us determine unambiguously the internal dynamics and mechanical properties of the proton.
  
%
%

%Finally,  using the formalism introduced in Ref.\cite{Kriesten:2019jep} based on helicity amplitudes, all the coefficients are expressed in an easy to use covariant form.

%evaluate them within the reggeized diquark model %\cite{Ahmad:2006gn,Ahmad:2007vw,GonzalezHernandez:2012jv}. 

\acknowledgements
We thank SURA and Jefferson Lab for the support from the 
%Commonwealth of Virginia funded 
Center for Nuclear Femtography, and in particular to our colleagues participating in the University of Virginia lead initiative, Pete Alonzi, Matthias Burkardt (NMSU), Gordon Cates, Donal Day, Joshua Hoskins. We are also grateful to Jian-Ping Chen, Xu Cao and Yuxiang Zhao for discussing the projected EicC kinematic settings. Comments from Marie Boer, Markus Diehl and Charles Hyde are also gratefully acknowledged. Finally, we thanks Kyle Shiells and Yushun Guo for a thorough check of many of our formulae.
This work was funded by DOE grants DOE grant DE-SC0016286 and in part by the DOE Topical Collaboration on TMDs (B.K. and S.L.).

\appendix
%%%%%% appendix 
%%%%%
%%%%%
\section{Cross section phase space factors} 
\label{app00}
The cross section for BKM and in this paper (VA) are written in terms of different kinematic variables as,
\begin{eqnarray}
(VA) \quad \rightarrow \quad    \frac{d \sigma}{dx d Q^2 d t d \phi} &= &\Gamma_{VA} \mid T \mid^2 \\
(BKM) \quad \rightarrow \quad     \frac{d \sigma}{dx d y d t d \phi} &= &\Gamma_{BKM} \mid T \mid^2 ,
\end{eqnarray}
where,
 \[ \Gamma_{VA} = \frac{1}{(s-M^2)^2 x}, \quad \quad \Gamma_{BKM} = \frac{xy}{Q^2} = \frac{1}{s-M^2}, \]
\[\frac{\Gamma_{VA}}{\Gamma_{BKM} }= \frac{Q^2}{(s-M^2)^2 x^2 y} = \frac{y}{Q^2} \quad\quad \]
%so that 
%\begin{equation}
% (2 M \epsilon_1 x)     \frac{d \sigma}{dx \, d Q^2 \, d t \, d \phi} =   \frac{d \sigma}{dx d y d t d \phi} 
%\end{equation}
%To extract the matrix element modulus squared one has,
%\begin{eqnarray}
%\mid T \mid^2_{VA} &=&  \frac{1}{\Gamma_{VA}}\left(\frac{d \sigma}{dx \, d Q^2 \, d t d \phi}\right) \\
%\mid T \mid^2_{BKM} &= &\frac{1}{\Gamma_{BKM}}\left( \frac{d \sigma}{dx \, dy \, d t d \phi}\right) \nonumber \\
%&=&\frac{1}{\Gamma_{BKM}} (2 M \epsilon_1 x)  \left(\frac{d \sigma}{dx \, d Q^2 \, d t d \phi}\right) 
%= \frac{\Gamma_{VA}}{\Gamma_{BKM}} (2 M \epsilon_1 x)  \mid T\mid^2_{VA} = \mid T\mid^2_{VA}
%\end{eqnarray}
%%%%

%%%%
%%%% KINEMATIC VARIABLES
\section{Kinematic variables in fixed target and collider frames}
\label{app0}
%%%%%%
%%%%%
%In the laboratory frame for fixed target experiments, t
To describe the reaction 
\[k + p \rightarrow  k' + p' + q' \]
one defines the lepton plane through the four-momenta $k$ and $k'$, with $q=k-k'$, and the initial proton four-momentum, $p$; the final photon momentum, $q'$, and the final proton momentum, $p'$ define the hadron scattering plane.

The cross section, Eq.(\ref{eq:xs5foldgeneral}), depends on the variables $k_o$ (initial electron energy), $Q^2$, $x_{Bj}=Q^2/2M \nu)$, $t$, and $\phi$.

\vspace{0.5cm}\noindent {\it Laboratory frame}

\noindent In the lab, or proton rest frame, the initial electron is aligned with the -z axis, and the final electron is scattered in the $x-z$ plane with components,
\label{eq:momenta}
\begin{align}
k & \equiv  (k_o;0, 0 , k_3= -\mid \vec{k} \mid \approx -k_o) 
\nonumber\\
p & \equiv  (M;0,0,0)  \nonumber \\
k^\prime & \equiv \Big( k'_o=k_o- \frac{Q^2}{2M x_{Bj}} ; \sqrt{k_o^{' 2}-k_3^{' 2}} , 0 , \frac{-Q^2 + 2k'_o k_o}{2 k_3} \Big) \nonumber \\
q & = k - k^\prime  \equiv \Big(\nu; - \sqrt{k_o^{' 2}-k_3^{' 2}}, 0,-\nu\Big(1+\frac{\gamma^2}{2y} \Big) \Big)
\end{align}
where, disregarding the electron mass, $k_o= \mid \vec{k} \mid = \epsilon_1$. The cross section calculations are done with $q$ aligned along the z axis, {\it i.e.} rotating from the lab frame by the angle,
\begin{equation}
    \alpha = \arctan (q_1/q_3) \nonumber
\end{equation}
in the $x - z$ plane. The rotated four-vectors in this frame are denoted by,
$k^R$, $k^{\prime R}$, $q^R$.

The outgoing photon components are derived from the relations,
\begin{eqnarray}
p_o'+q_o' &=& M + \nu \nonumber \\ 
p_3^{\prime R}+q_3^{\prime R} &=& q_3^R \nonumber 
\end{eqnarray}
and using the definitions,
\[ t = (p-p')^2 = 2M^2 - 2(pp'), \quad     t = (q'-q)^2 = -Q^2 - 2(qq') \]
One obtains,
\begin{eqnarray}
q^{\prime R}  \equiv   \Big(q_o'; q_o'  \sin\theta \cos\phi, q_o' \sin\theta \sin\phi, q_3'^R = q_o' \cos \theta \Big) \nonumber 
\end{eqnarray}
where,
\begin{eqnarray}
q_o' & = &   \nu + \frac{t}{2M} , \quad  q_3'^R = -\frac{t + Q^2 +  2 q_o q_o'}{2 \mid \vec{q}^{\, R} \mid }, \nonumber 
\\  \theta &= & \arccos (q_3^{\prime R}/q_o^{\prime}) . \nonumber 
\end{eqnarray}
From these vectors one defines $\Delta=q - q^{\prime}$, and $p'=p+\Delta$.

\vspace{0.5cm}\noindent {\it Collider frame} 

\noindent In the collider frame with the initial proton moving along the positive $z$ axis, one has,
\begin{eqnarray}
k & \equiv & (k_o;0, 0 , k_3= -\mid \vec{k} \mid \approx -k_o) 
\nonumber \\
p &\equiv& (p_o=\sqrt{p_3+M^2}; 0, 0, p_3)  \nonumber \\
k^\prime & \equiv & \Big( k'_o ; \sqrt{k_o^{' 2}-k_3^{' 2}} , 0 , \frac{-Q^2 + 2k'_o k_o}{2 k_3} \Big) \nonumber \\
q & = & k - k^\prime,  \nonumber
\end{eqnarray}
with,
\begin{eqnarray}
k_o' &=& k_o- \left[2 \tilde{\nu} - \sqrt{\tilde{\nu}^2 -  4 \Big(1- \Big(\frac{p_3}{p_o}\Big)^2\Big)\left(\tilde{\nu}^2- \frac{p_3}{p_o} \, Q^2  \right)} \right] 
\nonumber \\
& \times & \Big[2\Big(1- \Big(\frac{p_3}{p_o}\Big)\Big]^{-1} \nonumber \\
& \approx & k_o - \frac{\tilde{\nu}}{2}  + p_o x_{Bj} , \nonumber \\
  \tilde{\nu}  &=&   \frac{Q^2}{ 2 p_o x_{Bj}} \nonumber .
\end{eqnarray}  
%Notice that the relationship between the electron energy and $x_{Bj}$ changes, namely one has,
%\begin{equation}
%x_{Bj} = \frac{Q^2}{2(p_o q_o - p_3 q_3)},    \nonumber 
%\end{equation}
%therefore, using the relation, $Q^2 = -q_o^2 + q_3^2$, one derives the energy transfer, $q_o$, solving the equation,
%\begin{equation}
%\left( 1-\frac{p_3^2}{p_o^2} \right)    q_o^2 - 2 \tilde{\nu} q_o + \left[ \tilde{\nu}^2  - Q^2 \frac{p_3^2}{p_o^2}\right] = 0 \nonumber 
%\end{equation}
%with,
%\begin{equation}
%  \tilde{\nu}  =   \frac{Q^2}{ 2 p_o x_{Bj}} .\nonumber 
%\end{equation}
%The solutions is,
%\begin{equation}
%q_o = \frac{ 2 \tilde{\nu}
%- \left[ 4\tilde{\nu}^2 - 4 \left( 1-\displaystyle\frac{p_3^2}{p_o^2} \right) 
%\left(  \tilde{\nu}^2  - Q^2 \displaystyle\frac{p_3^2}{p_o^2}\right) \right]^{1/2} }{2 \left( 1-\displaystyle\frac{p_3^2}{p_o^2} \right) } \nonumber
%\end{equation}
%where the "-" sign between the first and second terms follows from energy conservation. 
%We then find $k_o' = k_o-q_o$.
%
The outgoing photon four-momentum components are derived by first boosting to the CoM frame with,
\begin{eqnarray}
\beta = -\frac{p_3}{p_o}, \quad 
\gamma = \frac{p_o}{M} \nonumber 
\end{eqnarray}
so that the components of the the boosted virtual photon $q^B$, are,
\begin{eqnarray}
q_o^B & = &\gamma (q_o + \beta q_3) \nonumber \\
q_{1,2}^B & = & q_{1,2} \nonumber \\ 
q_3^B & = & \gamma(q_3 + \beta q_o) \nonumber 
\end{eqnarray}
One now rotates $q^B$ to obtain a vector $q^R$ aligned with the $z$ axis using,
\begin{equation}
    \alpha = \arctan (q_1^B/q_3^B) , \nonumber
\end{equation}
The outgoing photon has components,
\begin{eqnarray}
q^{\prime R}  \equiv   \Big(q_o'; q_o'  \sin\theta \cos\phi, q_o' \sin\theta \sin\phi, q_3'^R \Big) \nonumber 
\end{eqnarray}
where,
\begin{eqnarray}
q_o' & = &   q_o^B + \frac{t}{2M} , \quad  q_3'^R = -\frac{t + Q^2 +  2 q_o^B q_o'}{2 \mid \vec{q}^{\, B} \mid }, \nonumber 
\\  \theta &= & \arccos (q_3^{\prime R}/q_o^{\prime}) . \nonumber 
\end{eqnarray}
$\Delta$ and $p'$ are reconstructed from these vectors. 

%%%%%%%
%%%%%%%
%%%%% appendix B
\section{Kinematic coefficients for BH and BH-DVCS interference terms}
\label{appa}
For convenience we write the expressions of the kinematic coefficients relevant for this paper. All coefficients were calculated in Ref.\cite{Kriesten:2019jep} 
in a covariant form using four vector products with notation, $(ab) = a_o b_o- \vec{a} \cdot \vec{b}$. The relevant four-vector components  for both the laboratory frame and the collider frame are given in Appendix \ref{app0}. 

The BH coefficients in Eq.(\ref{eq:siguuBH}) read,
\begin{widetext}
\begin{align}
A_{BH}   = &\frac{{8}\,M^2}{t \myprod{k}{q'} \myprod{k'}{q'}}
\Bigg[  
4 \tau 
 \Big(
 \myprod{k}{P}^2
 +
 \myprod{k'}{P}^2
 \Big)  
 -(\tau +1)
\Big(
 \myprod{k}{\Delta}^2
 +
 \myprod{k'}{\Delta}^2
\Big)
\Bigg] 
\label{eq:Aunpol}\\
B_{BH} = & 
\frac{{16} \,M^2}{t \myprod{k}{q'} \myprod{k'}{q'}}
\Big[
 \myprod{k}{\Delta}^2
 +
 \myprod{k'}{\Delta}^2
\Big]  \,,
\label{eq:Bunpol}
% \\
% \Gamma  =&\frac{\alpha^3}{16\pi^2 (s-M^2)2 \sqrt{1+\gamma^2} {\color{cyan}x_{Bj}} } 
% \label{Gammaunpol}
\end{align}
where  $P=(p+p')/2$,  $\tau=-t/4M^2$. 

\vspace{0.5cm}

\noindent The BH-DVCS interference cross section coefficients in Eq.(\ref{eq:siguuI}) read,
\begin{eqnarray}
  A_{UU}^{\cal I}  & = & 2 (P \Sigma_{S}^{\cos \phi})  \nonumber \\
 & = & -8D_{+} \Big[ (k'P) \Big(2 k_T^2 -k_T\cdot q_T' - 2(kq')\Big)  
 +  (kP) \Big( 2 k'_T \cdot k_{T} + k'_T \cdot q'_{T} +2(k'q')\Big) - (k_T \cdot P_T)  \Big(2(kk') + (k'q') \Big) \nonumber \\
 & + &  (k_T' \cdot P_T) (kq') \Big] \cos \phi 
 - 8D_{-} \Big[ (Pq') \Big(2 k_T \cdot k'_{T}  + 2(kk')\Big) 
  - (k_T \cdot P_T) (k'q') - (k_T' \cdot P_T)(kq')\nonumber \\
  &+& (P_T \cdot q_T') (kk')  \Big] \cos \phi  
%    \\
%A_{UU}^{\mathcal{I}} &=& -4D_{+}\Big[-(Pk')\Big(2(kk)_{T} - (kq')_{T} + 2(kq')\Big) \nonumber \\
%&& \hspace{0.5cm}-(Pk)\Big(2(kk)_{T} + (kq')_{T} - 2(k'q')\Big) \nonumber \\
%&& \hspace{0.5cm}+(kP)_{T}\Big(2(kk') + (kq') - (k'q')\Big)\Big]\cos{\phi}\nonumber \\
%&-& 4D_{-}\Big[-(Pq')\Big(2(kk)_{T} - 2(kk')\Big)\nonumber \\
%&& \hspace{0.5cm}+(k'q')(kP)_{T} - (kk')(q'P)_{T} + (kq')(kP)_{T}\Big]\cos{\phi} \nonumber \\
\\
% 
%A^{\cal I, OLD}_{UU} &=&\frac{1}{(kq')(k'q')}\Bigg\{ (Q^{2}+t)\bigg[\big((kq')_T - 2(kk)_T - 2(kq')\big)(Pk') + \big(2(k'q') - 2(k'k)_T - (k'q')_T\big)(Pk) + (k'q')(kP)_T \nonumber \\
%&+& (kq')(k'P)_T - 2(kk')(kP)_T\bigg] - (Q^{2}-t+4(k\Delta))\bigg[\big(2(kk') - (k'q')_T - (kk')_T\big)(Pq') + 2(kk')(Pq')_T \nonumber \\
%&-& (k'q')(kP)_T - (kq')(k'P)_T\bigg]\Bigg\} \cos{\phi}\\
%
 B_{UU}^{\cal I} & = &  \xi (\Delta \Sigma_{S}^{\cos \phi})  \nonumber \\
 & = & -4 \xi D_{+} \Big[ (k'\Delta) \Big(2 k_T^2 -k_T\cdot q_T' - 2(kq')\Big)  
 +  (k\Delta) \Big( 2 k'_T \cdot k_{T} + k'_T \cdot q'_{T} + 2(k'q')\Big) - (k_T \cdot \Delta_T)  \Big(2(kk') + (k'q') \Big) \nonumber \\
 & + &  (k_T' \cdot \Delta_T) (kq') \Big] \cos \phi 
 - 4\xi D_{-} \Big[ (\Delta q') \Big(2 k_T \cdot k'_{T}  + 2(kk')\Big) 
  - (k_T \cdot \Delta_T) (k'q') - (k_T' \cdot \Delta_T)(kq')\nonumber \\
  &+& (q_T' \cdot \Delta_T) (kk')  \Big] \cos \phi  \\
 C_{UU}^{\cal I} & = & \frac{1}{2P^{+}}\epsilon^{\mu \sigma \nu +}\Big(P_{\mu}\Delta_{\nu} - P_{\nu} \Delta_{\mu} + \frac{1}{2}\Delta_{\mu}\Delta_{\nu} \Big) g_{\sigma \rho}  \Big(\Sigma_{A}^{\cos \phi} \Big)^{\rho} \nonumber \\
 &=& 
 4\frac{D_{+}}{P^{+}}\Big[ -(k_{T}\cdot P_{T})(q'_{T} \cdot \Delta_{T})k^{\prime +} + (k_{T} \cdot \Delta_{T})(P_{T}\cdot q'_{T})k^{\prime +} + (k'_{T}\cdot P_{T})(q'_{T} \cdot \Delta_{T})k^{+} \nonumber \\
 &-&(k'_{T} \cdot \Delta_{T})(P_{T} \cdot q'_{T})k^{+} - 2(kk')(k'_{T} \cdot P_{T})\Delta^{+} -(kq')(k'_{T} \cdot P_{T})\Delta^{+} \nonumber \\
 &+& (k'q')(k_{T} \cdot P_{T})\Delta^{+} + 2(kk')(k'_{T} \cdot \Delta_{T})P^{+} + (kq')(k'_{T} \cdot \Delta_{T})P^{+} - (k'q')(k_{T} \cdot \Delta_{T})P^{+}\Big] \cos \phi  \nonumber \\
&+& 4\frac{D_{-}}{P^{+}}\Big[(k'q')(k_{T} \cdot P_{T})\Delta^{+} - (kk')(q'_{T} \cdot P_{T})\Delta^{+} + (kq')(k'_{T} \cdot P_{T})\Delta^{+} \nonumber \\
&-& (k'q')(k_{T} \cdot \Delta_{T})P^{+} + (kk')(q'_{T} \cdot \Delta_{T})P^{+} - (kq')(k'_{T} \cdot \Delta_{T})P^{+}\Big] \cos \phi 
\end{eqnarray}
where $(\Sigma_{S}^{\cos \phi})^{\rho}$ and $(\Sigma_{A}^{\cos \phi})^\rho$ are linear combinations of the kinematic four-vectors, $k^\rho, k'^\rho, q'^\rho$, describing the BH-DVCS interference lepton tensor\cite{Kriesten:2019jep},   
\begin{eqnarray}
\left(\Sigma_S^{\cos \phi}\right)^\rho & = &  \Big\{ k'^\rho   \left[ \left(2 k_\mu k_\nu - q'_\mu k_\nu\right) g_T^{\mu\nu}    + 2 (kq') \right] 2 D_+   
%%%%
+  k^\rho    \left[ \left(2 k'_\mu k_\nu + q'_\mu k'_\nu\right) g_T^{\mu\nu}    - 2 (k'q') \right]  2 D_+  
%%%%
%
\nonumber \\
&+& q'^\rho    \left[ 2k_\mu k'_\nu  g_T^{\mu\nu}   -  2 (kk') \right] D_-   
%%%%%%
 + g_T^{\rho \mu}  \left[ \left(-2 (k k') + (k'q') \right) k_{\mu}  - (kq') k'_{\mu}  \right] 2 D_+  \nonumber \\
& + &  g_T^{\rho \mu} \left[ (kk') q'_\mu - (k'q')  k_\mu   - (kq') k'_\mu \right]  2 D_- \Big\} \cos \phi ,
\end{eqnarray}
and,
\begin{eqnarray}
\left(\Sigma_A^{\cos \phi}\right)^\rho & = &  \Big\{ k'^\rho   \left[ \left(2 k_\mu k_\nu - q'_\mu k_\nu\right) \epsilon_T^{\mu\nu}    + 2 (kq') \right] 2 D_+   
%%%%
+  k^\rho    \left[ \left(2 k'_\mu k_\nu + q'_\mu k'_\nu\right) \epsilon_T^{\mu\nu}    - 2 (k'q') \right]  2 D_+  
%%%%
%
\nonumber \\
&+& q'^\rho    \left[ 2k_\mu k'_\nu  \epsilon_T^{\mu\nu}   -  2 (kk') \right] D_-   
%%%%%%
 + \epsilon_T^{\rho \mu}  \left[ \left(-2 (k k') + (k'q') \right) k_{\mu}  - (kq') k'_{\mu}  \right] 2 D_+  \nonumber \\
& + &  \epsilon_T^{\rho \mu} \left[ (kk') q'_\mu - (k'q')  k_\mu   - (kq') k'_\mu \right]  2 D_- \Big\} \cos \phi 
\end{eqnarray}
$D^+$ and $D^-$ are defined as,
\begin{equation} 
D^+ = \frac{(kq')-(kq')}{2(k'q')(kq')}, \;\;\;\; \quad\quad\quad D^- = -\frac{(kq')+(kq')}{2(k'q')(kq')}
\end{equation}

We also calculate leading terms that restore electromagnetic gauge invariance. We introduce these expressions below.

\begin{eqnarray}
A_{UU}^{\mathcal{I} \prime} &=& A_{UU}^{\mathcal{I}} - 8D_{-}\Big[ (Pk')(k_{T} \cdot q'_{T}) + (Pk)(k'_{T} \cdot q'_{T}) - (kk')(P_{T} \cdot q'_{T})\Big] \cos \phi\\
B_{UU}^{\mathcal{I} \prime} &=& B_{UU}^{\mathcal{I}} - 4\xi D_{-}\Big[(k'\Delta)(k_{T} \cdot q'_{T}) + (k\Delta)(k'_{T} \cdot q'_{T}) - (kk') (q'_{T} \cdot \Delta_{T}) \Big]\cos\phi \nonumber \\
&-& 2\frac{t}{P^{+}}D_{+}\Big[ k^{\prime +}\Big(2(k_{T} \cdot k_{T}) - (k_{T} \cdot q'_{T}) - 2(kq')\Big) +k^{ +}\Big( 2(k'_{T} \cdot k'_{T}) + (k'_{T} \cdot q'_{T}) + 2(k'q')\Big)\Big]\cos \phi \nonumber \\
&-& 2\frac{t}{P^{+}}D_{-} \Big[ q^{\prime +}\Big(2(k_{T} \cdot k'_{T}) + 2(kk') \Big) + k^{\prime +}(k_{T} \cdot q'_{T}) + k^{+}(k'_{T} \cdot q'_{T})\Big] \cos \phi\\
C_{UU}^{\mathcal{I} \prime} &=& C_{UU}^{\mathcal{I}}+ 4 \frac{D_{-}}{P^{+}}\Big[-2(P_{T} \cdot q'_{T})(k_{T} \cdot k'_{T})\Delta^{+} + (P_{T} \cdot q'_{T})(k_{T} \cdot \Delta_{T})k^{\prime +} + (P_{T} \cdot q'_{T})(k'_{T} \cdot \Delta_{T})k^{+} \nonumber \\
&+& 2(k'_{T} \cdot q'_{T})(k_{T} \cdot P_{T})\Delta^{+} - 2(k'_{T} \cdot q'_{T})(k_{T} \cdot \Delta_{T})P^{+} - (q'_{T} \cdot \Delta_{T})(k_{T} \cdot P_{T})k^{\prime +} - (q'_{T} \cdot \Delta_{T})(P_{T} \cdot k'_{T})k^{+} \nonumber \\
&+& 2(q'_{T}\cdot \Delta_{T})(k_{T} \cdot k'_{T})P^{+} - (kk')(P_{T} \cdot q'_{T})\Delta^{+} + (kk')(q'_{T} \cdot \Delta_{T})P^{+}\Big]\cos \phi
\end{eqnarray}

%%%%%% Appendix BKM
%\section{BKM expressions for $A_{UU}^{\cal I}$, $B_{UU}^{\cal I}$ and
%$C_{UU}^{\cal I}$}
%\label{appa}
\vspace{1.0cm}
\noindent The coefficients $A_{UU}^{\cal I}$, $B_{UU}^{\cal I}$ and $C_{UU}^{\cal I}$ can be re-evalauted using the expressions in BKM \cite{Belitsky:2001ns},
%\subsection{ BKM'01 Coefficients \cite{Belitsky:2001ns}}
\begin{eqnarray}
A_{UU}^{\mathcal{I}} &=& \frac{1}{x_{Bj}y^{3}t \mathcal{P}_{1}(\phi) \mathcal{P}_{2}(\phi)}\Big[ -8\frac{(2-y)^{3}}{1-y}K^{2} 
- 8(2-y)\frac{t}{Q^{2}}(1-y)(2-x_{Bj}) 
- 8K(2-2y+y^{2})\cos{\phi} \Big] 
\\
B_{UU}^{\mathcal{I}} &=& \frac{\xi^{2}}{x_{Bj}y^{3}t \mathcal{P}_{1}(\phi) \mathcal{P}_{2}\phi}\Bigg[8(2-y)\frac{t}{Q^{2}}(1-y)(2-x_{Bj}) \Bigg] 
\\
C_{UU}^{\mathcal{I}} &=& \frac{\xi}{x_{Bj}y^{3}t \mathcal{P}_{1}(\phi) \mathcal{P}_{2}(\phi)} \Big[ 8\frac{(2-y)^{3}}{(1-y)}K^{2} 
+ 8 K(2-y+y^{2})\cos{\phi} \Big]
\end{eqnarray}
where,
\begin{eqnarray}
\mathcal{P}_{1}(\phi) &=& -\frac{1}{y(1+\gamma^{2})}\{J + 2K\cos{\phi} \} , \quad\quad
\mathcal{P}_{2}(\phi) = 1 + \frac{t}{Q^{2}} + \frac{1}{y(1+\gamma^{2})}\{J + 2K\cos{\phi} \} \\
K^{2} &=& -\frac{t}{Q^{2}}(1-x_{Bj})\Big(1-y-\frac{y^{2}\gamma^{2}}{4}\Big)\Big(1 - \frac{t_{0}}{t} \Big)\Big(\sqrt{1+\gamma^{2}} 
+ \frac{4x_{Bj}(1-x_{Bj})+\gamma^{2}}{4(1-x_{Bj})}\frac{t-t_{0}}{Q^{2}} \Big) \\
J &= &\Big(1-y-\frac{y\gamma^{2}}{2} \Big)\Big(1 + \frac{t}{Q^{2}} \Big)- (1-x_{Bj})(2-y)\frac{t}{Q^{2}} 
\end{eqnarray}
with $y=Q^2/(x_{Bj}(s-M^2))$, $\gamma^2=4M^2x_{Bj}^2/Q^2$, $t_o= - 2\xi  M^2/(1-\xi^2)$, $\xi = x_{Bj}/(2-x_{Bj})$. 

\vspace{1.0cm}
\noindent The updated ``BKM'10" expressions accounting for $1/Q^2$ power corrections are given in the following form \cite{Belitsky:2010jw},
\begin{eqnarray}
A_{UU}^{\mathcal{I}} &=& \frac{1}{x_{Bj}y^{3}t \mathcal{P}_{1}(\phi) \mathcal{P}_{2}(\phi)} \Bigg\{ \sum_{n = 0}^{3} C_{++}^{\text{unp}}(n)\cos{(n\phi)} \Bigg\} \\
B_{UU}^{\mathcal{I}} &=& \frac{\xi}{x_{Bj}y^{3}t \mathcal{P}_{1}(\phi) \mathcal{P}_{2}(\phi)} \Bigg\{  \sum_{n = 0}^{3} C_{++}^{\text{unp},V}(n)\cos{(n\phi)} \Bigg\} \\
C_{UU}^{\mathcal{I}} &=& \frac{-\xi}{x_{Bj}y^{3}t \mathcal{P}_{1}(\phi) \mathcal{P}_{2}(\phi)} \Bigg\{ \sum_{n = 0}^{3} \Big( C_{+ + 0}^{\text{unp,A}}(n) + C_{++}^{\text{unp}}(n) \Big)\cos{(n\phi)}\Bigg\}
\end{eqnarray}
where the expressions for $C_{++}^{\rm unp}(n)$, $C_{++}^{\rm unp, V}(n)$, $C_{++ 0}^{\rm unp, A}(n)$ can be found in Ref.\cite{Belitsky:2010jw}. 

The coefficients for the longitudinally polarized electron beam, Eq.(\ref{eq:sigLUI}), are the same as for the unpolarized case:  $A_{LU}^{\cal I} = A_{UU}^{\cal I}$, $B_{LU}^{\cal I} = B_{UU}^{\cal I}$, and $C_{LU}^{\cal I} = C_{UU}^{\cal I}$. 
\end{widetext}

\bibliography{DVCS_BH_bib}

\begin{thebibliography}{54}%
\makeatletter
\providecommand \@ifxundefined [1]{%
 \@ifx{#1\undefined}
}%
\providecommand \@ifnum [1]{%
 \ifnum #1\expandafter \@firstoftwo
 \else \expandafter \@secondoftwo
 \fi
}%
\providecommand \@ifx [1]{%
 \ifx #1\expandafter \@firstoftwo
 \else \expandafter \@secondoftwo
 \fi
}%
\providecommand \natexlab [1]{#1}%
\providecommand \enquote  [1]{``#1''}%
\providecommand \bibnamefont  [1]{#1}%
\providecommand \bibfnamefont [1]{#1}%
\providecommand \citenamefont [1]{#1}%
\providecommand \href@noop [0]{\@secondoftwo}%
\providecommand \href [0]{\begingroup \@sanitize@url \@href}%
\providecommand \@href[1]{\@@startlink{#1}\@@href}%
\providecommand \@@href[1]{\endgroup#1\@@endlink}%
\providecommand \@sanitize@url [0]{\catcode `\\12\catcode `\$12\catcode
  `\&12\catcode `\#12\catcode `\^12\catcode `\_12\catcode `\%12\relax}%
\providecommand \@@startlink[1]{}%
\providecommand \@@endlink[0]{}%
\providecommand \url  [0]{\begingroup\@sanitize@url \@url }%
\providecommand \@url [1]{\endgroup\@href {#1}{\urlprefix }}%
\providecommand \urlprefix  [0]{URL }%
\providecommand \Eprint [0]{\href }%
\providecommand \doibase [0]{http://dx.doi.org/}%
\providecommand \selectlanguage [0]{\@gobble}%
\providecommand \bibinfo  [0]{\@secondoftwo}%
\providecommand \bibfield  [0]{\@secondoftwo}%
\providecommand \translation [1]{[#1]}%
\providecommand \BibitemOpen [0]{}%
\providecommand \bibitemStop [0]{}%
\providecommand \bibitemNoStop [0]{.\EOS\space}%
\providecommand \EOS [0]{\spacefactor3000\relax}%
\providecommand \BibitemShut  [1]{\csname bibitem#1\endcsname}%
\let\auto@bib@innerbib\@empty
%</preamble>
\bibitem [{\citenamefont {Ji}(1997{\natexlab{a}})}]{Ji:1996ek}%
  \BibitemOpen
  \bibfield  {author} {\bibinfo {author} {\bibfnamefont {X.-D.}\ \bibnamefont
  {Ji}},\ }\href {\doibase 10.1103/PhysRevLett.78.610} {\bibfield  {journal}
  {\bibinfo  {journal} {Phys. Rev. Lett.}\ }\textbf {\bibinfo {volume} {78}},\
  \bibinfo {pages} {610} (\bibinfo {year} {1997}{\natexlab{a}})},\ \Eprint
  {http://arxiv.org/abs/hep-ph/9603249} {arXiv:hep-ph/9603249 [hep-ph]}
  \BibitemShut {NoStop}%
%%CITATION = HEP-PH/9603249;%%
\bibitem [{\citenamefont {Ji}(1997{\natexlab{b}})}]{Ji:1996nm}%
  \BibitemOpen
  \bibfield  {author} {\bibinfo {author} {\bibfnamefont {X.-D.}\ \bibnamefont
  {Ji}},\ }\href {\doibase 10.1103/PhysRevD.55.7114} {\bibfield  {journal}
  {\bibinfo  {journal} {Phys. Rev. D}\ }\textbf {\bibinfo {volume} {55}},\
  \bibinfo {pages} {7114} (\bibinfo {year} {1997}{\natexlab{b}})},\ \Eprint
  {http://arxiv.org/abs/hep-ph/9609381} {arXiv:hep-ph/9609381} \BibitemShut
  {NoStop}%
\bibitem [{\citenamefont {Collins}\ and\ \citenamefont
  {Freund}(1999)}]{Collins:1998be}%
  \BibitemOpen
  \bibfield  {author} {\bibinfo {author} {\bibfnamefont {J.~C.}\ \bibnamefont
  {Collins}}\ and\ \bibinfo {author} {\bibfnamefont {A.}~\bibnamefont
  {Freund}},\ }\href {\doibase 10.1103/PhysRevD.59.074009} {\bibfield
  {journal} {\bibinfo  {journal} {Phys. Rev.}\ }\textbf {\bibinfo {volume}
  {D59}},\ \bibinfo {pages} {074009} (\bibinfo {year} {1999})},\ \Eprint
  {http://arxiv.org/abs/hep-ph/9801262} {arXiv:hep-ph/9801262 [hep-ph]}
  \BibitemShut {NoStop}%
%%CITATION = HEP-PH/9801262;%%
\bibitem [{\citenamefont {Ji}\ and\ \citenamefont
  {Osborne}(1998{\natexlab{a}})}]{Ji:1997nk}%
  \BibitemOpen
  \bibfield  {author} {\bibinfo {author} {\bibfnamefont {X.-D.}\ \bibnamefont
  {Ji}}\ and\ \bibinfo {author} {\bibfnamefont {J.}~\bibnamefont {Osborne}},\
  }\href {\doibase 10.1103/PhysRevD.57.R1337} {\bibfield  {journal} {\bibinfo
  {journal} {Phys. Rev. D}\ }\textbf {\bibinfo {volume} {57}},\ \bibinfo
  {pages} {1337} (\bibinfo {year} {1998}{\natexlab{a}})},\ \Eprint
  {http://arxiv.org/abs/hep-ph/9707254} {arXiv:hep-ph/9707254} \BibitemShut
  {NoStop}%
\bibitem [{\citenamefont {Ji}\ and\ \citenamefont
  {Osborne}(1998{\natexlab{b}})}]{Ji:1998xh}%
  \BibitemOpen
  \bibfield  {author} {\bibinfo {author} {\bibfnamefont {X.-D.}\ \bibnamefont
  {Ji}}\ and\ \bibinfo {author} {\bibfnamefont {J.}~\bibnamefont {Osborne}},\
  }\href {\doibase 10.1103/PhysRevD.58.094018} {\bibfield  {journal} {\bibinfo
  {journal} {Phys. Rev. D}\ }\textbf {\bibinfo {volume} {58}},\ \bibinfo
  {pages} {094018} (\bibinfo {year} {1998}{\natexlab{b}})},\ \Eprint
  {http://arxiv.org/abs/hep-ph/9801260} {arXiv:hep-ph/9801260} \BibitemShut
  {NoStop}%
\bibitem [{\citenamefont {Diehl}(2003)}]{Diehl:2003ny}%
  \BibitemOpen
  \bibfield  {author} {\bibinfo {author} {\bibfnamefont {M.}~\bibnamefont
  {Diehl}},\ }\href {\doibase 10.1016/j.physrep.2003.08.002} {\bibfield
  {journal} {\bibinfo  {journal} {Phys. Rept.}\ }\textbf {\bibinfo {volume}
  {388}},\ \bibinfo {pages} {41} (\bibinfo {year} {2003})},\ \Eprint
  {http://arxiv.org/abs/hep-ph/0307382} {arXiv:hep-ph/0307382 [hep-ph]}
  \BibitemShut {NoStop}%
%%CITATION = HEP-PH/0307382;%%
\bibitem [{\citenamefont {Belitsky}\ and\ \citenamefont
  {Radyushkin}(2005)}]{Belitsky:2005qn}%
  \BibitemOpen
  \bibfield  {author} {\bibinfo {author} {\bibfnamefont {A.}~\bibnamefont
  {Belitsky}}\ and\ \bibinfo {author} {\bibfnamefont {A.}~\bibnamefont
  {Radyushkin}},\ }\href {\doibase 10.1016/j.physrep.2005.06.002} {\bibfield
  {journal} {\bibinfo  {journal} {Phys.Rept.}\ }\textbf {\bibinfo {volume}
  {418}},\ \bibinfo {pages} {1} (\bibinfo {year} {2005})},\ \Eprint
  {http://arxiv.org/abs/hep-ph/0504030} {arXiv:hep-ph/0504030 [hep-ph]}
  \BibitemShut {NoStop}%
%CITATION = HEP-PH/0504030;%%
\bibitem [{\citenamefont {Kumericki}\ \emph {et~al.}(2016)\citenamefont
  {Kumericki}, \citenamefont {Liuti},\ and\ \citenamefont
  {Moutarde}}]{Kumericki:2016ehc}%
  \BibitemOpen
  \bibfield  {author} {\bibinfo {author} {\bibfnamefont {K.}~\bibnamefont
  {Kumericki}}, \bibinfo {author} {\bibfnamefont {S.}~\bibnamefont {Liuti}}, \
  and\ \bibinfo {author} {\bibfnamefont {H.}~\bibnamefont {Moutarde}},\ }\href
  {\doibase 10.1140/epja/i2016-16157-3} {\bibfield  {journal} {\bibinfo
  {journal} {Eur. Phys. J.}\ }\textbf {\bibinfo {volume} {A52}},\ \bibinfo
  {pages} {157} (\bibinfo {year} {2016})},\ \Eprint
  {http://arxiv.org/abs/1602.02763} {arXiv:1602.02763 [hep-ph]} \BibitemShut
  {NoStop}%
%%CITATION = ARXIV:1602.02763;%%
\bibitem [{\citenamefont {Kriesten}\ \emph {et~al.}(2020)\citenamefont
  {Kriesten}, \citenamefont {Liuti}, \citenamefont {Calero-Diaz}, \citenamefont
  {Keller}, \citenamefont {Meyer}, \citenamefont {Goldstein},\ and\
  \citenamefont {Osvaldo Gonzalez-Hernandez}}]{Kriesten:2019jep}%
  \BibitemOpen
  \bibfield  {author} {\bibinfo {author} {\bibfnamefont {B.}~\bibnamefont
  {Kriesten}}, \bibinfo {author} {\bibfnamefont {S.}~\bibnamefont {Liuti}},
  \bibinfo {author} {\bibfnamefont {L.}~\bibnamefont {Calero-Diaz}}, \bibinfo
  {author} {\bibfnamefont {D.}~\bibnamefont {Keller}}, \bibinfo {author}
  {\bibfnamefont {A.}~\bibnamefont {Meyer}}, \bibinfo {author} {\bibfnamefont
  {G.~R.}\ \bibnamefont {Goldstein}}, \ and\ \bibinfo {author} {\bibfnamefont
  {J.}~\bibnamefont {Osvaldo Gonzalez-Hernandez}},\ }\href {\doibase
  10.1103/PhysRevD.101.054021} {\bibfield  {journal} {\bibinfo  {journal}
  {Phys. Rev. D}\ }\textbf {\bibinfo {volume} {101}},\ \bibinfo {pages}
  {054021} (\bibinfo {year} {2020})},\ \Eprint
  {http://arxiv.org/abs/1903.05742} {arXiv:1903.05742 [hep-ph]} \BibitemShut
  {NoStop}%
\bibitem [{\citenamefont {Perdrisat}\ \emph {et~al.}(2007)\citenamefont
  {Perdrisat}, \citenamefont {Punjabi},\ and\ \citenamefont
  {Vanderhaeghen}}]{Perdrisat:2006hj}%
  \BibitemOpen
  \bibfield  {author} {\bibinfo {author} {\bibfnamefont {C.}~\bibnamefont
  {Perdrisat}}, \bibinfo {author} {\bibfnamefont {V.}~\bibnamefont {Punjabi}},
  \ and\ \bibinfo {author} {\bibfnamefont {M.}~\bibnamefont {Vanderhaeghen}},\
  }\href {\doibase 10.1016/j.ppnp.2007.05.001} {\bibfield  {journal} {\bibinfo
  {journal} {Prog. Part. Nucl. Phys.}\ }\textbf {\bibinfo {volume} {59}},\
  \bibinfo {pages} {694} (\bibinfo {year} {2007})},\ \Eprint
  {http://arxiv.org/abs/hep-ph/0612014} {arXiv:hep-ph/0612014} \BibitemShut
  {NoStop}%
\bibitem [{\citenamefont {Gao}(2003)}]{Gao:2003ag}%
  \BibitemOpen
  \bibfield  {author} {\bibinfo {author} {\bibfnamefont {H.-y.}\ \bibnamefont
  {Gao}},\ }\href {\doibase 10.1142/S021830130300117X} {\bibfield  {journal}
  {\bibinfo  {journal} {Int. J. Mod. Phys.}\ }\textbf {\bibinfo {volume}
  {E12}},\ \bibinfo {pages} {1} (\bibinfo {year} {2003})},\ \bibinfo {note}
  {[Erratum: Int. J. Mod. Phys.E12,567(2003)]},\ \Eprint
  {http://arxiv.org/abs/nucl-ex/0301002} {arXiv:nucl-ex/0301002 [nucl-ex]}
  \BibitemShut {NoStop}%
%%CITATION = NUCL-EX/0301002;%%
\bibitem [{\citenamefont {Arens}\ \emph {et~al.}(1997)\citenamefont {Arens},
  \citenamefont {Nachtmann}, \citenamefont {Diehl},\ and\ \citenamefont
  {Landshoff}}]{Arens:1996xw}%
  \BibitemOpen
  \bibfield  {author} {\bibinfo {author} {\bibfnamefont {T.}~\bibnamefont
  {Arens}}, \bibinfo {author} {\bibfnamefont {O.}~\bibnamefont {Nachtmann}},
  \bibinfo {author} {\bibfnamefont {M.}~\bibnamefont {Diehl}}, \ and\ \bibinfo
  {author} {\bibfnamefont {P.~V.}\ \bibnamefont {Landshoff}},\ }\href {\doibase
  10.1007/s002880050430} {\bibfield  {journal} {\bibinfo  {journal} {Z. Phys.}\
  }\textbf {\bibinfo {volume} {C74}},\ \bibinfo {pages} {651} (\bibinfo {year}
  {1997})},\ \Eprint {http://arxiv.org/abs/hep-ph/9605376}
  {arXiv:hep-ph/9605376 [hep-ph]} \BibitemShut {NoStop}%
%%CITATION = HEP-PH/9605376;%%
\bibitem [{\citenamefont {Diehl}\ and\ \citenamefont
  {Sapeta}(2005)}]{Diehl:2005pc}%
  \BibitemOpen
  \bibfield  {author} {\bibinfo {author} {\bibfnamefont {M.}~\bibnamefont
  {Diehl}}\ and\ \bibinfo {author} {\bibfnamefont {S.}~\bibnamefont {Sapeta}},\
  }\href {\doibase 10.1140/epjc/s2005-02242-9} {\bibfield  {journal} {\bibinfo
  {journal} {Eur. Phys. J.}\ }\textbf {\bibinfo {volume} {C41}},\ \bibinfo
  {pages} {515} (\bibinfo {year} {2005})},\ \Eprint
  {http://arxiv.org/abs/hep-ph/0503023} {arXiv:hep-ph/0503023 [hep-ph]}
  \BibitemShut {NoStop}%
%%CITATION = HEP-PH/0503023;%%
\bibitem [{\citenamefont {Belitsky}\ \emph {et~al.}(2001)\citenamefont
  {Belitsky}, \citenamefont {Mueller}, \citenamefont {Niedermeier},\ and\
  \citenamefont {Schafer}}]{Belitsky:2000gz}%
  \BibitemOpen
  \bibfield  {author} {\bibinfo {author} {\bibfnamefont {A.~V.}\ \bibnamefont
  {Belitsky}}, \bibinfo {author} {\bibfnamefont {D.}~\bibnamefont {Mueller}},
  \bibinfo {author} {\bibfnamefont {L.}~\bibnamefont {Niedermeier}}, \ and\
  \bibinfo {author} {\bibfnamefont {A.}~\bibnamefont {Schafer}},\ }\href
  {\doibase 10.1016/S0550-3213(00)00588-5} {\bibfield  {journal} {\bibinfo
  {journal} {Nucl. Phys. B}\ }\textbf {\bibinfo {volume} {593}},\ \bibinfo
  {pages} {289} (\bibinfo {year} {2001})},\ \Eprint
  {http://arxiv.org/abs/hep-ph/0004059} {arXiv:hep-ph/0004059} \BibitemShut
  {NoStop}%
\bibitem [{\citenamefont {Belitsky}\ \emph {et~al.}(2002)\citenamefont
  {Belitsky}, \citenamefont {Mueller},\ and\ \citenamefont
  {Kirchner}}]{Belitsky:2001ns}%
  \BibitemOpen
  \bibfield  {author} {\bibinfo {author} {\bibfnamefont {A.~V.}\ \bibnamefont
  {Belitsky}}, \bibinfo {author} {\bibfnamefont {D.}~\bibnamefont {Mueller}}, \
  and\ \bibinfo {author} {\bibfnamefont {A.}~\bibnamefont {Kirchner}},\ }\href
  {\doibase 10.1016/S0550-3213(02)00144-X} {\bibfield  {journal} {\bibinfo
  {journal} {Nucl. Phys.}\ }\textbf {\bibinfo {volume} {B629}},\ \bibinfo
  {pages} {323} (\bibinfo {year} {2002})},\ \Eprint
  {http://arxiv.org/abs/hep-ph/0112108} {arXiv:hep-ph/0112108 [hep-ph]}
  \BibitemShut {NoStop}%
%%CITATION = HEP-PH/0112108;%%
\bibitem [{\citenamefont {Belitsky}\ and\ \citenamefont
  {Mueller}(2010)}]{Belitsky:2010jw}%
  \BibitemOpen
  \bibfield  {author} {\bibinfo {author} {\bibfnamefont {A.~V.}\ \bibnamefont
  {Belitsky}}\ and\ \bibinfo {author} {\bibfnamefont {D.}~\bibnamefont
  {Mueller}},\ }\href {\doibase 10.1103/PhysRevD.82.074010} {\bibfield
  {journal} {\bibinfo  {journal} {Phys. Rev.}\ }\textbf {\bibinfo {volume}
  {D82}},\ \bibinfo {pages} {074010} (\bibinfo {year} {2010})},\ \Eprint
  {http://arxiv.org/abs/1005.5209} {arXiv:1005.5209 [hep-ph]} \BibitemShut
  {NoStop}%
%%CITATION = ARXIV:1005.5209;%%
\bibitem [{\citenamefont {Belitsky}\ \emph {et~al.}(2014)\citenamefont
  {Belitsky}, \citenamefont {Müller},\ and\ \citenamefont
  {Ji}}]{Belitsky:2012ch}%
  \BibitemOpen
  \bibfield  {author} {\bibinfo {author} {\bibfnamefont {A.~V.}\ \bibnamefont
  {Belitsky}}, \bibinfo {author} {\bibfnamefont {D.}~\bibnamefont {Müller}}, \
  and\ \bibinfo {author} {\bibfnamefont {Y.}~\bibnamefont {Ji}},\ }\href
  {\doibase 10.1016/j.nuclphysb.2013.11.014} {\bibfield  {journal} {\bibinfo
  {journal} {Nucl. Phys.}\ }\textbf {\bibinfo {volume} {B878}},\ \bibinfo
  {pages} {214} (\bibinfo {year} {2014})},\ \Eprint
  {http://arxiv.org/abs/1212.6674} {arXiv:1212.6674 [hep-ph]} \BibitemShut
  {NoStop}%
%%CITATION = ARXIV:1212.6674;%%
\bibitem [{\citenamefont {Burkardt}\ \emph {et~al.}(2010)\citenamefont
  {Burkardt}, \citenamefont {Miller},\ and\ \citenamefont
  {Nowak}}]{Burkardt:2008jw}%
  \BibitemOpen
  \bibfield  {author} {\bibinfo {author} {\bibfnamefont {M.}~\bibnamefont
  {Burkardt}}, \bibinfo {author} {\bibfnamefont {C.}~\bibnamefont {Miller}}, \
  and\ \bibinfo {author} {\bibfnamefont {W.}~\bibnamefont {Nowak}},\ }\href
  {\doibase 10.1088/0034-4885/73/1/016201} {\bibfield  {journal} {\bibinfo
  {journal} {Rept.Prog.Phys.}\ }\textbf {\bibinfo {volume} {73}},\ \bibinfo
  {pages} {016201} (\bibinfo {year} {2010})},\ \Eprint
  {http://arxiv.org/abs/0812.2208} {arXiv:0812.2208 [hep-ph]} \BibitemShut
  {NoStop}%
%%CITATION = ARXIV:0812.2208;%%
\bibitem [{\citenamefont {Kriesten}\ and\ \citenamefont
  {Liuti}(2020)}]{Kriesten:2020apm}%
  \BibitemOpen
  \bibfield  {author} {\bibinfo {author} {\bibfnamefont {B.}~\bibnamefont
  {Kriesten}}\ and\ \bibinfo {author} {\bibfnamefont {S.}~\bibnamefont
  {Liuti}},\ }\href@noop {} {\  (\bibinfo {year} {2020})},\ \Eprint
  {http://arxiv.org/abs/2011.04484} {arXiv:2011.04484 [hep-ph]} \BibitemShut
  {NoStop}%
\bibitem [{\citenamefont {Guichon}\ and\ \citenamefont
  {Vanderhaeghen}(1998)}]{Guichon:1998xv}%
  \BibitemOpen
  \bibfield  {author} {\bibinfo {author} {\bibfnamefont {P.~A.~M.}\
  \bibnamefont {Guichon}}\ and\ \bibinfo {author} {\bibfnamefont
  {M.}~\bibnamefont {Vanderhaeghen}},\ }\href {\doibase
  10.1016/S0146-6410(98)00056-8} {\bibfield  {journal} {\bibinfo  {journal}
  {Prog. Part. Nucl. Phys.}\ }\textbf {\bibinfo {volume} {41}},\ \bibinfo
  {pages} {125} (\bibinfo {year} {1998})},\ \Eprint
  {http://arxiv.org/abs/hep-ph/9806305} {arXiv:hep-ph/9806305 [hep-ph]}
  \BibitemShut {NoStop}%
%%CITATION = HEP-PH/9806305;%%
\bibitem [{\citenamefont {Vanderhaeghen}\ \emph {et~al.}(1998)\citenamefont
  {Vanderhaeghen}, \citenamefont {Guichon},\ and\ \citenamefont
  {Guidal}}]{Vanderhaeghen:1998uc}%
  \BibitemOpen
  \bibfield  {author} {\bibinfo {author} {\bibfnamefont {M.}~\bibnamefont
  {Vanderhaeghen}}, \bibinfo {author} {\bibfnamefont {P.~A.}\ \bibnamefont
  {Guichon}}, \ and\ \bibinfo {author} {\bibfnamefont {M.}~\bibnamefont
  {Guidal}},\ }\href {\doibase 10.1103/PhysRevLett.80.5064} {\bibfield
  {journal} {\bibinfo  {journal} {Phys. Rev. Lett.}\ }\textbf {\bibinfo
  {volume} {80}},\ \bibinfo {pages} {5064} (\bibinfo {year}
  {1998})}\BibitemShut {NoStop}%
\bibitem [{\citenamefont {Vanderhaeghen}\ \emph {et~al.}(1999)\citenamefont
  {Vanderhaeghen}, \citenamefont {Guichon},\ and\ \citenamefont
  {Guidal}}]{Vanderhaeghen:1999cas}%
  \BibitemOpen
  \bibfield  {author} {\bibinfo {author} {\bibfnamefont {M.}~\bibnamefont
  {Vanderhaeghen}}, \bibinfo {author} {\bibfnamefont {P.}~\bibnamefont
  {Guichon}}, \ and\ \bibinfo {author} {\bibfnamefont {M.}~\bibnamefont
  {Guidal}},\ }\href {\doibase 10.1016/S0375-9474(00)88509-7} {\bibfield
  {journal} {\bibinfo  {journal} {Nucl. Phys. A}\ }\textbf {\bibinfo {volume}
  {654}},\ \bibinfo {pages} {602c} (\bibinfo {year} {1999})}\BibitemShut
  {NoStop}%
\bibitem [{\citenamefont {Braun}\ and\ \citenamefont
  {Manashov}(2012)}]{Braun:2011dg}%
  \BibitemOpen
  \bibfield  {author} {\bibinfo {author} {\bibfnamefont {V.}~\bibnamefont
  {Braun}}\ and\ \bibinfo {author} {\bibfnamefont {A.}~\bibnamefont
  {Manashov}},\ }\href {\doibase 10.1007/JHEP01(2012)085} {\bibfield  {journal}
  {\bibinfo  {journal} {JHEP}\ }\textbf {\bibinfo {volume} {01}},\ \bibinfo
  {pages} {085} (\bibinfo {year} {2012})},\ \Eprint
  {http://arxiv.org/abs/1111.6765} {arXiv:1111.6765 [hep-ph]} \BibitemShut
  {NoStop}%
\bibitem [{\citenamefont {Braun}\ \emph {et~al.}(2012)\citenamefont {Braun},
  \citenamefont {Manashov},\ and\ \citenamefont {Pirnay}}]{Braun:2012hq}%
  \BibitemOpen
  \bibfield  {author} {\bibinfo {author} {\bibfnamefont {V.}~\bibnamefont
  {Braun}}, \bibinfo {author} {\bibfnamefont {A.}~\bibnamefont {Manashov}}, \
  and\ \bibinfo {author} {\bibfnamefont {B.}~\bibnamefont {Pirnay}},\ }\href
  {\doibase 10.1103/PhysRevLett.109.242001} {\bibfield  {journal} {\bibinfo
  {journal} {Phys. Rev. Lett.}\ }\textbf {\bibinfo {volume} {109}},\ \bibinfo
  {pages} {242001} (\bibinfo {year} {2012})},\ \Eprint
  {http://arxiv.org/abs/1209.2559} {arXiv:1209.2559 [hep-ph]} \BibitemShut
  {NoStop}%
\bibitem [{\citenamefont {Braun}\ \emph {et~al.}(2014)\citenamefont {Braun},
  \citenamefont {Manashov}, \citenamefont {Müller},\ and\ \citenamefont
  {Pirnay}}]{Braun:2014sta}%
  \BibitemOpen
  \bibfield  {author} {\bibinfo {author} {\bibfnamefont {V.~M.}\ \bibnamefont
  {Braun}}, \bibinfo {author} {\bibfnamefont {A.~N.}\ \bibnamefont {Manashov}},
  \bibinfo {author} {\bibfnamefont {D.}~\bibnamefont {Müller}}, \ and\
  \bibinfo {author} {\bibfnamefont {B.~M.}\ \bibnamefont {Pirnay}},\ }\href
  {\doibase 10.1103/PhysRevD.89.074022} {\bibfield  {journal} {\bibinfo
  {journal} {Phys. Rev.}\ }\textbf {\bibinfo {volume} {D89}},\ \bibinfo {pages}
  {074022} (\bibinfo {year} {2014})},\ \Eprint {http://arxiv.org/abs/1401.7621}
  {arXiv:1401.7621 [hep-ph]} \BibitemShut {NoStop}%
%%CITATION = ARXIV:1401.7621;%%
\bibitem [{\citenamefont {Bacchetta}\ \emph {et~al.}(2007)\citenamefont
  {Bacchetta}, \citenamefont {Diehl}, \citenamefont {Goeke}, \citenamefont
  {Metz}, \citenamefont {Mulders} \emph {et~al.}}]{Bacchetta:2006tn}%
  \BibitemOpen
  \bibfield  {author} {\bibinfo {author} {\bibfnamefont {A.}~\bibnamefont
  {Bacchetta}}, \bibinfo {author} {\bibfnamefont {M.}~\bibnamefont {Diehl}},
  \bibinfo {author} {\bibfnamefont {K.}~\bibnamefont {Goeke}}, \bibinfo
  {author} {\bibfnamefont {A.}~\bibnamefont {Metz}}, \bibinfo {author}
  {\bibfnamefont {P.~J.}\ \bibnamefont {Mulders}},  \emph {et~al.},\ }\href
  {\doibase 10.1088/1126-6708/2007/02/093} {\bibfield  {journal} {\bibinfo
  {journal} {JHEP}\ }\textbf {\bibinfo {volume} {0702}},\ \bibinfo {pages}
  {093} (\bibinfo {year} {2007})},\ \Eprint
  {http://arxiv.org/abs/hep-ph/0611265} {arXiv:hep-ph/0611265 [hep-ph]}
  \BibitemShut {NoStop}%
%%CITATION = HEP-PH/0611265;%%
\bibitem [{\citenamefont {Bogasz}(2021)}]{Bogasz}%
  \BibitemOpen
  \bibfield  {author} {\bibinfo {author} {\bibfnamefont {A.}~\bibnamefont
  {Bogasz}},\ }\href
  {https://www.jlab.org/accelerator-seminar-alex-bogacz-remote} {\enquote
  {\bibinfo {title}
  {https://www.jlab.org/accelerator-seminar-alex-bogacz-remote},}\ } (\bibinfo
  {year} {2021})\BibitemShut {NoStop}%
\bibitem [{\citenamefont {Anderle}\ \emph {et~al.}(2021)\citenamefont {Anderle}
  \emph {et~al.}}]{Anderle:2021wcy}%
  \BibitemOpen
  \bibfield  {author} {\bibinfo {author} {\bibfnamefont {D.~P.}\ \bibnamefont
  {Anderle}} \emph {et~al.},\ }\href@noop {} {\  (\bibinfo {year} {2021})},\
  \Eprint {http://arxiv.org/abs/2102.09222} {arXiv:2102.09222 [nucl-ex]}
  \BibitemShut {NoStop}%
\bibitem [{\citenamefont {Khalek}\ \emph {et~al.}(2021)\citenamefont {Khalek}
  \emph {et~al.}}]{eicyellowreport}%
  \BibitemOpen
  \bibfield  {author} {\bibinfo {author} {\bibfnamefont {R.~A.}\ \bibnamefont
  {Khalek}} \emph {et~al.},\ }\href@noop {} {\  (\bibinfo {year} {2021})},\
  \Eprint {http://arxiv.org/abs/2103.05419} {arXiv:2103.05419 [hep-ph]}
  \BibitemShut {NoStop}%
\bibitem [{\citenamefont {Dmitrasinovic}\ and\ \citenamefont
  {Gross}(1989)}]{Dmitrasinovic:1989bf}%
  \BibitemOpen
  \bibfield  {author} {\bibinfo {author} {\bibfnamefont {V.}~\bibnamefont
  {Dmitrasinovic}}\ and\ \bibinfo {author} {\bibfnamefont {F.}~\bibnamefont
  {Gross}},\ }\href {\doibase 10.1103/PhysRevC.40.2479,
  10.1103/PhysRevC.43.1495} {\bibfield  {journal} {\bibinfo  {journal} {Phys.
  Rev.}\ }\textbf {\bibinfo {volume} {C40}},\ \bibinfo {pages} {2479} (\bibinfo
  {year} {1989})},\ \bibinfo {note} {[Erratum: Phys.
  Rev.C43,1495(1991)]}\BibitemShut {NoStop}%
%%CITATION = PHRVA,C40,2479;%%
\bibitem [{\citenamefont {Boffi}\ \emph {et~al.}(1993)\citenamefont {Boffi},
  \citenamefont {Giusti},\ and\ \citenamefont {Pacati}}]{Boffi:1993gs}%
  \BibitemOpen
  \bibfield  {author} {\bibinfo {author} {\bibfnamefont {S.}~\bibnamefont
  {Boffi}}, \bibinfo {author} {\bibfnamefont {C.}~\bibnamefont {Giusti}}, \
  and\ \bibinfo {author} {\bibfnamefont {F.~D.}\ \bibnamefont {Pacati}},\
  }\href {\doibase 10.1016/0370-1573(93)90132-W} {\bibfield  {journal}
  {\bibinfo  {journal} {Phys. Rept.}\ }\textbf {\bibinfo {volume} {226}},\
  \bibinfo {pages} {1} (\bibinfo {year} {1993})}\BibitemShut {NoStop}%
%%CITATION = PRPLC,226,1;%%
\bibitem [{\citenamefont {Golec-Biernat}\ and\ \citenamefont
  {Martin}(1999)}]{GolecBiernat:1998ja}%
  \BibitemOpen
  \bibfield  {author} {\bibinfo {author} {\bibfnamefont {K.~J.}\ \bibnamefont
  {Golec-Biernat}}\ and\ \bibinfo {author} {\bibfnamefont {A.~D.}\ \bibnamefont
  {Martin}},\ }\href {\doibase 10.1103/PhysRevD.59.014029} {\bibfield
  {journal} {\bibinfo  {journal} {Phys. Rev. D}\ }\textbf {\bibinfo {volume}
  {59}},\ \bibinfo {pages} {014029} (\bibinfo {year} {1999})},\ \Eprint
  {http://arxiv.org/abs/hep-ph/9807497} {arXiv:hep-ph/9807497} \BibitemShut
  {NoStop}%
\bibitem [{\citenamefont {Goldstein}\ \emph {et~al.}(2012)\citenamefont
  {Goldstein}, \citenamefont {Hernandez},\ and\ \citenamefont
  {Liuti}}]{Goldstein:2012az}%
  \BibitemOpen
  \bibfield  {author} {\bibinfo {author} {\bibfnamefont {G.~R.}\ \bibnamefont
  {Goldstein}}, \bibinfo {author} {\bibfnamefont {J.~O.~G.}\ \bibnamefont
  {Hernandez}}, \ and\ \bibinfo {author} {\bibfnamefont {S.}~\bibnamefont
  {Liuti}},\ }\href {\doibase 10.1088/0954-3899/39/11/115001} {\bibfield
  {journal} {\bibinfo  {journal} {J.Phys.}\ }\textbf {\bibinfo {volume}
  {G39}},\ \bibinfo {pages} {115001} (\bibinfo {year} {2012})},\ \Eprint
  {http://arxiv.org/abs/1201.6088} {arXiv:1201.6088 [hep-ph]} \BibitemShut
  {NoStop}%
%%CITATION = ARXIV:1201.6088;%%
\bibitem [{\citenamefont {Meissner}\ \emph {et~al.}(2009)\citenamefont
  {Meissner}, \citenamefont {Metz},\ and\ \citenamefont
  {Schlegel}}]{Meissner:2009ww}%
  \BibitemOpen
  \bibfield  {author} {\bibinfo {author} {\bibfnamefont {S.}~\bibnamefont
  {Meissner}}, \bibinfo {author} {\bibfnamefont {A.}~\bibnamefont {Metz}}, \
  and\ \bibinfo {author} {\bibfnamefont {M.}~\bibnamefont {Schlegel}},\ }\href
  {\doibase 10.1088/1126-6708/2009/08/056} {\bibfield  {journal} {\bibinfo
  {journal} {JHEP}\ }\textbf {\bibinfo {volume} {0908}},\ \bibinfo {pages}
  {056} (\bibinfo {year} {2009})},\ \Eprint {http://arxiv.org/abs/0906.5323}
  {arXiv:0906.5323 [hep-ph]} \BibitemShut {NoStop}%
%%CITATION = ARXIV:0906.5323;%%
\bibitem [{\citenamefont {Rajan}\ \emph {et~al.}(2018)\citenamefont {Rajan},
  \citenamefont {Engelhardt},\ and\ \citenamefont {Liuti}}]{Raja:2017xlo}%
  \BibitemOpen
  \bibfield  {author} {\bibinfo {author} {\bibfnamefont {A.}~\bibnamefont
  {Rajan}}, \bibinfo {author} {\bibfnamefont {M.}~\bibnamefont {Engelhardt}}, \
  and\ \bibinfo {author} {\bibfnamefont {S.}~\bibnamefont {Liuti}},\ }\href
  {\doibase 10.1103/PhysRevD.98.074022} {\bibfield  {journal} {\bibinfo
  {journal} {Phys. Rev.}\ }\textbf {\bibinfo {volume} {D98}},\ \bibinfo {pages}
  {074022} (\bibinfo {year} {2018})},\ \Eprint
  {http://arxiv.org/abs/1709.05770} {arXiv:1709.05770 [hep-ph]} \BibitemShut
  {NoStop}%
%%CITATION = ARXIV:1709.05770;%%
\bibitem [{\citenamefont {Rajan}\ \emph {et~al.}(2016)\citenamefont {Rajan},
  \citenamefont {Courtoy}, \citenamefont {Engelhardt},\ and\ \citenamefont
  {Liuti}}]{Rajan:2016tlg}%
  \BibitemOpen
  \bibfield  {author} {\bibinfo {author} {\bibfnamefont {A.}~\bibnamefont
  {Rajan}}, \bibinfo {author} {\bibfnamefont {A.}~\bibnamefont {Courtoy}},
  \bibinfo {author} {\bibfnamefont {M.}~\bibnamefont {Engelhardt}}, \ and\
  \bibinfo {author} {\bibfnamefont {S.}~\bibnamefont {Liuti}},\ }\href
  {\doibase 10.1103/PhysRevD.94.034041} {\bibfield  {journal} {\bibinfo
  {journal} {Phys. Rev.}\ }\textbf {\bibinfo {volume} {D94}},\ \bibinfo {pages}
  {034041} (\bibinfo {year} {2016})},\ \Eprint
  {http://arxiv.org/abs/1601.06117} {arXiv:1601.06117 [hep-ph]} \BibitemShut
  {NoStop}%
%%CITATION = ARXIV:1601.06117;%%
\bibitem [{\citenamefont {Jaffe}\ and\ \citenamefont
  {Ji}(1992)}]{Jaffe:1992ra}%
  \BibitemOpen
  \bibfield  {author} {\bibinfo {author} {\bibfnamefont {R.}~\bibnamefont
  {Jaffe}}\ and\ \bibinfo {author} {\bibfnamefont {X.-D.}\ \bibnamefont {Ji}},\
  }\href {\doibase 10.1016/0550-3213(92)90110-W} {\bibfield  {journal}
  {\bibinfo  {journal} {Nucl.Phys.}\ }\textbf {\bibinfo {volume} {B375}},\
  \bibinfo {pages} {527} (\bibinfo {year} {1992})}\BibitemShut {NoStop}%
%%CITATION = NUPHA,B375,527;%%
\bibitem [{\citenamefont {Mulders}\ and\ \citenamefont
  {Tangerman}(1996)}]{Mulders:1995dh}%
  \BibitemOpen
  \bibfield  {author} {\bibinfo {author} {\bibfnamefont {P.}~\bibnamefont
  {Mulders}}\ and\ \bibinfo {author} {\bibfnamefont {R.}~\bibnamefont
  {Tangerman}},\ }\href {\doibase 10.1016/0550-3213(95)00632-X} {\bibfield
  {journal} {\bibinfo  {journal} {Nucl. Phys.}\ }\textbf {\bibinfo {volume}
  {B461}},\ \bibinfo {pages} {197} (\bibinfo {year} {1996})},\ \Eprint
  {http://arxiv.org/abs/hep-ph/9510301} {arXiv:hep-ph/9510301 [hep-ph]}
  \BibitemShut {NoStop}%
%%CITATION = HEP-PH/9510301;%%
\bibitem [{\citenamefont {Donnelly}\ and\ \citenamefont
  {Raskin}(1986)}]{Donnelly:1985ry}%
  \BibitemOpen
  \bibfield  {author} {\bibinfo {author} {\bibfnamefont {T.~W.}\ \bibnamefont
  {Donnelly}}\ and\ \bibinfo {author} {\bibfnamefont {A.~S.}\ \bibnamefont
  {Raskin}},\ }\href {\doibase 10.1016/0003-4916(86)90173-9} {\bibfield
  {journal} {\bibinfo  {journal} {Annals Phys.}\ }\textbf {\bibinfo {volume}
  {169}},\ \bibinfo {pages} {247} (\bibinfo {year} {1986})}\BibitemShut
  {NoStop}%
%%CITATION = APNYA,169,247;%%
\bibitem [{\citenamefont {Gastmans}\ and\ \citenamefont {Wu}(1990)}]{Gastmans}%
  \BibitemOpen
  \bibfield  {author} {\bibinfo {author} {\bibfnamefont {R.}~\bibnamefont
  {Gastmans}}\ and\ \bibinfo {author} {\bibfnamefont {T.}~\bibnamefont {Wu}},\
  }\href@noop {} {\emph {\bibinfo {title} {The Ubiquitos Photon}}}\ (\bibinfo
  {publisher} {Clarendon Press},\ \bibinfo {address} {Oxford},\ \bibinfo {year}
  {1990})\BibitemShut {NoStop}%
\bibitem [{\citenamefont {Georges}(2018)}]{Georges:2018kyi}%
  \BibitemOpen
  \bibfield  {author} {\bibinfo {author} {\bibfnamefont {F.}~\bibnamefont
  {Georges}},\ }\emph {\bibinfo {title} {{Deeply virtual Compton scattering at
  Jefferson Lab}}},\ \href@noop {} {Ph.D. thesis},\ \bibinfo  {school}
  {Institut de Physique Nucléaire d'Orsay, France} (\bibinfo {year}
  {2018})\BibitemShut {NoStop}%
\bibitem [{\citenamefont {Gonzalez-Hernandez}\ \emph
  {et~al.}(2013)\citenamefont {Gonzalez-Hernandez}, \citenamefont {Liuti},
  \citenamefont {Goldstein},\ and\ \citenamefont
  {Kathuria}}]{GonzalezHernandez:2012jv}%
  \BibitemOpen
  \bibfield  {author} {\bibinfo {author} {\bibfnamefont {J.~O.}\ \bibnamefont
  {Gonzalez-Hernandez}}, \bibinfo {author} {\bibfnamefont {S.}~\bibnamefont
  {Liuti}}, \bibinfo {author} {\bibfnamefont {G.~R.}\ \bibnamefont
  {Goldstein}}, \ and\ \bibinfo {author} {\bibfnamefont {K.}~\bibnamefont
  {Kathuria}},\ }\href {\doibase 10.1103/PhysRevC.88.065206} {\bibfield
  {journal} {\bibinfo  {journal} {Phys. Rev.}\ }\textbf {\bibinfo {volume}
  {C88}},\ \bibinfo {pages} {065206} (\bibinfo {year} {2013})},\ \Eprint
  {http://arxiv.org/abs/1206.1876} {arXiv:1206.1876 [hep-ph]} \BibitemShut
  {NoStop}%
%%CITATION = ARXIV:1206.1876;%%
\bibitem [{\citenamefont {Kriesten}\ \emph {et~al.}(2021)\citenamefont
  {Kriesten}, \citenamefont {Velie}, \citenamefont {Yeats}, \citenamefont
  {Lopez},\ and\ \citenamefont {Liuti}}]{Kriesten:2021sqc}%
  \BibitemOpen
  \bibfield  {author} {\bibinfo {author} {\bibfnamefont {B.}~\bibnamefont
  {Kriesten}}, \bibinfo {author} {\bibfnamefont {P.}~\bibnamefont {Velie}},
  \bibinfo {author} {\bibfnamefont {E.}~\bibnamefont {Yeats}}, \bibinfo
  {author} {\bibfnamefont {F.~Y.}\ \bibnamefont {Lopez}}, \ and\ \bibinfo
  {author} {\bibfnamefont {S.}~\bibnamefont {Liuti}},\ }\href@noop {} {\
  (\bibinfo {year} {2021})},\ \Eprint {http://arxiv.org/abs/2101.01826}
  {arXiv:2101.01826 [hep-ph]} \BibitemShut {NoStop}%
\bibitem [{\citenamefont {Defurne}\ \emph {et~al.}(2015)\citenamefont {Defurne}
  \emph {et~al.}}]{Defurne:2015kxq}%
  \BibitemOpen
  \bibfield  {author} {\bibinfo {author} {\bibfnamefont {M.}~\bibnamefont
  {Defurne}} \emph {et~al.} (\bibinfo {collaboration} {Jefferson Lab Hall A}),\
  }\href {\doibase 10.1103/PhysRevC.92.055202} {\bibfield  {journal} {\bibinfo
  {journal} {Phys. Rev.}\ }\textbf {\bibinfo {volume} {C92}},\ \bibinfo {pages}
  {055202} (\bibinfo {year} {2015})},\ \Eprint
  {http://arxiv.org/abs/1504.05453} {arXiv:1504.05453 [nucl-ex]} \BibitemShut
  {NoStop}%
%%CITATION = ARXIV:1504.05453;%%
\bibitem [{\citenamefont {Defurne}\ \emph {et~al.}(2017)\citenamefont {Defurne}
  \emph {et~al.}}]{Defurne:2017paw}%
  \BibitemOpen
  \bibfield  {author} {\bibinfo {author} {\bibfnamefont {M.}~\bibnamefont
  {Defurne}} \emph {et~al.},\ }\href {\doibase 10.1038/s41467-017-01819-3}
  {\bibfield  {journal} {\bibinfo  {journal} {Nature Commun.}\ }\textbf
  {\bibinfo {volume} {8}},\ \bibinfo {pages} {1408} (\bibinfo {year} {2017})},\
  \Eprint {http://arxiv.org/abs/1703.09442} {arXiv:1703.09442 [hep-ex]}
  \BibitemShut {NoStop}%
\bibitem [{\citenamefont {Musatov}\ and\ \citenamefont
  {Radyushkin}(2000)}]{Musatov:1999xp}%
  \BibitemOpen
  \bibfield  {author} {\bibinfo {author} {\bibfnamefont {I.}~\bibnamefont
  {Musatov}}\ and\ \bibinfo {author} {\bibfnamefont {A.}~\bibnamefont
  {Radyushkin}},\ }\href {\doibase 10.1103/PhysRevD.61.074027} {\bibfield
  {journal} {\bibinfo  {journal} {Phys. Rev. D}\ }\textbf {\bibinfo {volume}
  {61}},\ \bibinfo {pages} {074027} (\bibinfo {year} {2000})},\ \Eprint
  {http://arxiv.org/abs/hep-ph/9905376} {arXiv:hep-ph/9905376} \BibitemShut
  {NoStop}%
\bibitem [{\citenamefont {Goldstein}\ \emph {et~al.}(2011)\citenamefont
  {Goldstein}, \citenamefont {Gonzalez-Hernandez},\ and\ \citenamefont
  {Liuti}}]{Goldstein:2010gu}%
  \BibitemOpen
  \bibfield  {author} {\bibinfo {author} {\bibfnamefont {G.~R.}\ \bibnamefont
  {Goldstein}}, \bibinfo {author} {\bibfnamefont {J.~O.}\ \bibnamefont
  {Gonzalez-Hernandez}}, \ and\ \bibinfo {author} {\bibfnamefont
  {S.}~\bibnamefont {Liuti}},\ }\href {\doibase 10.1103/PhysRevD.84.034007}
  {\bibfield  {journal} {\bibinfo  {journal} {Phys. Rev.}\ }\textbf {\bibinfo
  {volume} {D84}},\ \bibinfo {pages} {034007} (\bibinfo {year} {2011})},\
  \Eprint {http://arxiv.org/abs/1012.3776} {arXiv:1012.3776 [hep-ph]}
  \BibitemShut {NoStop}%
%%CITATION = ARXIV:1012.3776;%%
\bibitem [{\citenamefont {Kumericki}\ \emph {et~al.}(2011)\citenamefont
  {Kumericki}, \citenamefont {Lautenschlager}, \citenamefont {Mueller},
  \citenamefont {Passek-Kumericki}, \citenamefont {Schaefer},\ and\
  \citenamefont {Meskauskas}}]{Kumericki:2011zc}%
  \BibitemOpen
  \bibfield  {author} {\bibinfo {author} {\bibfnamefont {K.}~\bibnamefont
  {Kumericki}}, \bibinfo {author} {\bibfnamefont {T.}~\bibnamefont
  {Lautenschlager}}, \bibinfo {author} {\bibfnamefont {D.}~\bibnamefont
  {Mueller}}, \bibinfo {author} {\bibfnamefont {K.}~\bibnamefont
  {Passek-Kumericki}}, \bibinfo {author} {\bibfnamefont {A.}~\bibnamefont
  {Schaefer}}, \ and\ \bibinfo {author} {\bibfnamefont {M.}~\bibnamefont
  {Meskauskas}},\ }\href@noop {} {\  (\bibinfo {year} {2011})},\ \Eprint
  {http://arxiv.org/abs/1105.0899} {arXiv:1105.0899 [hep-ph]} \BibitemShut
  {NoStop}%
\bibitem [{\citenamefont {Cates}\ \emph {et~al.}(2011)\citenamefont {Cates},
  \citenamefont {de~Jager}, \citenamefont {Riordan},\ and\ \citenamefont
  {Wojtsekhowski}}]{Cates:2011pz}%
  \BibitemOpen
  \bibfield  {author} {\bibinfo {author} {\bibfnamefont {G.~D.}\ \bibnamefont
  {Cates}}, \bibinfo {author} {\bibfnamefont {C.~W.}\ \bibnamefont {de~Jager}},
  \bibinfo {author} {\bibfnamefont {S.}~\bibnamefont {Riordan}}, \ and\
  \bibinfo {author} {\bibfnamefont {B.}~\bibnamefont {Wojtsekhowski}},\ }\href
  {\doibase 10.1103/PhysRevLett.106.252003} {\bibfield  {journal} {\bibinfo
  {journal} {Phys. Rev. Lett.}\ }\textbf {\bibinfo {volume} {106}},\ \bibinfo
  {pages} {252003} (\bibinfo {year} {2011})},\ \Eprint
  {http://arxiv.org/abs/1103.1808} {arXiv:1103.1808 [nucl-ex]} \BibitemShut
  {NoStop}%
%%CITATION = ARXIV:1103.1808;%%
\bibitem [{\citenamefont {Schindler}\ and\ \citenamefont
  {Scherer}(2007)}]{Schindler:2006jq}%
  \BibitemOpen
  \bibfield  {author} {\bibinfo {author} {\bibfnamefont {M.}~\bibnamefont
  {Schindler}}\ and\ \bibinfo {author} {\bibfnamefont {S.}~\bibnamefont
  {Scherer}},\ }\href {\doibase 10.1140/epja/i2006-10403-3} {\bibfield
  {journal} {\bibinfo  {journal} {Eur. Phys. J. A}\ }\textbf {\bibinfo {volume}
  {32}},\ \bibinfo {pages} {429} (\bibinfo {year} {2007})},\ \Eprint
  {http://arxiv.org/abs/hep-ph/0608325} {arXiv:hep-ph/0608325} \BibitemShut
  {NoStop}%
\bibitem [{\citenamefont {Gorringe}\ and\ \citenamefont
  {Fearing}(2004)}]{Gorringe:2002xx}%
  \BibitemOpen
  \bibfield  {author} {\bibinfo {author} {\bibfnamefont {T.}~\bibnamefont
  {Gorringe}}\ and\ \bibinfo {author} {\bibfnamefont {H.~W.}\ \bibnamefont
  {Fearing}},\ }\href {\doibase 10.1103/RevModPhys.76.31} {\bibfield  {journal}
  {\bibinfo  {journal} {Rev. Mod. Phys.}\ }\textbf {\bibinfo {volume} {76}},\
  \bibinfo {pages} {31} (\bibinfo {year} {2004})},\ \Eprint
  {http://arxiv.org/abs/nucl-th/0206039} {arXiv:nucl-th/0206039} \BibitemShut
  {NoStop}%
\bibitem [{\citenamefont {Ahmad}\ \emph {et~al.}(2007)\citenamefont {Ahmad},
  \citenamefont {Honkanen}, \citenamefont {Liuti},\ and\ \citenamefont
  {Taneja}}]{Ahmad:2006gn}%
  \BibitemOpen
  \bibfield  {author} {\bibinfo {author} {\bibfnamefont {S.}~\bibnamefont
  {Ahmad}}, \bibinfo {author} {\bibfnamefont {H.}~\bibnamefont {Honkanen}},
  \bibinfo {author} {\bibfnamefont {S.}~\bibnamefont {Liuti}}, \ and\ \bibinfo
  {author} {\bibfnamefont {S.~K.}\ \bibnamefont {Taneja}},\ }\href {\doibase
  10.1103/PhysRevD.75.094003} {\bibfield  {journal} {\bibinfo  {journal} {Phys.
  Rev. D}\ }\textbf {\bibinfo {volume} {75}},\ \bibinfo {pages} {094003}
  (\bibinfo {year} {2007})},\ \Eprint {http://arxiv.org/abs/hep-ph/0611046}
  {arXiv:hep-ph/0611046} \BibitemShut {NoStop}%
\bibitem [{\citenamefont {Ahmad}\ \emph {et~al.}(2009)\citenamefont {Ahmad},
  \citenamefont {Honkanen}, \citenamefont {Liuti},\ and\ \citenamefont
  {Taneja}}]{Ahmad:2007vw}%
  \BibitemOpen
  \bibfield  {author} {\bibinfo {author} {\bibfnamefont {S.}~\bibnamefont
  {Ahmad}}, \bibinfo {author} {\bibfnamefont {H.}~\bibnamefont {Honkanen}},
  \bibinfo {author} {\bibfnamefont {S.}~\bibnamefont {Liuti}}, \ and\ \bibinfo
  {author} {\bibfnamefont {S.~K.}\ \bibnamefont {Taneja}},\ }\href {\doibase
  10.1140/epjc/s10052-009-1073-4} {\bibfield  {journal} {\bibinfo  {journal}
  {Eur. Phys. J. C}\ }\textbf {\bibinfo {volume} {63}},\ \bibinfo {pages} {407}
  (\bibinfo {year} {2009})},\ \Eprint {http://arxiv.org/abs/0708.0268}
  {arXiv:0708.0268 [hep-ph]} \BibitemShut {NoStop}%
\bibitem [{\citenamefont {Kumeri\v~cki}\ and\ \citenamefont
  {Müller}(2016)}]{Kumericki:2015lhb}%
  \BibitemOpen
  \bibfield  {author} {\bibinfo {author} {\bibfnamefont {K.~s.}\ \bibnamefont
  {Kumeri\v~cki}}\ and\ \bibinfo {author} {\bibfnamefont {D.}~\bibnamefont
  {Müller}},\ }\href {\doibase 10.1051/epjconf/201611201012} {\bibfield
  {journal} {\bibinfo  {journal} {EPJ Web Conf.}\ }\textbf {\bibinfo {volume}
  {112}},\ \bibinfo {pages} {01012} (\bibinfo {year} {2016})},\ \Eprint
  {http://arxiv.org/abs/1512.09014} {arXiv:1512.09014 [hep-ph]} \BibitemShut
  {NoStop}%
\end{thebibliography}%
\end{document}